\documentclass[conference,twoside,web]{ieeecolor}
\usepackage{xcolor}
\definecolor{nblue}{RGB}{0,0,128}  
\definecolor{nblue}{HTML}{000080}
\usepackage{jsen}
\usepackage{cite}
\usepackage{amsmath,amssymb,amsfonts}
\usepackage{algorithmic}
\usepackage{textcomp}
\usepackage{wrapfig}
\usepackage{fontawesome}
\usepackage{pifont}
\usepackage{hyperref}
\usepackage{adjustbox}
\usepackage{array}
\usepackage{multirow}
\usepackage{colortbl}
\usepackage{cuted}
\usepackage{times}

\def\BibTeX{{\rm B\kern-.05em{\sc i\kern-.025em b}\kern-.08em
    T\kern-.1667em\lower.7ex\hbox{E}\kern-.125emX}}
\definecolor{abstractbg}{rgb}{0.89804,0.94510,0.83137}
\begin{document}
\title{Reliability Assessment of Low-Cost PM Sensors under High Humidity and High PM Level Outdoor Conditions}
\author{Gulshan Kumar, Prasannaa Kumar D., Jay Dhariwal, and Seshan Srirangarajan
\thanks{This work was partly supported by the Industrial Research and Development (IRD) unit of IIT Delhi through research project no. MI02086G. }
\thanks{Gulshan Kumar, Prasannaa Kumar D., and Jay Dhariwal are with the Department of Design, Indian Institute of Technology Delhi, New Delhi, India. Seshan Srirangarajan is with the Department of Electrical Engineering, Indian Institute of Technology Delhi, New Delhi, India.}
\thanks{Corresponding Author: Seshan Srirangarajan (seshan@ee.iitd.ac.in)}
}

\IEEEtitleabstractindextext{%
\fcolorbox{abstractbg}{abstractbg}{%
\begin{minipage}{\textwidth}%
\begin{wrapfigure}[17]{r}{2.7in}%
\includegraphics[width=2.5in]{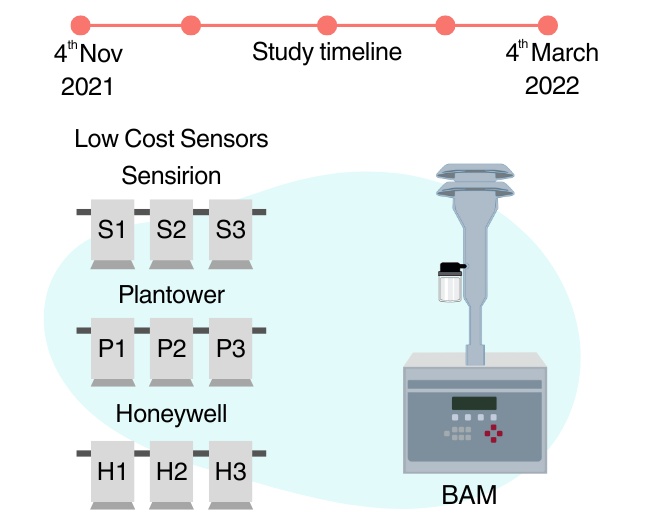}%
\end{wrapfigure}%
\begin{abstract}
Low-cost particulate matter (PM) sensors have become increasingly popular due to their compact size, low power consumption, and cost-effective installation and maintenance. While several studies have explored the effects of meteorological conditions and pollution exposure on low-cost sensor (LCS) performance, few have addressed the combined impact of high PM concentration and high humidity levels. In contrast to most evaluation studies, which generally report $\text{PM}_{2.5}$ levels below $150~\mu\text{g/m}^3$, our study observed hourly average $\text{PM}_{2.5}$ concentrations ranging from $6-611~\mu\text{g/m}^3$ (mean value of $137~\mu\text{g/m}^3$), with relative humidity between $25-95\%$ (mean value of $72\%$), and temperature varying from $6-29^\circ$C (mean value of $16^\circ$C). We evaluate three LCS models (SPS30, PMS7003, HPMA115C0-004) in outdoor conditions during the winter season in New Delhi, India, deployed alongside a reference-grade beta attenuation monitor (BAM). The results indicate a strong correlation between LCS and BAM measurements (${R^2} > 90\%$). The RMSE increases with increasing PM concentration and humidity levels but the narrow $95\%$ confidence interval range of LCS as a function of the reference BAM suggests the importance of LCS in air pollution monitoring. Among the evaluated LCS models, SPS30 showed the highest overall accuracy. Overall, the study demonstrates that LCS can effectively monitor air quality in regions with high PM and high humidity levels, provided appropriate correction models are applied.
\end{abstract}

\end{minipage}}}

\maketitle
\section{Introduction}
\IEEEPARstart{C}lean air is a fundamental right for all living beings, yet widespread air pollution has denied this right to a significant portion of the world's population. Air pollution has been recognized as the most serious environmental health concern globally with seven million or 10\% of all the premature annual fatalities attributed to it~\cite{Iqair2020}. In India, for instance, air pollution ranks among the top three causes of death~\cite{Pandey2019,Iqair2021}. Air pollution is also becoming a major threat due to the surge in wildfires attributed to rise in temperatures from human-induced climate crisis and land use change, affecting millions of people across the world~\cite{Popescu2022,Liu2022}.

High resolution spatiotemporal data of air pollutants is crucial for understanding their patterns and formulating effective mitigation policies~\cite{Brauer2019,li2021}. Traditionally, reference-grade monitors such as tapered element oscillating microbalance (TEOM) and beta attenuation monitor (BAM) have been used for air pollution monitoring~\cite{Zheng2019,Duvall2021,Dryer2020}. However, their high installation and maintenance costs along with their bulky size, limit their widespread deployment~\cite{Snyder2013}. The United States Environmental Protection Agency (USEPA) has introduced performance metrics to evaluate low-cost sensors (LCS) against reference monitors, however, a comprehensive evaluation of the accuracy and reliability of LCS data is lacking in the literature~\cite{Bulot2019,Brauer2019,Duvall2021,Snyder2013,Morawska2018,Shindler2021,Forehead2020,kang2022performance,demanega2021performance,connolly2022long,bulot2023characterisation}. Using regression analysis between LCS and reference monitor measurements, we can estimate the confidence interval for the mean reference monitor measurements and predict the range of LCS measurements. This approach helps assess the uncertainty in LCS measurements ~\cite{Montgomery2006,sahu2020validation,yue2020stronger}. The air quality sensor performance evaluation center (AQSPEC) reports LCS performance under outdoor and controlled conditions, however their experimental particulate matter (PM) concentration levels are mostly below $150~\mu\text{g/m}^{3}$~\cite{AQSPEC}. Several research articles address LCS performance at high humidity levels~\cite{Badura2018,chu2020}, however studies under both high PM concentrations and high humidity levels are limited.

In our study, conducted between 4 Nov 2021 and 4 Mar 2022 (winter season) in New Delhi, we deployed three PM sensor models from different manufacturers — Sensirion, Plantower, and Honeywell, collocated with a reference BAM. During the study, $\text{PM}_{2.5}$ and RH levels ranged between $6$ - $611~\mu\text{g/m}^{3}$ and $25$ - $95\%$, respectively. It is noted that $63\%$ and $19\%$ of the data has $\text{PM}_{2.5}$ levels above $100~\mu\text{g/m}^{3}$  and $200~\mu\text{g/m}^{3}$, respectively. In addition, $42\%$ of the data has RH levels above $80\%$. These conditions provide a unique opportunity to evaluate the LCS performance. Performance metrics such as precision, linearity, error, and the influence of meteorological factors have been investigated in this study. To gain deeper insights, the data was further divided into smaller bins based on humidity and PM concentration levels, offering a more detailed analysis as compared to examining the dataset as a whole.
One of the objectives of this study is to identify the most suitable LCS for deployment in similar conditions at other locations. 
\section{Methodology}
\subsection{Low-Cost Particulate Matter (PM) Sensors}
The low-cost PM sensors are based on the Mie scattering principle~\cite{Giordano2021}. According to Mie scattering, when light is incident on a particle, it is diffracted, refracted, and/or absorbed. The amount of scattering that occurs depends on the particle’s composition, size, and refractive index. This scattered light is measured and converted to particle concentration using an analyser~\cite{Alfano2020}. The low-cost PM sensors can be classified into two categories~\cite{Morawska2018}:
\begin{itemize}
\item Nephelometer: An instrument used to detect particles as a group, by measuring the light scattered by the particles at various angles, typically between $7^{\circ}$ and $173^{\circ}$ to avoid measuring only forward or backward scattering. The total scattered light intensity is compared to the mass measurement from a reference instrument to determine the particle concentration.
\item Optical particle counter (OPC): Instrument used to detect particles, one at a time, and measure their number and size. Each particle scatters light, which is measured by the instrument, and the resulting signal is analyzed to assign the particle to a specific bin size based on the light intensity. This process results in a particle size distribution which can be used to determine the mass loading of particles in the air. The low-cost OPCs are generally more expensive than the nephelometers. For example, the Alphasense OPC-N3 is priced at approximately $350$~USD, whereas the nephelometers (such as SPS30, PMS7003, HPMA115S0) are available for less than $100$~USD.
\end{itemize}

LCS have the potential to revolutionize the way air pollution is monitored; however, their use is currently limited due to a lack of understanding of their data quality and reliability. Additionally, there are no guidelines on suitability of LCS, based on their type, for a particular application. For example, a PM sensor used to detect traffic-related pollution must be capable of detecting finer particles, whereas the sensors used to detect construction dust must be able to detect coarser particles. In other words, the sensors must be suitable for the intended use and the task must be adequately defined~\cite{Morawska2018}. The low-cost PM sensors are often calibrated in the laboratory with a limited variety of particulate matter, which cannot simulate real outdoor pollution conditions~\cite{Badura2018,Kuula2020}. In addition, it is necessary to understand the effect of varying weather conditions on LCS measurements. Even though LCS are factory calibrated, their calibration data is not publicly available and they tend to over-/under-estimate the reference measurement under different exposure conditions~\cite{macias2023effect,aix2023calibration,molina2023size}. Excessive humidity may result in hygroscopic growth of particles and failure of electronic components, which can change the pattern of scattering, reducing the accuracy of the sensor. At humidity levels close to $100\%$, fog and mist may form inside the sensor~\cite{Badura2018,Tan2017}. To address these issues, methods based on machine learning (ML)~\cite{Alfano2020,Kuula2020,Park2021,aix2023calibration} and correction factors based on physical parameters such as humidity have been used in the literature~\cite{crilley2020}. Whereas, hardware modifications such as the addition of a dehumidifier have not been examined widely~\cite{Kuula2020}. Several studies have also highlighted concerns about the claimed efficiency of LCS. For instance, a sensor designed for $\text{PM}_{2.5}$ measurement ought to possess the capability to identify all particles that are smaller than $2.5~\mu\text{m}$. However, it is worth noting that several low-cost $\text{PM}_{2.5}$ sensors have a restricted capacity to detect particles of certain sizes~\cite{Kuula2020}. In this study, we aim to investigate the performance of low-cost PM sensors considering the above-mentioned issues.
\subsection{Sensor Selection Criteria}
We refer the manufacturer’s datasheet and previous studies investigating the performance of low-cost PM sensors under different conditions in order to select the sensor models for our study. Table~\ref{Table:sensors} lists the most commonly used low-cost PM sensors along with their measurement and operating RH range, working principle, and key features.

Based on our literature review, we propose the following sensor selection criteria:
\begin{itemize}
\item[1.] Sensor should include a fan to regulate the airflow rate.
\item[2.] Sensor should be based on light scattering with the light source being a laser.
\item[3.] PM measurement range should be up to at least $1000~\mu\text{g/m}^{3}$.
\item[4.] Upper limit of the sensors' operating RH range should be $90\%$ or higher. 
\end{itemize}

Numerous studies have recommended the use of a fan for improved performance~\cite{Bulot2019,Forehead2020,Thamban2021,Giordano2021,Jayaratne2020,Cross2017}. The second criterion is included since the performance of laser-based sensors~\cite{Alfano2020,Duvall2021,Jiao2016} has been found to be better compared to the light-emitting diode (LED)-based sensors~\cite{Alfano2020,Duvall2021,Jiao2016}. The third and fourth criteria are based on the high PM and high RH levels encountered in many Indian cities including New Delhi. Another factor in the selection process is the sensor unit price; we have set the maximum price at 50 USD per unit. Based on the above selection criteria, Plantower PMS7003, Sensirion SPS30, and Honeywell HPMA115C0-004 have been chosen for our study. Sensors from Shinyei, Sharp, and Omron are not selected as they do not include a laser and a fan. Sensor from Nova is not selected as its operating humidity range is less than $70\%$, while PurpleAir and Alphasense are not selected due to their higher unit price. Of all the sensors listed in Table~\ref{Table:sensors}, PurpleAir is a sensor package consisting of two Plantower PMS5003 sensors.
\begin{table*}  
\centering
\begin{adjustbox}{width=\textwidth}
\begin{tabular}{|c|c|c|c|c|c|c|c|}
\hline
S. No. & Sensor manufacturer & Sensor model & \multicolumn{1}{p{1.7cm}|}{\centering PM measurement\\range $(\mu\text{g/m}^{3})$} & \multicolumn{1}{p{1.7cm}|}{Operating relative humidity range (\%)} & Light source & \multicolumn{1}{p{2.5cm}|}{Provision of a fan to maintain airflow (Yes/No)} & \multicolumn{1}{p{2cm}|}{Approximate cost of sensor in India (USD)}\\
\hline
1 & Plantower~\cite{Badura2018,Bulot2019,Zheng2019,Duvall2021} & PMS7003 & 0-1000 & 0-99 & Laser & Yes & 24\\ 
\hline
2 & Sensirion~\cite{Kuula2020,Park2021,Duvall2021} & SPS30 & 0-1000 & 0-95 & Laser & Yes & 58\\

\hline
3 & Honeywell~\cite{Bulot2019,Alfano2020} & HPMA115S0 & 0-1000 & 0-95 & Laser & Yes & 72\\
\hline
4 & Honeywell~\cite{Giordano2021,Alfano2020} & HPMA115C0-004 & 0-1000 & 0-99 & Laser & Yes & 81\\
\hline
5 & Shinyei~\cite{Alfano2020} & PPD71 & 0-500 & 0-95 & LED & No & 13.5\\
\hline
6 & Alphasense~\cite{AQSPEC} & OPC-N2 & 0-10000 & 0-95 & Laser & Yes & 400\\
\hline
7 & Alphasense~\cite{AQSPEC} & OPC-N3 & 0-10000 & 0-95 & Laser & Yes & 457\\
\hline
8 & Sharp~\cite{Bulot2019,Hagan2020} & GP2Y101AU0F & 0-500 & NA & Infrared LED & No & 3.5\\
\hline
9 & Nova~\cite{Morawska2018,Kuula2020,Zhang2020} & SDS011 & 0-999 & 0-70 & Laser & Yes & 25\\
\hline
10 & Omron~\cite{Kuula2020} & B5W-LD0101 & NA & NA & LED & No & 19\\
\hline
11 & PurpleAir~\cite{PurpleAir} & PA-II-SD & 0-1000 & 0-100 & Laser & Yes & 330\\
\hline
\end{tabular}
\end{adjustbox}
\caption{List of commonly used low-cost PM sensors available in the market (``NA'' stands for ``Not available'').}
\label{Table:sensors}
\end{table*}
\subsection{Reference Grade Monitor}
The USEPA specifies methods for measuring ambient concentration of air pollutants under Federal Regulation Title 40, Part 53 (40 CFR Part 53)~\cite{USEPA_Federal_Regulation}. The equipment that meet these standards are known as federal reference monitor (FRM) or federal equivalent method (FEM). Typically, the TEOM and BAM are used as a reference~\cite{Badura2018,Bulot2019,Hong2021}. The BAM is preferred over TEOM because of its suitability for continuous long-term measurement, higher accuracy and precision, reliability for all particle sizes, and simpler operating principle~\cite{Shukla2022}. However, recent research has pointed out potential influences on BAM measurement, including high mass loading, and elevated humidity levels $(\geq 80 \%)$~\cite{Shukla2021}. For the present study, BAM is considered the ideal reference monitor, and an in-depth investigation into BAM accuracy is outside the scope of this work. For this study, the reference grade BAM (Met One Instruments, BAM 1022) site is located within the IIT Delhi campus (28°32\textquotesingle 31.5\textquotesingle\textquotesingle N 77°11\textquotesingle 27.5\textquotesingle\textquotesingle E). The reference monitor recorded data for $\text{PM}_{2.5}$, temperature, and humidity with a time resolution of $15$~minutes. Regular reviews of error logs and additional calibration tests were conducted at scheduled intervals to ensure proper functioning of the BAM.
\subsection{Design and Development of Low-Cost $\text{PM}_{2.5}$ Sensing Monitor}
\begin{figure}
\centerline{\includegraphics[width=\columnwidth]{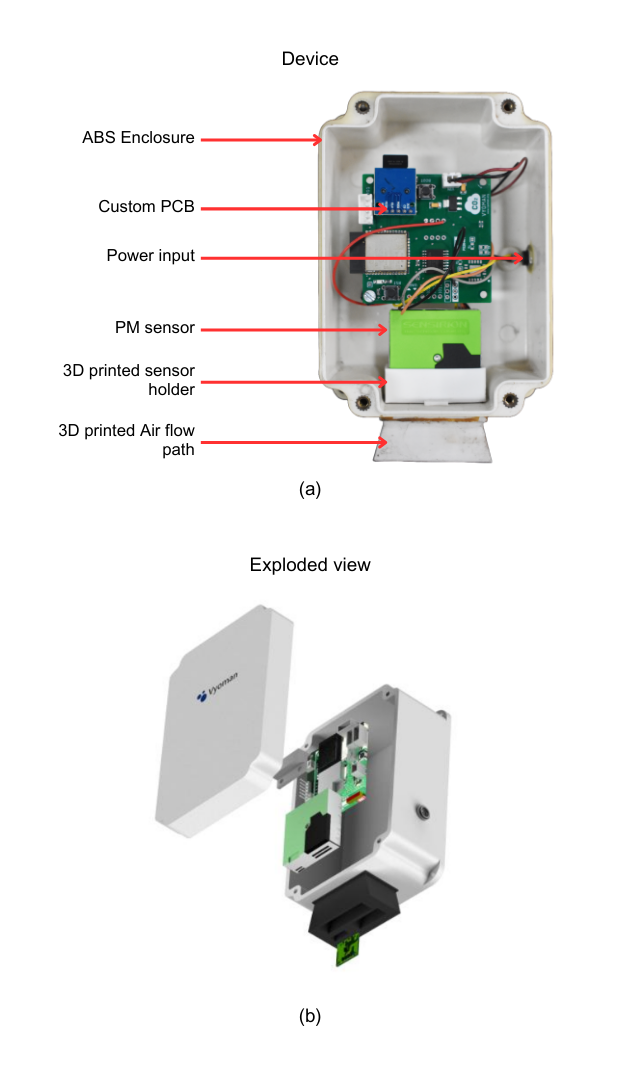}}
\caption{Layout of the sensing monitor used in this study. (a) Different parts of the monitor, and (b) exploded view of the monitor along with orientation of the parts.}
\label{devicefig}
\end{figure}
We have designed and developed a sensing monitor consisting of a low-cost PM sensor, a memory card to store the data, and a communication interface to transfer the data in real-time. The data interface for communicating with a microcontroller on most low-cost sensors is via universal asynchronous receiver-transmitter (UART). A generic printed circuit board (PCB) was designed for data collection, storage, and communicating data to a cloud platform. The PCB was designed using a microcontroller with a $240$~MHz clock and integrated Wi-Fi and Bluetooth. A real-time clock (RTC), connected to the microcontroller via inter-integrated circuit ($\text{I}^{2}\text{C}$) interface, is used to provide the timestamp for local data storage in an SD card. The SD card is interfaced to the microcontroller via serial peripheral interface (SPI). The PCB includes connectors for UART and $\text{I}^{2}\text{C}$ protocols. The board is connected via UART to Plantower PMS7003 and Honeywell HPMA115C0-004 whereas $\text{I}^{2}\text{C}$ is used to communicate with Sensirion SPS30. 

The sensing monitor is operated in two modes: active sampling and idle modes. After every $14$ minutes, the device switches to the active sampling mode where the microcontroller records data from the sensor. Twenty samples are taken at $1$~s interval and averaged. In active sampling mode, the monitor spends a total of one minute to read data, process it, and transmit it to the cloud. After completion of these tasks, the device switches to the idle mode. Locally stored averaged data is transferred to the cloud via a Wi-Fi network. An enclosure of dimension $138\times 98\times 68$~mm is used to house the PCB and sensor, and mounts were created and 3D printed for air inlet and outlet. The sensors were oriented as per the guidelines provided by the manufacturers~\cite{sensiriondesign,honeywelldesign}. Due to the absence of specific guidelines from the Plantower's manufacturer, the sensor was assumed to be similar to the other models based on its physical design. Each monitor is equipped with a DC connector for powering the PCB and the sensor. It is powered using an adapter rated at $5$~V and $1$~A. The sensing monitor draws $230$-$260$~mA depending on the sensor used in the monitor.
\subsection{Site Description}
The study was conducted within the IIT Delhi campus in New Delhi, India. The reference monitor was located on the top of a four-storey building and the set of LCS were deployed at a distance of $500$~m. Between the reference monitor and LCS site, there are a few four-storey buildings and a playground, but no significant sources of $\text{PM}_{2.5}$. The LCS were installed at a height of $3$-$5$~m from the ground. While mounting the LCS on the poles, identical sensors have been kept at a minimum distance of $1$~m. The three units of each type of LCS are referred by the sensor brand name and number, e.g., Sensirion SPS30’s three units will be referred to as Sensirion-1, Sensirion-2, and Sensirion-3. Similarly, Plantower-1, Plantower-2, and Plantower-3 will refer to the three units of Plantower PMS7003, while Honeywell-1, Honeywell-2, and Honeywell-3 refer to the three units of Honeywell HPMA115C0-004. 
\subsection{Data Collection}
From 4 Nov 2021 to 4 Mar 2022, data from the reference grade BAM and LCS were collected. Temperature and humidity measurements were captured by the reference BAM's internal sensors. The data from both LCS and the reference BAM were recorded at $15$-minutes interval and later converted into hourly and daily averages for performance evaluation. The raw data collected as part of this study has been made available (refer Data Availability section of the supplementary material). The LCS uptime was $85\%$ for hourly averaged data and $90\%$ for daily averaged data, which is acceptable as per the USEPA guidelines~\cite{Duvall2021}. We had a few missing data points due to power interruptions (refer Supplementary Fig.~S1). 

As the data was collected during the winter season in New Delhi, the mean humidity was $72\%$, and the mean temperature was $16~^{\circ}$C. The average $\text{PM}_{2.5}$ concentration reported by the BAM was $137~\mu\text{g/m}^{3}$. On 5 Nov 2021, the highest $\text{PM}_{2.5}$ level of the study duration was recorded due to the fireworks on the occasion of the Diwali festival. The lowest and the highest temperatures recorded during the study were $6~^{\circ}$C and $29~^{\circ}$C, respectively. The humidity levels varied between $25\%$ and $95\%$. Only $91$ out of $2904$ data points had humidity levels below $40\%$. Table~\ref{tab:Daily_hourly_range_uptime} summarises the sensor data in terms of the maximum, minimum, mean values, and the sensor uptime.
\begin{table*}[]
\resizebox{\textwidth}{!}{
\centering
\begin{tabular}{|cc|c|c|c|c|c|c|c|c|c|c|c|c|}
\hline
\multicolumn{2}{|c|}{} & \begin{tabular}[c]{@{}c@{}}BAM\\$(\mu\text{g/m}^3)$\end{tabular} & \begin{tabular}[c]{@{}c@{}}Temperature\\$(^{\circ}\text{C})$\end{tabular} & \begin{tabular}[c]{@{}c@{}}Humidity\\$(\%)$\end{tabular} & \begin{tabular}[c]{@{}c@{}}Sensirion-1\\$(\mu\text{g/m}^3)$\end{tabular} & \begin{tabular}[c]{@{}c@{}}Sensirion-2\\$(\mu\text{g/m}^3)$\end{tabular} & \begin{tabular}[c]{@{}c@{}}Sensirion-3\\$(\mu\text{g/m}^3)$\end{tabular} & \begin{tabular}[c]{@{}c@{}}Plantower-1\\$(\mu\text{g/m}^3)$\end{tabular} & \begin{tabular}[c]{@{}c@{}}Plantower-2\\$(\mu\text{g/m}^3)$\end{tabular} & \begin{tabular}[c]{@{}c@{}}Plantower-3\\$(\mu\text{g/m}^3)$\end{tabular} & \begin{tabular}[c]{@{}c@{}}Honeywell-1\\$(\mu\text{g/m}^3)$\end{tabular} & \begin{tabular}[c]{@{}c@{}}Honeywell-2\\$(\mu\text{g/m}^3)$\end{tabular} & \begin{tabular}[c]{@{}c@{}}Honeywell-3\\$(\mu\text{g/m}^3)$\end{tabular}\\ 
\hline
\multicolumn{1}{|c|}{\multirow{4}{*}{Hourly}} & Minimum & 6 & 6 & 25 & 8 & 9 & 2 & 15 & 15 & 13 & 44 & 42 & 64\\ 
\cline{2-14} 
\multicolumn{1}{|c|}{} & Maximum & 611 & 29 & 95 & 1230 & 1333 & 871 & 695 & 737 & 670 & 939 & 950 & 950\\ 
\cline{2-14} 
\multicolumn{1}{|c|}{} & Mean & 137 & 16 & 73 & 215 & 222 & 140 & 210 & 233 & 188 & 244 & 288 & 336\\ 
\cline{2-14} 
\multicolumn{1}{|c|}{} & Uptime & $96\%$ & $100\%$ & $100\%$ & $94\%$ & $94\%$ & $90\%$ & $95\%$ & $90\%$ & $86\%$ & $94\%$ & $95\%$ & $95\%$\\ 
\hline
\multicolumn{1}{|c|}{\multirow{4}{*}{\begin{tabular}[c]{@{}c@{}}Daily\\ (24-Hour)\end{tabular}}} & Minimum & 18 & 10 & 52 & 23 & 23 & 6 & 34 & 32 & 30 & 61 & 68 & 92\\
\cline{2-14} 
\multicolumn{1}{|c|}{} & Maximum & 344 & 22 & 93 & 643 & 690 & 645 & 439 & 506 & 420 & 693 & 737 & 828\\ 
\cline{2-14} 
\multicolumn{1}{|c|}{} & Mean & 135 & 16 & 73 & 213 & 219 & 143 & 207 & 232 & 187 & 242 & 285 & 332\\ 
\cline{2-14} 
\multicolumn{1}{|c|}{} & Uptime & $100\%$ & $100\%$ & $100\%$ & $98\%$ & $98\%$ & $94\%$ & $98\%$ & $95\%$ & $91\%$ & $98\%$ & $98\%$ & $98\%$\\ 
\hline
\end{tabular}
}
\caption{Summary of the minimum, maximum, and average values of meteorological parameters alongside the reference (BAM) and each LCS $\text{PM}_{2.5}$ measurement, including uptime, for both hourly and daily averaged data.}
\label{tab:Daily_hourly_range_uptime}
\end{table*}
\begin{table*}
\resizebox{\textwidth}{!}{
\centering
\begin{tabular}{|
>{\columncolor[HTML]{FFFFFF}}c 
>{\columncolor[HTML]{FFFFFF}}c 
>{\columncolor[HTML]{FFFFFF}}c 
>{\columncolor[HTML]{FFFFFF}}c 
>{\columncolor[HTML]{FFFFFF}}c 
>{\columncolor[HTML]{FFFFFF}}c 
>{\columncolor[HTML]{FFFFFF}}c 
>{\columncolor[HTML]{FFFFFF}}c 
>{\columncolor[HTML]{FFFFFF}}c 
>{\columncolor[HTML]{FFFFFF}}c 
>{\columncolor[HTML]{FFFFFF}}c |}
\hline
\multicolumn{11}{|c|}{\cellcolor[HTML]{FFFFFF}{\textbf{\textbf{Daily Average (24 hours)}}}}\\ 
\hline
\multicolumn{1}{|c|}{\cellcolor[HTML]{FFFFFF}{\color[HTML]{000000}}} & \multicolumn{1}{c|}{\cellcolor[HTML]{FFFFFF}{Sensirion-1}} & \multicolumn{1}{c|}{\cellcolor[HTML]{FFFFFF}{Sensirion-2}} & \multicolumn{1}{c|}{\cellcolor[HTML]{FFFFFF}{Sensirion-3}} & \multicolumn{1}{c|}{\cellcolor[HTML]{FFFFFF}{Plantower-1}} & \multicolumn{1}{c|}{\cellcolor[HTML]{FFFFFF}{Plantower-2}} & \multicolumn{1}{c|}{\cellcolor[HTML]{FFFFFF}{Plantower-3}} & \multicolumn{1}{c|}{\cellcolor[HTML]{FFFFFF}{Honeywell-1}} & \multicolumn{1}{c|}{\cellcolor[HTML]{FFFFFF}{Honeywell-2}} & \multicolumn{1}{c|}{\cellcolor[HTML]{FFFFFF}{Honeywell-3}} & {\begin{tabular}[c]{@{}c@{}}USEPA\\ Guideline\end{tabular}}\\ 
\hline
\multicolumn{1}{|c|}{\cellcolor[HTML]{FFFFFF}{\color[HTML]{000000} $R^2$}}                  & \multicolumn{1}{c|}{\cellcolor[HTML]{FFFFFF}{\color[HTML]{000000} 0.9569}}      & \multicolumn{1}{c|}{\cellcolor[HTML]{FFFFFF}{\color[HTML]{000000} 0.9552}}      & \multicolumn{1}{c|}{\cellcolor[HTML]{FFFFFF}{\color[HTML]{000000} 0.7006}}      & \multicolumn{1}{c|}{\cellcolor[HTML]{FFFFFF}{\color[HTML]{000000} 0.9219}}      & \multicolumn{1}{c|}{\cellcolor[HTML]{FFFFFF}{\color[HTML]{000000} 0.8222}}      & \multicolumn{1}{c|}{\cellcolor[HTML]{FFFFFF}{\color[HTML]{000000} 0.9033}}      & \multicolumn{1}{c|}{\cellcolor[HTML]{FFFFFF}{\color[HTML]{000000} 0.9396}}       & \multicolumn{1}{c|}{\cellcolor[HTML]{FFFFFF}{\color[HTML]{000000} 0.9507}}      & \multicolumn{1}{c|}{\cellcolor[HTML]{FFFFFF}{\color[HTML]{000000} 0.9647}}      & {\color[HTML]{000000} \textbf{$\geq 0.7$}}\\
\hline
\multicolumn{1}{|c|}{\cellcolor[HTML]{FFFFFF}{\color[HTML]{000000} Intercept $(b)$}}        & \multicolumn{1}{c|}{\cellcolor[HTML]{FFFFFF}{\color[HTML]{000000} -25.47}}      & \multicolumn{1}{c|}{\cellcolor[HTML]{FFFFFF}{\color[HTML]{000000} -30.59}}      & \multicolumn{1}{c|}{\cellcolor[HTML]{FFFFFF}{\color[HTML]{000000} -62.50}}      & \multicolumn{1}{c|}{\cellcolor[HTML]{FFFFFF}{\color[HTML]{000000} 33.60}}       & \multicolumn{1}{c|}{\cellcolor[HTML]{FFFFFF}{\color[HTML]{000000} 40.47}}       & \multicolumn{1}{c|}{\cellcolor[HTML]{FFFFFF}{\color[HTML]{000000} 26.09}}       & \multicolumn{1}{c|}{\cellcolor[HTML]{FFFFFF}{\color[HTML]{000000} 10.17}}       & \multicolumn{1}{c|}{\cellcolor[HTML]{FFFFFF}{\color[HTML]{000000} 18.87}}       & \multicolumn{1}{c|}{\cellcolor[HTML]{FFFFFF}{\color[HTML]{000000} 38.98}}       & {\color[HTML]{000000} \textbf{$0 \pm 5$}}\\ \hline
\multicolumn{1}{|c|}{\cellcolor[HTML]{FFFFFF}{\color[HTML]{000000} Slope $(m)$}}            & \multicolumn{1}{c|}{\cellcolor[HTML]{FFFFFF}{\color[HTML]{000000} 1.77}}        & \multicolumn{1}{c|}{\cellcolor[HTML]{FFFFFF}{\color[HTML]{000000} 1.84}}        & \multicolumn{1}{c|}{\cellcolor[HTML]{FFFFFF}{\color[HTML]{000000} 1.57}}        & \multicolumn{1}{c|}{\cellcolor[HTML]{FFFFFF}{\color[HTML]{000000} 1.28}}        & \multicolumn{1}{c|}{\cellcolor[HTML]{FFFFFF}{\color[HTML]{000000} 1.40}}        & \multicolumn{1}{c|}{\cellcolor[HTML]{FFFFFF}{\color[HTML]{000000} 1.18}}        & \multicolumn{1}{c|}{\cellcolor[HTML]{FFFFFF}{\color[HTML]{000000} 1.71}}        & \multicolumn{1}{c|}{\cellcolor[HTML]{FFFFFF}{\color[HTML]{000000} 1.96}}        & \multicolumn{1}{c|}{\cellcolor[HTML]{FFFFFF}{\color[HTML]{000000} 2.16}}        & {\color[HTML]{000000} \textbf{$1 \pm 0.35$}}\\ \hline
\multicolumn{1}{|c|}{\cellcolor[HTML]{FFFFFF}{\color[HTML]{000000} MAE $(\mu\text{g/m}^3)$}}      & \multicolumn{1}{c|}{\cellcolor[HTML]{FFFFFF}{\color[HTML]{000000} 78.97}}       & \multicolumn{1}{c|}{\cellcolor[HTML]{FFFFFF}{\color[HTML]{000000} 84.07}}       & \multicolumn{1}{c|}{\cellcolor[HTML]{FFFFFF}{\color[HTML]{000000} 53.64}}       & \multicolumn{1}{c|}{\cellcolor[HTML]{FFFFFF}{\color[HTML]{000000} 71.97}}       & \multicolumn{1}{c|}{\cellcolor[HTML]{FFFFFF}{\color[HTML]{000000} 95.45}}       & \multicolumn{1}{c|}{\cellcolor[HTML]{FFFFFF}{\color[HTML]{000000} 53.42}}       & \multicolumn{1}{c|}{\cellcolor[HTML]{FFFFFF}{\color[HTML]{000000} 106.46}}      & \multicolumn{1}{c|}{\cellcolor[HTML]{FFFFFF}{\color[HTML]{000000} 149.95}}      & \multicolumn{1}{c|}{\cellcolor[HTML]{FFFFFF}{\color[HTML]{000000} 196.46}}      & {\color[HTML]{000000}}\\ 
\hline
\multicolumn{1}{|c|}{\cellcolor[HTML]{FFFFFF}{\color[HTML]{000000} $\text{RMSE}_{\text{group}}~(\mu\text{g/m}^3)$}}   & \multicolumn{3}{c|}{\cellcolor[HTML]{FFFFFF}{\color[HTML]{000000} $100.49^*$}} & \multicolumn{3}{c|}{\cellcolor[HTML]{FFFFFF}{\color[HTML]{000000} 84.05}} & \multicolumn{3}{c|}{\cellcolor[HTML]{FFFFFF}{\color[HTML]{000000} 170.06}} & {\color[HTML]{000000} \textbf{$\leq 7$}}\\ 
\hline
\multicolumn{1}{|c|}{\cellcolor[HTML]{FFFFFF}{\color[HTML]{000000} $\text{NRMSE}_{\text{mean}}~(\%)$}}  & \multicolumn{3}{c|}{\cellcolor[HTML]{FFFFFF}{\color[HTML]{000000} $74.76^*$}} & \multicolumn{3}{c|}{\cellcolor[HTML]{FFFFFF}{\color[HTML]{000000} 61.51}} & \multicolumn{3}{c|}{\cellcolor[HTML]{FFFFFF}{\color[HTML]{000000} 125.83}} & {\color[HTML]{000000} \textbf{$\leq 30\%$}}\\ 
\hline
\multicolumn{1}{|c|}{\cellcolor[HTML]{FFFFFF}{\color[HTML]{000000} $\text{NRMSE}_{\text{range}}~(\%)$}} & \multicolumn{3}{c|}{\cellcolor[HTML]{FFFFFF}{\color[HTML]{000000} $30.82^*$}} & \multicolumn{3}{c|}{\cellcolor[HTML]{FFFFFF}{\color[HTML]{000000} 25.78}} & \multicolumn{3}{c|}{\cellcolor[HTML]{FFFFFF}{\color[HTML]{000000} 52.16}} & {\color[HTML]{000000} \textbf{$\leq 30\%$}}\\ 
\hline
\multicolumn{1}{|c|}{\cellcolor[HTML]{FFFFFF}{\color[HTML]{000000} SD $(\mu\text{g/m}^3)$}}     & \multicolumn{3}{c|}{\cellcolor[HTML]{FFFFFF}{\color[HTML]{000000} $7.47^*$}} & \multicolumn{3}{c|}{\cellcolor[HTML]{FFFFFF}{\color[HTML]{000000} 27.89}} & \multicolumn{3}{c|}{\cellcolor[HTML]{FFFFFF}{\color[HTML]{000000} 40.37}} & {\color[HTML]{000000} \textbf{$\leq 5$}}\\ 
\hline
\multicolumn{1}{|c|}{\cellcolor[HTML]{FFFFFF}{\color[HTML]{000000} CV $(\%)$}}            & \multicolumn{3}{c|}{\cellcolor[HTML]{FFFFFF}{\color[HTML]{000000} $3.46^*$}} & \multicolumn{3}{c|}{\cellcolor[HTML]{FFFFFF}{\color[HTML]{000000} 13.33}} & \multicolumn{3}{c|}{\cellcolor[HTML]{FFFFFF}{\color[HTML]{000000} 14.11}} & {\color[HTML]{000000} \textbf{$\leq 30\%$}}\\ 
\hline
\multicolumn{11}{|c|}{\cellcolor[HTML]{FFFFFF}{\color[HTML]{000000} \textbf{\textbf{Hourly Average}}}}\\ 
\hline
\multicolumn{1}{|c|}{\cellcolor[HTML]{FFFFFF}{\color[HTML]{000000} }}                       & \multicolumn{1}{c|}{\cellcolor[HTML]{FFFFFF}{\color[HTML]{000000} Sensirion-1}} & \multicolumn{1}{c|}{\cellcolor[HTML]{FFFFFF}{\color[HTML]{000000} Sensirion-2}} & \multicolumn{1}{c|}{\cellcolor[HTML]{FFFFFF}{\color[HTML]{000000} Sensirion-3}} & \multicolumn{1}{c|}{\cellcolor[HTML]{FFFFFF}{\color[HTML]{000000} Plantower-1}} & \multicolumn{1}{c|}{\cellcolor[HTML]{FFFFFF}{\color[HTML]{000000} Plantower-2}} & \multicolumn{1}{c|}{\cellcolor[HTML]{FFFFFF}{\color[HTML]{000000} Plantower-3}} & \multicolumn{1}{c|}{\cellcolor[HTML]{FFFFFF}{\color[HTML]{000000} Honeywell-1}} & \multicolumn{1}{c|}{\cellcolor[HTML]{FFFFFF}{\color[HTML]{000000} Honeywell-2}} & \multicolumn{1}{c|}{\cellcolor[HTML]{FFFFFF}{\color[HTML]{000000} Honeywell-3}} & {\color[HTML]{000000} USEPA} \\ 
\hline
\multicolumn{1}{|c|}{\cellcolor[HTML]{FFFFFF}{\color[HTML]{000000} $R^2$}} & \multicolumn{1}{c|}{\cellcolor[HTML]{FFFFFF}{\color[HTML]{000000} 0.9179}}      & \multicolumn{1}{c|}{\cellcolor[HTML]{FFFFFF}{\color[HTML]{000000} 0.9138}}      & \multicolumn{1}{c|}{\cellcolor[HTML]{FFFFFF}{\color[HTML]{000000} 0.6856}}      & \multicolumn{1}{c|}{\cellcolor[HTML]{FFFFFF}{\color[HTML]{000000} 0.8699}}      & \multicolumn{1}{c|}{\cellcolor[HTML]{FFFFFF}{\color[HTML]{000000} 0.7768}}      & \multicolumn{1}{c|}{\cellcolor[HTML]{FFFFFF}{\color[HTML]{000000} 0.8839}}      & \multicolumn{1}{c|}{\cellcolor[HTML]{FFFFFF}{\color[HTML]{000000} 0.9106}}       & \multicolumn{1}{c|}{\cellcolor[HTML]{FFFFFF}{\color[HTML]{000000} 0.9058}}      & \multicolumn{1}{c|}{\cellcolor[HTML]{FFFFFF}{\color[HTML]{000000} 0.8970}}      & \cellcolor[HTML]{FFFFFF}{\color[HTML]{000000}}\\ 
\cline{1-10}
\multicolumn{1}{|c|}{\cellcolor[HTML]{FFFFFF}{\color[HTML]{000000} Intercept $(b)$}}        & \multicolumn{1}{c|}{\cellcolor[HTML]{FFFFFF}{\color[HTML]{000000} -26.78}}      & \multicolumn{1}{c|}{\cellcolor[HTML]{FFFFFF}{\color[HTML]{000000} -33.67}}      & \multicolumn{1}{c|}{\cellcolor[HTML]{FFFFFF}{\color[HTML]{000000} -50.18}}      & \multicolumn{1}{c|}{\cellcolor[HTML]{FFFFFF}{\color[HTML]{000000} 26.48}}       & \multicolumn{1}{c|}{\cellcolor[HTML]{FFFFFF}{\color[HTML]{000000} 33.25}}       & \multicolumn{1}{c|}{\cellcolor[HTML]{FFFFFF}{\color[HTML]{000000} 20.44}}       & \multicolumn{1}{c|}{\cellcolor[HTML]{FFFFFF}{\color[HTML]{000000} 10.98}}       & \multicolumn{1}{c|}{\cellcolor[HTML]{FFFFFF}{\color[HTML]{000000} 17.91}}       & \multicolumn{1}{c|}{\cellcolor[HTML]{FFFFFF}{\color[HTML]{000000} 41.21}}       & \cellcolor[HTML]{FFFFFF}{\color[HTML]{000000}}\\ 
\cline{1-10}
\multicolumn{1}{|c|}{\cellcolor[HTML]{FFFFFF}{\color[HTML]{000000} Slope $(m)$}}            & \multicolumn{1}{c|}{\cellcolor[HTML]{FFFFFF}{\color[HTML]{000000} 1.78}}        & \multicolumn{1}{c|}{\cellcolor[HTML]{FFFFFF}{\color[HTML]{000000} 1.86}}        & \multicolumn{1}{c|}{\cellcolor[HTML]{FFFFFF}{\color[HTML]{000000} 1.46}}        & \multicolumn{1}{c|}{\cellcolor[HTML]{FFFFFF}{\color[HTML]{000000} 1.33}}        & \multicolumn{1}{c|}{\cellcolor[HTML]{FFFFFF}{\color[HTML]{000000} 1.45}}        & \multicolumn{1}{c|}{\cellcolor[HTML]{FFFFFF}{\color[HTML]{000000} 1.23}}        & \multicolumn{1}{c|}{\cellcolor[HTML]{FFFFFF}{\color[HTML]{000000} 1.69}}        & \multicolumn{1}{c|}{\cellcolor[HTML]{FFFFFF}{\color[HTML]{000000} 1.97}}        & \multicolumn{1}{c|}{\cellcolor[HTML]{FFFFFF}{\color[HTML]{000000} 2.14}}        & \cellcolor[HTML]{FFFFFF}{\color[HTML]{000000}}\\ 
\cline{1-10}
\multicolumn{1}{|c|}{\cellcolor[HTML]{FFFFFF}{\color[HTML]{000000} MAE $(\mu\text{g/m}^3)$}}      & \multicolumn{1}{c|}{\cellcolor[HTML]{FFFFFF}{\color[HTML]{000000} 79.42}}       & \multicolumn{1}{c|}{\cellcolor[HTML]{FFFFFF}{\color[HTML]{000000} 84.78}}       & \multicolumn{1}{c|}{\cellcolor[HTML]{FFFFFF}{\color[HTML]{000000} 54.98}}       & \multicolumn{1}{c|}{\cellcolor[HTML]{FFFFFF}{\color[HTML]{000000} 72.22}}       & \multicolumn{1}{c|}{\cellcolor[HTML]{FFFFFF}{\color[HTML]{000000} 95.54}}       & \multicolumn{1}{c|}{\cellcolor[HTML]{FFFFFF}{\color[HTML]{000000} 53.01}}       & \multicolumn{1}{c|}{\cellcolor[HTML]{FFFFFF}{\color[HTML]{000000} 105.57}}      & \multicolumn{1}{c|}{\cellcolor[HTML]{FFFFFF}{\color[HTML]{000000} 149.77}}      & \multicolumn{1}{c|}{\cellcolor[HTML]{FFFFFF}{\color[HTML]{000000} 196.52}}      & \cellcolor[HTML]{FFFFFF}{\color[HTML]{000000}}\\ 
\cline{1-10}
\multicolumn{1}{|c|}{\cellcolor[HTML]{FFFFFF}{\color[HTML]{000000} $\text{RMSE}_\text{single}~(\mu\text{g/m}^3)$}}     & \multicolumn{1}{c|}{\cellcolor[HTML]{FFFFFF}{\color[HTML]{000000} 108.78}}      & \multicolumn{1}{c|}{\cellcolor[HTML]{FFFFFF}{\color[HTML]{000000} 117.59}}      & \multicolumn{1}{c|}{\cellcolor[HTML]{FFFFFF}{\color[HTML]{000000} 79.80}}       & \multicolumn{1}{c|}{\cellcolor[HTML]{FFFFFF}{\color[HTML]{000000} 87.07}}       & \multicolumn{1}{c|}{\cellcolor[HTML]{FFFFFF}{\color[HTML]{000000} 119.33}}      & \multicolumn{1}{c|}{\cellcolor[HTML]{FFFFFF}{\color[HTML]{000000} 65.93}}       & \multicolumn{1}{c|}{\cellcolor[HTML]{FFFFFF}{\color[HTML]{000000} 125.78}}      & \multicolumn{1}{c|}{\cellcolor[HTML]{FFFFFF}{\color[HTML]{000000} 175.13}}      & \multicolumn{1}{c|}{\cellcolor[HTML]{FFFFFF}{\color[HTML]{000000} 223.06}}      & \cellcolor[HTML]{FFFFFF}{\color[HTML]{000000}}\\ 
\cline{1-10}
\multicolumn{1}{|c|}{\cellcolor[HTML]{FFFFFF}{\color[HTML]{000000} $\text{RMSE}_{\text{group}}~(\mu\text{g/m}^3)$}}   & \multicolumn{3}{c|}{\cellcolor[HTML]{FFFFFF}{\color[HTML]{000000} $113.53^*$}} & \multicolumn{3}{c|}{\cellcolor[HTML]{FFFFFF}{\color[HTML]{000000} 92.14}} & \multicolumn{3}{c|}{\cellcolor[HTML]{FFFFFF}{\color[HTML]{000000} 179.52}} & \cellcolor[HTML]{FFFFFF}{\color[HTML]{000000}}\\ 
\cline{1-10}
\multicolumn{1}{|c|}{\cellcolor[HTML]{FFFFFF}{\color[HTML]{000000} $\text{NRMSE}_{\text{mean}}~(\%)$}}   & \multicolumn{1}{c|}{\cellcolor[HTML]{FFFFFF}{\color[HTML]{000000} 81}} & \multicolumn{1}{c|}{\cellcolor[HTML]{FFFFFF}{\color[HTML]{000000} 86}}          & \multicolumn{1}{c|}{\cellcolor[HTML]{FFFFFF}{\color[HTML]{000000} 62}}          & \multicolumn{1}{c|}{\cellcolor[HTML]{FFFFFF}{\color[HTML]{000000} 64}}          & \multicolumn{1}{c|}{\cellcolor[HTML]{FFFFFF}{\color[HTML]{000000} 88}}          & \multicolumn{1}{c|}{\cellcolor[HTML]{FFFFFF}{\color[HTML]{000000} 49}}          & \multicolumn{1}{c|}{\cellcolor[HTML]{FFFFFF}{\color[HTML]{000000} 93}}          & \multicolumn{1}{c|}{\cellcolor[HTML]{FFFFFF}{\color[HTML]{000000} 130}}         & \multicolumn{1}{c|}{\cellcolor[HTML]{FFFFFF}{\color[HTML]{000000} 164}}         & \cellcolor[HTML]{FFFFFF}{\color[HTML]{000000}}\\ 
\cline{1-10}
\multicolumn{1}{|c|}{\cellcolor[HTML]{FFFFFF}{\color[HTML]{000000} $\text{NRMSE}_{\text{mean}}~(\%)$}}   & \multicolumn{3}{c|}{\cellcolor[HTML]{FFFFFF}{\color[HTML]{000000} $83.72^*$}} & \multicolumn{3}{c|}{\cellcolor[HTML]{FFFFFF}{\color[HTML]{000000} 67.52}} & \multicolumn{3}{c|}{\cellcolor[HTML]{FFFFFF}{\color[HTML]{000000} 131.65}} & \cellcolor[HTML]{FFFFFF}{\color[HTML]{000000}}\\ 
\cline{1-10}
\multicolumn{1}{|c|}{\cellcolor[HTML]{FFFFFF}{\color[HTML]{000000} $\text{NRMSE}_{\text{range}}~(\%)$}}   & \multicolumn{1}{c|}{\cellcolor[HTML]{FFFFFF}{\color[HTML]{000000} 18}} & \multicolumn{1}{c|}{\cellcolor[HTML]{FFFFFF}{\color[HTML]{000000} 19}}          & \multicolumn{1}{c|}{\cellcolor[HTML]{FFFFFF}{\color[HTML]{000000} 19}}          & \multicolumn{1}{c|}{\cellcolor[HTML]{FFFFFF}{\color[HTML]{000000} 14}}          & \multicolumn{1}{c|}{\cellcolor[HTML]{FFFFFF}{\color[HTML]{000000} 20}}          & \multicolumn{1}{c|}{\cellcolor[HTML]{FFFFFF}{\color[HTML]{000000} 11}}          & \multicolumn{1}{c|}{\cellcolor[HTML]{FFFFFF}{\color[HTML]{000000} 21}}          & \multicolumn{1}{c|}{\cellcolor[HTML]{FFFFFF}{\color[HTML]{000000} 29}}          & \multicolumn{1}{c|}{\cellcolor[HTML]{FFFFFF}{\color[HTML]{000000} 37}}          & \cellcolor[HTML]{FFFFFF}{\color[HTML]{000000}}\\ 
\cline{1-10}
\multicolumn{1}{|c|}{\cellcolor[HTML]{FFFFFF}$\text{NRMSE}_{\text{range}}~(\%)$} & \multicolumn{3}{c|}{\cellcolor[HTML]{FFFFFF}$18.75^*$} & \multicolumn{3}{c|}{\cellcolor[HTML]{FFFFFF}15.21} & \multicolumn{3}{c|}{\cellcolor[HTML]{FFFFFF}29.91} & \cellcolor[HTML]{FFFFFF}{\color[HTML]{000000}}\\ 
\cline{1-10}
\multicolumn{1}{|c|}{\cellcolor[HTML]{FFFFFF}{\color[HTML]{000000} SD $(\mu\text{g/m}^3)$}}     & \multicolumn{3}{c|}{\cellcolor[HTML]{FFFFFF}{\color[HTML]{000000} $7.55^*$}} & \multicolumn{3}{c|}{\cellcolor[HTML]{FFFFFF}{\color[HTML]{000000} 30.14}} & \multicolumn{3}{c|}{\cellcolor[HTML]{FFFFFF}{\color[HTML]{000000} 43.49}} & \cellcolor[HTML]{FFFFFF}{\color[HTML]{000000}}\\ 
\cline{1-10}
\multicolumn{1}{|c|}{\cellcolor[HTML]{FFFFFF}{\color[HTML]{000000} CV $(\%)$}}            & \multicolumn{3}{c|}{\cellcolor[HTML]{FFFFFF}{\color[HTML]{000000} $3.47^*$}} & \multicolumn{3}{c|}{\cellcolor[HTML]{FFFFFF}{\color[HTML]{000000} 14.48}} & \multicolumn{3}{c|}{\cellcolor[HTML]{FFFFFF}{\color[HTML]{000000} 15.14}} & \multirow{-12}{*}{\cellcolor[HTML]{FFFFFF}{\color[HTML]{000000} Not available}}\\ 
\hline
\end{tabular}
}
\caption{Performance evaluation metrics for hourly and daily averaged data. Recommended levels of performance metrics by the USEPA is listed only for daily averaged data as no target values are recommended for hourly averaged data. (\textbf{$^*$} represents metric values without the Sensirion-3 unit)}
\label{tab:Daily_Hourly_entire_analysis}
\end{table*}
\section{Results}
\subsection{Performance Comparison of LCS with Reference Monitor}
\begin{figure}
    \centering \includegraphics[width=1.00\linewidth,height=0.7\linewidth]{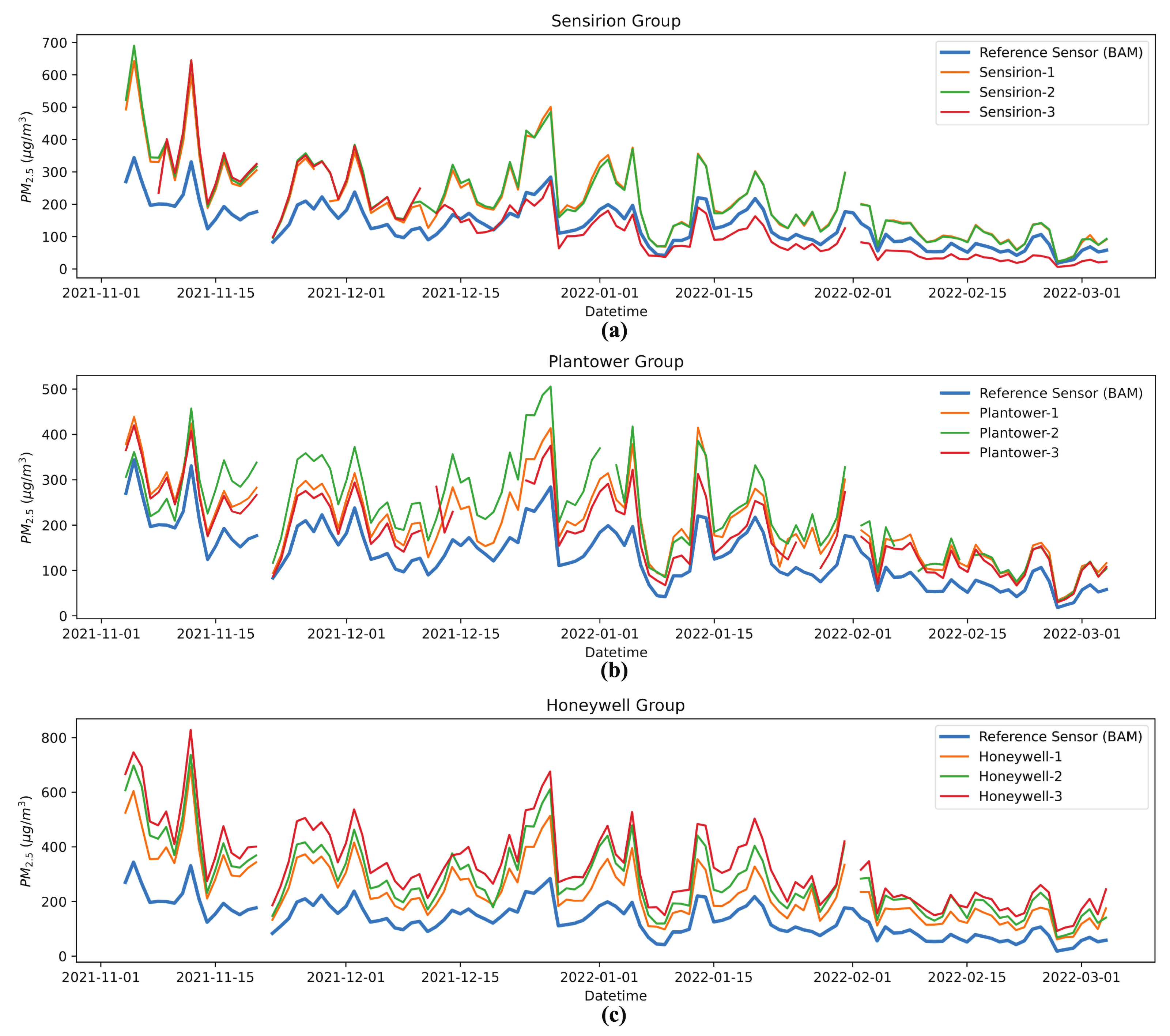}
    \caption{Time-series plots comparing the reference BAM with the three LCS models for daily averaged data: (a) Sensirion, (b) Plantower, and (c) Honeywell. With the exception of Sensirion-3 (which had cobweb growth), all LCS units can be seen to overestimate $\text{PM}_{2.5}$ level compared to the reference BAM while following a similar trend.}
    \label{fig:Daily Average Comparison}
\end{figure}

\begin{figure*}
    \centering
    \includegraphics[width=0.95\linewidth]{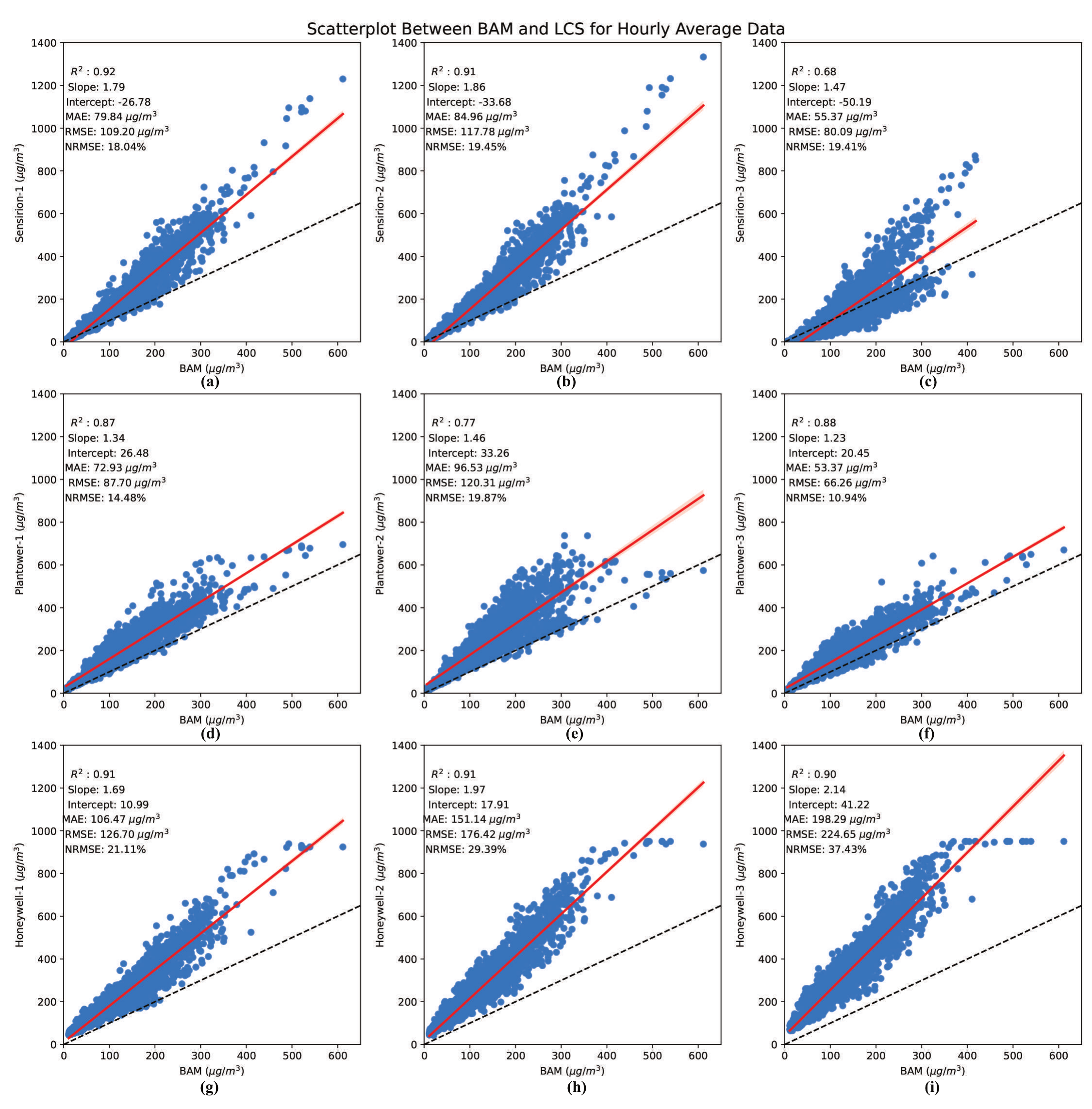}
    \caption{Scatter plots depicting the relationship between the reference BAM ($\text{PM}_{2.5}$) and different units of LCS for hourly averaged data: (a)-(c) Sensirion units 1, 2 and 3, (d)-(f) Plantower units 1, 2, and 3, and (g)-(i) Honeywell units 1, 2 and 3. Additionally, each plot includes different parameters or metrics for the respective LCS unit. The red line in each subplot represents the regression line fit to the data points. The black dashed line indicates a 1:1 relationship between BAM and the respective LCS measurements.}
    \label{fig:Scatterplot Hourly}
\end{figure*}

Prior to evaluating the LCS against the reference BAM, each LCS underwent a thorough examination to identify any physical damage. Given the study's four-month duration, which encompassed episodes of rain and harsh cold conditions, we conducted inspections for potential water ingress in the LCS units. Additionally, we examined the recorded data for anomalies such as abrupt spikes in the measurements. Upon examination, it was discovered that the Sensirion-3 unit had cobwebs at the inlet of its airflow path. Further investigation revealed cobweb accumulation inside the sensor (refer Supplementary Fig.~S3). Despite the Sensirion-3's uptime aligning with other LCS, a noticeable decline in its measurements can be attributed to the growth of cobweb (refer Fig.~\ref{fig:Daily Average Comparison}), resulting in reduced airflow. Consequently, Sensirion-3 was excluded from further analysis. The remaining LCS were assessed against the BAM for hourly and daily averaged data. Based on the scatter plots between the BAM and LCS data shown in Fig.~\ref{fig:Scatterplot Hourly}, it is seen that each sensor type is able to achieve $R^2$ above $85\%$ for one or more sensors for the hourly averaged data. Each of the scatter plots has a diverging funnel shape. The Sensirion-3 scatter plot has two funnel-shaped regions corresponding to different air flow patterns representing before and after the cobweb event. The $R^2$ values between the BAM and LCS improve further for the daily averaged data (refer Supplementary Fig.~S4). Fig.~\ref{fig:Daily Average Comparison} shows that each sensor type is able to capture the trend of the reference BAM on a daily basis for the entire duration of the study. 

\begin{figure}
    \centering
    \includegraphics[width=1\linewidth]{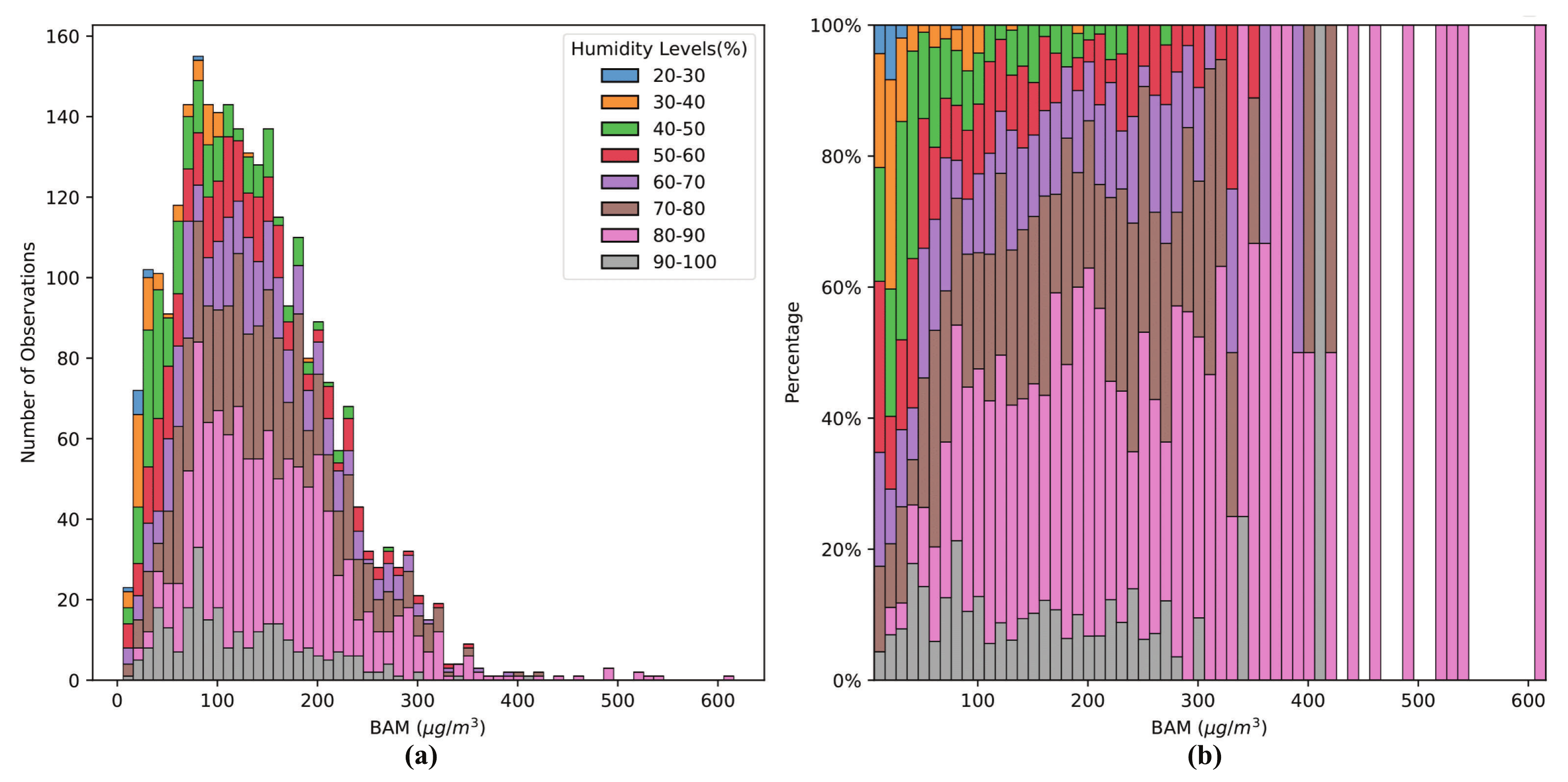}
    \caption{Distribution of $\text{PM}_{2.5}$ measurements across different humidity levels for the reference BAM data. (a) Number of observations for different BAM $\text{PM}_{2.5}$ concentrations, stacked by RH levels, (b) BAM data distribution in percentage for each humidity bin across $\text{PM}_{2.5}$ concentrations. In these plots, the $\text{PM}_{2.5}$ concentration is divided into bins of size $10~\mu\text{g/m}^{3}$. It is seen that high $\text{PM}_{2.5}$ concentration episodes were predominantly accompanied by high RH levels.} 
    \label{fig:DATA count}
\end{figure}

The data is further divided into bins based on the $\text{PM}_{2.5}$ and RH levels using bin sizes of $100~\mu\text{g/m}^{3}$ and $10\%$ for $\text{PM}_{2.5}$ and RH, respectively. This binning strategy will provide a more detailed understanding of the data. The analysis of the data by dividing them into bins will be referred to as granular analysis in the subsequent sections. For granular analysis, a minimum data collection period of $10$ hours was considered for evaluation. Fig.~\ref{fig:DATA count} shows the data distribution across $\text{PM}_{2.5}$ and RH levels for the reference BAM (refer Supplementary Fig.~S6 for a similar plot for the Sensirion-1 data). It is seen that largest fraction of measurements are in the $100$-$200~\mu\text{g/m}^3$ range and RH levels between $80$-$90\%$. 
\subsection{Reliability Assessment of LCS as per USEPA Guidelines}
After understanding the overall characteristics of the data for LCS and the reference BAM, LCS accuracy is measured against USEPA recommended ranges using hourly and daily averaged data. The accuracy assessment is based on the following linear regression-based performance metrics: slope ($m$), intercept ($b$), coefficient of determination ($R^2$), mean absolute error (MAE), root mean square error (RMSE), and normalized mean square error (NRMSE).

For evaluating precision among different units of a given sensor model, standard deviation (SD) and coefficient of variation (CV) are used. The USEPA states that the performance metrics are designed for daily (24-hour) averaged data. However, analyzing data at an hourly level offers increased granularity, providing insights into diurnal variations, aiding in comprehending daily patterns, and potentially indicating pollution sources. Additionally, it is important to note that some of the performance metrics might be misinterpreted at high $\text{PM}_{2.5}$ concentrations~\cite{zimmerman2022tutorial}. Therefore, employing the binning strategy for analysis can prove to be an effective option. Next, we provide a brief description of these performance metrics. 
\subsubsection*{Slope}
A linear regression analysis between the BAM and each LCS data is carried out to determine their relationship. Using the BAM data as the independent variable ($x$) and LCS data as the dependent variable ($y$), we compare performance of different sensor types with the BAM. The slope ($m$) for each linear fit is determined using~\eqref{eq:slope}. 
\begin{equation}
m= \frac{\sum_{i=1}^{n}(y_i-\Bar{y})(x_i-\Bar{x})}{\sum_{i=1}^{n}{(x_i-\Bar{x})^2}}
\label{eq:slope}
\end{equation}
where $y_i$, $\Bar{y}$, $x_i$, $\Bar{x}$, and $n$ are the $i^{\text{th}}$ LCS measurement, mean LCS measurement, $i^{\text{th}}$ BAM measurement, mean BAM measurement, and number of measurements, respectively. 
The USEPA guidelines suggest a slope in the range $1.0 \pm 0.35$ as acceptable for daily averaged data. In this four-month experiment, only Plantower-1 and Plantower-3 were found to be within the recommended slope range. The Sensirion units have slope values between $1.57$ and $1.84$. The Honeywell units displayed the most deviation, with slope values ranging from $1.71$ to $2.16$ (refer Fig.~\ref{fig:Scatterplot Hourly}, Supplementary Fig.~S4, and Table~\ref{tab:Daily_Hourly_entire_analysis})
\subsubsection*{Intercept}
The intercept ($b$) is computed as $b= \Bar{y}- m\Bar{x}$. The USEPA guidelines suggest an intercept value between $-5$ and $+5$ for daily averaged data. However, in our study where the average BAM exposure was $135~\mu\text{g/m}^3$ and LCS average exposure ranged between $142$-$332~\mu\text{g/m}^3$, none of the LCS could meet this recommended intercept level (refer Table~\ref{tab:Daily_Hourly_entire_analysis}).
\subsubsection*{Coefficient of Determination}
The coefficient of determination ($R^2$) quantifies the fraction of the dependent variable that is explained by the independent variable and is computed as:
\begin{equation}
  R^2 = \frac{\sum_{i=1}^{n}(y_i- x_i)^2}{\sum_{i=1}^{n}(y_i-\Bar{y})^2}  
\end{equation}
In the ideal case, $R^2 = 1$. For daily averaged data, the USEPA recommends $R^2 \geq 0.7$. In our study, $8$ out of $9$ LCS units achieved $R^2 \geq 0.9$ for daily averaged data. Granular analysis using hourly averaged data results in a decrease in $R^2$ values as short-term variations in $\text{PM}_{2.5}$ and RH levels come into play (refer Fig.~\ref{fig: Granular Analysis}).
\subsubsection*{Mean Absolute Error (MAE)}
The mean absolute error (MAE) is computed as:
\begin{equation}
    \text{MAE}= \frac{1}{n}\sum_{i=1}^{n}|x_i-y_i|
\end{equation}

Although the MAE is not included in the USEPA's performance evaluation parameters, it is an important metric for understanding the overall error magnitude between the measured and the reference values. We note that lower MAE values are achieved at lower $\text{PM}_{2.5}$ and RH levels. Interestingly, for the same $\text{PM}_{2.5}$ concentration bin of the reference BAM, a notable change in MAE is observed with increasing RH. As $\text{PM}_{2.5}$ exposure levels rise, higher MAE was observed under similar RH conditions. The Sensirion LCS group achieved the lowest MAE, followed by Plantower, while the Honeywell LCS group recorded the highest MAE (refer Supplementary Fig.~S2, Supplementary Table S2). 
\subsubsection*{Mean Bias Error (MBE)}
\begin{figure}
    \centering
    \includegraphics[width=1\linewidth]{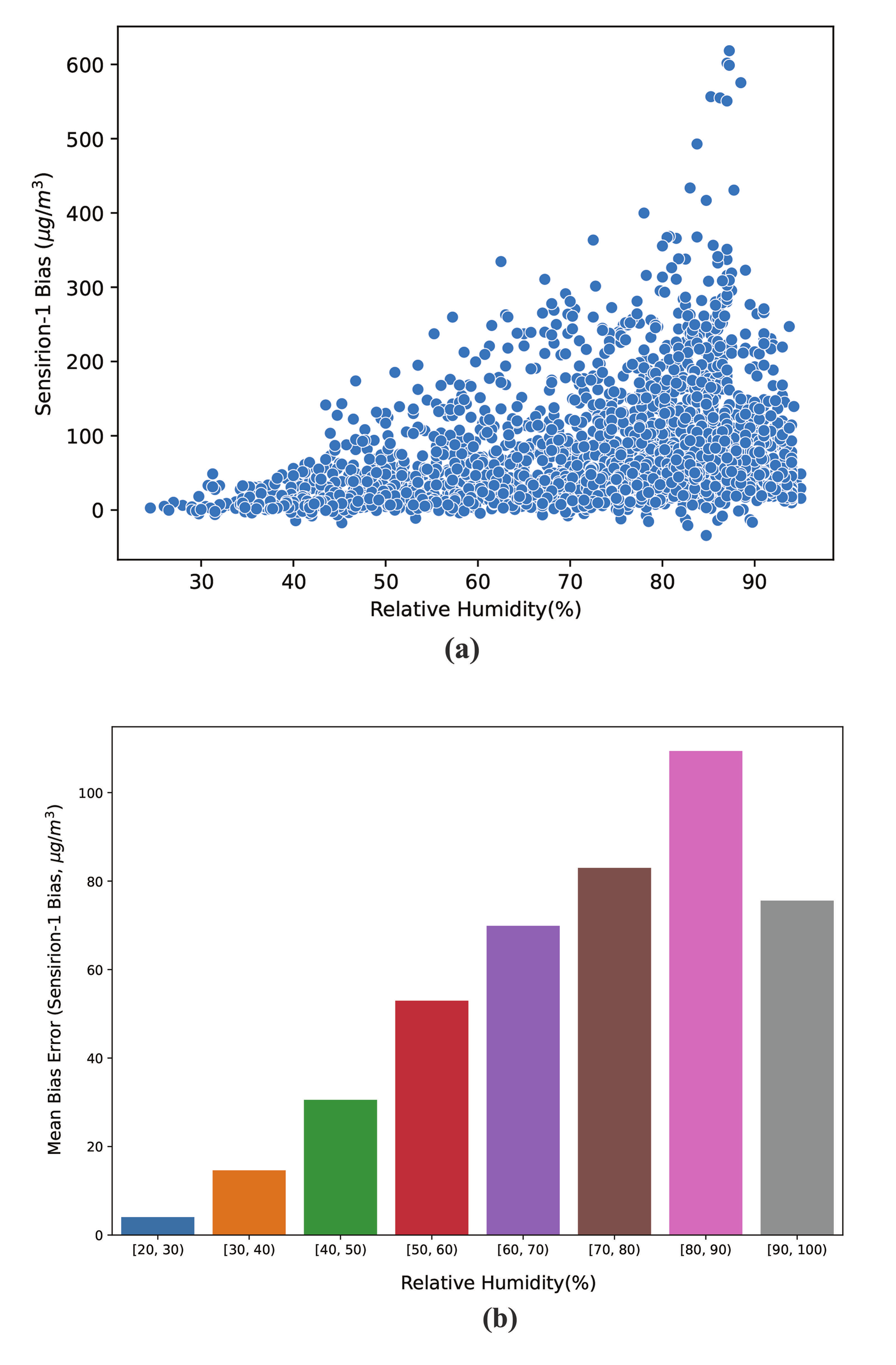}
    \caption{(a) Sensirion-1 $\text{PM}_{2.5}$ measurement bias ($\text{PM}_{2.5, \text{Sensirion-1}}-\text{PM}_{2.5, \text{BAM}}$) as a function of relative humidity, (b) MBE of Sensirion-1 as a function of relative humidity, demonstrating a similar trend of increasing mean bias as humidity levels increase. Beyond $90\%$ relative humidity, a decrease in MBE is observed, potentially due to the BAM's heater being less effective in removing moisture, leading to higher attenuation and higher $\text{PM}_{2.5}$ measurements reported by BAM.}
    \label{fig:mbe}
\end{figure}

\begin{equation}
    \text{MBE}= \frac{1}{n}\sum_{i=1}^{n}(x_i-y_i)
\end{equation}
We also compute the mean bias error (MBE), with respect to variation in relative humidity, to investigate potential bias in the sensor measurements. Figure~\ref{fig:mbe}(a) illustrates the bias of Sensirion-1 ($\text{PM}_{2.5, \text{Sensirion-1}}-\text{PM}_{2.5, \text{BAM}}$) as a function of relative humidity. It is evident that the bias increases as the relative humidity increases. Similarly, Figure~\ref{fig:mbe}(b) shows the MBE of Sensirion-1 as a function of relative humidity. A similar pattern of increasing bias with increasing humidity has been observed for other LCS. This analysis highlights the importance of accounting for humidity when interpreting LCS data, as elevated humidity levels can introduce bias in the measurements. We also observe a decrease in MBE beyond $90\%$ relative humidity. A likely explanation for this reduction in MBE is that, at such high humidity levels, the BAM's heater becomes less effective in removing moisture from the air sample. This results in greater attenuation in BAM and, consequently, higher $\text{PM}_{2.5}$ readings.

\subsubsection*{Root Mean Square Error (RMSE)}
The root mean square error (RMSE) metrics are computed as shown below.

\begin{align}
\label{eq:RMSE}
    \text{RMSE}_{\text{single}} & = \sqrt{\frac{1}{N}\sum_{i=1}^{N}(y_i-x_i)^2}\\
    \text{RMSE}_{\text{group}} & = \sqrt{\frac{1}{N\times M} \sum_{j=1}^{M}\sum_{d=1}^{N}(x_{dj}-R_d)^2}
\end{align}
%
where $N$ is the number of 24-hour (daily) periods during which an LCS unit reported valid measurements, $M$ is the number of LCS units of a given sensor model, $x_{dj}$ is daily average $\text{PM}_{2.5}$ concentration for $d^{\text{th}}$ day and $j^{\text{th}}$ LCS unit, and $R_d$ is the reference BAM daily average for $d^{\text{th}}$ day.

The USEPA recommends $\text{RMSE} \leq 7~\mu\text{g/m}^3$ or $\text{NRMSE} \leq 30\%$ for daily averaged data. We compute RMSE for each LCS unit (given as $\text{RMSE}_{\text{single}}$ in~\eqref{eq:RMSE}) and also for every sensor model group (given as $\text{RMSE}_{\text{group}}$ in~\eqref{eq:RMSE}). RMSE values based on daily averaged data are $100.41~\mu\text{g/m}^3$, $84.05~\mu\text{g/m}^3$, and $170.06~\mu\text{g/m}^3$ for the Sensirion, Plantower, and Honeywell LCS unit groups, respectively, over the study duration. The $\text{RMSE}_{\text{single}}$ values based on hourly averaged data range between $66$-$223~\mu\text{g/m}^3$ (refer Table~\ref{tab:Daily_Hourly_entire_analysis}). The lowest RMSE values are achieved for $\text{PM}_{2.5}$ concentrations in the range $0$-$100~\mu\text{g/m}^3$ and RH below $80\%$ (refer Fig.~\ref{fig: Granular Analysis}), indicating better LCS performance at lower $\text{PM}_{2.5}$ concentrations and lower RH (refer Supplementary Table S3). None of the LCS tested in our study meet the USEPA recommended RMSE limit under the high $\text{PM}_{2.5}$ concentrations and high RH conditions.
\subsubsection*{Normalized Root Mean Square Error (NRMSE)}
The normalized root mean square error (NRMSE) is computed as:
\begin{equation}
\label{eq:NRMSE}
    \text{NRMSE}_{\text{mean}} = \frac{\text{RMSE}}{\Bar{R}_d}\times 100 \; \text{and} \; \text{NRMSE}_{\text{range}} = \frac{\text{RMSE}}{R_{\text{range}}}\times 100
\end{equation}
where $\Bar{R}_d$ is the average reference $\text{PM}_{2.5}$ concentration over the entire testing period and $R_{\text{range}}$ is the difference between the maximum and minimum daily averaged $\text{PM}_{2.5}$ data recorded by the reference BAM. NRMSE helps to normalize errors, particularly when $\text{PM}_{2.5}$ concentrations are beyond the typical conditions considered by the USEPA~\cite{zimmerman2022tutorial,Duvall2021}.

The USEPA recommended threshold is $\text{NRMSE} \leq30\%$ for daily averaged data. The NRMSE based on daily averaged data are $74.46\%$, $61.51\%$, and $125.83\%$ for the Sensirion, Plantower, and Honeywell LCS group, respectively (refer Table \ref{tab:Daily_Hourly_entire_analysis}). The skewness in the distribution of the data can be observed in Fig.~\ref{fig:DATA count}. Thus, the NRMSE is normalized based on the range of the data instead of the mean of the daily averaged reference BAM data and is denoted as $\text{NRMSE}_{\text{range}}$ in~\eqref{eq:NRMSE}. The $\text{NRMSE}_{\text{range}}$ values 
are $30.82\%$, $25.72\%$, and $52.16\%$ for the Sensirion, Plantower, and Honeywell LCS group, respectively. Granular analysis of the hourly averaged LCS data reveals an adverse impact of increasing RH and $\text{PM}_{2.5}$ levels on the NRMSE values (refer Supplementary Table S4).
\subsubsection*{Standard Deviation (SD)}
Standard deviation (SD) is an important metric that measures the precision among LCS units of the same sensor model, with lower SD values indicating closely clustered measurements. SD is computed as:
\begin{equation}\label{eq:SD}
    \text{SD} = \sqrt{\frac{1}{(NM-1)}\sum_{j=1}^{M}\sum_{d=1}^{N}(x_{dj}-\bar{x}_d)^2}
\end{equation}
where $\bar{x}_d$ is daily average LCS $\text{PM}_{2.5}$ concentration for the $d^{\text{th}}$ day. As per the USEPA, recommended limit is $\text{SD} \leq 5~\mu\text{g/m}^3$ for daily averaged data. In Table~\ref{tab:Daily_Hourly_entire_analysis}, we see that SD for daily averaged data is $7.47~\mu\text{g/m}^3$, $27.89~\mu\text{g/m}^3$, and $40.37~\mu\text{g/m}^3$ for the Sensirion, Plantower, and Honeywell LCS group, respectively. Granular analysis using hourly averaged data show that SD for the Sensirion LCS group is less than $2~\mu\text{g/m}^3$ with $0$-$100~\mu\text{g/m}^3$ $\text{PM}_{2.5}$ concentrations and RH levels up to $80\%$. Granular analysis also shows that the precision of LCS groups is adversely affected with an increase in $\text{PM}_{2.5}$ and RH levels. (refer Supplementary Table S5).
\begin{figure}
    \centering
    \includegraphics[width=\linewidth]{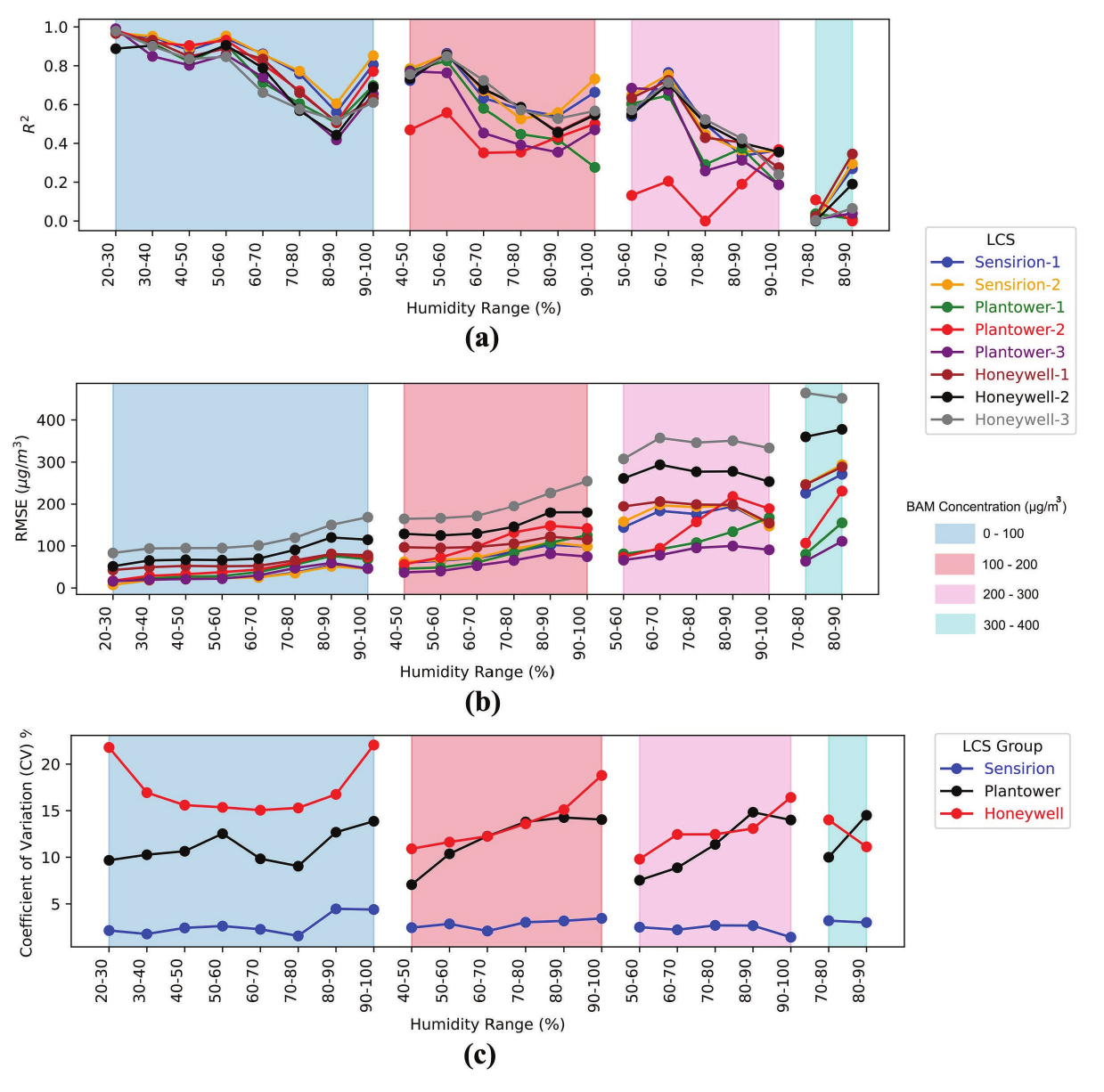}
    \caption{(a) Coefficient of determination ($R^2$) between each LCS unit (except Sensirion-3) and the reference BAM, (b) RMSE between each LCS unit and BAM, (c) coefficient of variation (CV) for each LCS group. Each sub-plot has different colored sections representing different $\text{PM}_{2.5}$ exposure levels, with the RH levels indicated for each of these sections. Subplots (a) and (b) show the impact of $\text{PM}_{2.5}$ and RH levels on the accuracy of LCS, while subplot (c) shows the effect on LCS precision. The precision of LCS tends to decline as $\text{PM}_{2.5}$ and RH levels increase. From subplot (b) it is seen that data collected at similar humidity levels but with higher $\text{PM}_{2.5}$ concentrations exhibits larger error. Precision of the Sensirion LCS group is relatively consistent and best among the three LCS groups.}
    \label{fig: Granular Analysis}
\end{figure}
\subsubsection*{Coefficient of Variation (CV)}
The coefficient of variation (CV) is computed as:
\begin{equation}
    \text{CV} = \frac{\text{SD}}{\Bar{y}} \times 100
\end{equation}
As per the USEPA guidelines, $\text{CV} \leq 30\%$ for daily averaged data. In our study, the CV value for each LCS group is well within this limit, with the CV values being $3.46\%$, $13.33\%$, and $14.11\%$ for the Sensirion, Plantower, and Honeywell LCS group, respectively, for daily averaged data (refer Table~\ref{tab:Daily_Hourly_entire_analysis}). The granular analysis using hourly averaged data indicates similar results (refer Fig.~\ref{fig: Granular Analysis} and Supplementary Table S6).
\subsection{Observations about the LCS Models}
\subsubsection*{Sensirion SPS30}
During our study, growth of cobweb was found in the airflow path of the Sensirion-3 unit and thus data from this sensor unit was excluded from the analysis. This underscores the need for redesigning the airflow path so to mitigate the possibility of such undesired occurrences, especially in outdoor applications. The sensor uptime for the remaining two Sensirion units was more than $89\%$ for hourly averaged data and $97\%$ for daily averaged data. The missing data points can be attributed to power failures and maintenance.
The $R^2$ values are the best for the Sensirion-1 and Sensirion-2 units among all the LCS units being studied, for both hourly and daily averaged data. 
Based on granular analysis, a notable decline in $R^2$ values is observed with increasing $\text{PM}_{2.5}$ and RH levels (refer Fig.~\ref{fig: Granular Analysis}). The Sensirion units perform very well with $\text{PM}_{2.5}$ exposure in the range $0$-$100~\mu\text{g/m}^3$ and RH below $70\%$. This performance is consistent based on the other metrics such as MAE, RMSE, and NRMSE. The Sensirion units achieved the highest precision among all LCS units tested. The granular analysis indicates that Sensirion is the only LCS group achieving $\text{CV} \leq 5\%$ for the entire range of $\text{PM}_{2.5}$ and RH levels encountered in this study (refer Fig.~\ref{fig: Granular Analysis}). 

The Sensirion SPS30 sensors incorporate a low-power sleep mode suitable for battery-powered applications and standalone units. Additionally, these units feature an autocleaning function which is achieved by increasing the fan speed of the sensor to clean the airflow path. For our study, the Sensirion LCS units were configured to carry out autocleaning once in four days. The current consumption of these LCS units during the sensing operation is $80$~mA.
\subsubsection*{Plantower PMS7003}
The Plantower PMS7003 is one of the widely used PM sensors~\cite{Badura2018,Alfano2020,Chen2018}. While numerous studies have used these sensors, to the best of our knowledge, there have not been any studies looking into their performance in outdoor conditions with $\text{PM}_{2.5}$ levels up to $600~\mu\text{g/m}^3$ and RH levels up to $95\%$. The Plantower LCS group achieved sensor uptime of more than $90\%$ and $86\%$ for daily and hourly averaged data, respectively. Considering the entire study duration, the $R^2$ values for the Plantower LCS units is the lowest among all the LCS groups, although it remains within acceptable limits (refer Table~\ref{tab:Daily_Hourly_entire_analysis}). Additionally, highest variation in $R^2$ values is observed for the Plantower LCS group. Evaluation parameters related to measurement accuracy are the highest accuracy for this LCS group. In terms of precision, Plantower ranked as the second-best LCS after the Sensirion group. Granular analysis further reveals that an increase in $\text{PM}_{2.5}$ exposure and RH levels result in a significant drop in accuracy and precision for this sensor group.
\subsubsection*{Honeywell HPMA115C0-003}
During our study, the Honeywell LCS group achieved the highest sensor uptime among all LCS units. The $R^2$ values for the Honeywell LCS units are found to be more than $0.89$ and $0.93$ for hourly and daily averaged data, respectively. However, the accuracy metrics such as MAE, RMSE, and NRMSE are the worst for this group among the LCS groups being compared. The precision metrics are within the USEPA recommended limits, although they are highest among the three LCS groups. However, the sensing limit is a concern for the Honeywell LCS. Despite the manufacturer datasheet stating a detection limit of $0$-$1000~\mu\text{g/m}^3$, the measurements appear to saturate at $950~\mu\text{g/m}^3$. When the reference BAM measurements exceed $400~\mu\text{g/m}^3$, the Honeywell LCS units consistently report $950~\mu\text{g/m}^3$ (refer Fig.~\ref{fig:Scatterplot Hourly}(g)-(i)). This issue has been detected during hourly analysis and could have been easily overlooked using only the daily average analysis.
\subsection{Confidence and Prediction Intervals for LCS Measurement}
\begin{figure*}
    \centering
    \includegraphics[width=\linewidth]{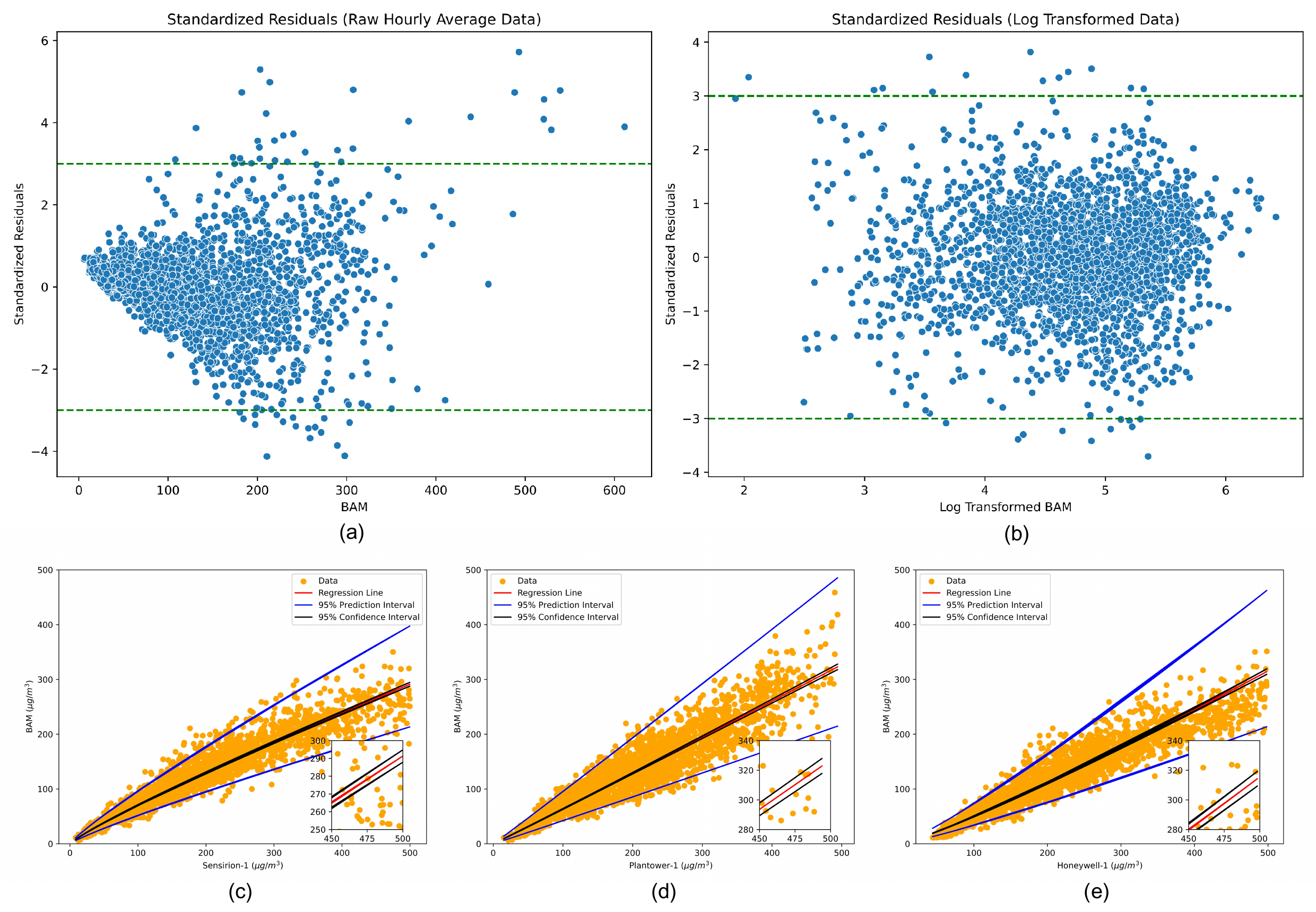}
    \caption{(a) Standardized residual plot between BAM and Sensirion-1 unit raw data, (b) standardized residual plot between log-transformed BAM and Sensirion-1 unit data. (c)-(e) Scatter plots between three LCS units and the reference BAM with $95\%$ confidence and prediction interval lines. (a) and (b) depict the treatment of heteroscedasticity in the data.}
    \label{fig:95CI}
\end{figure*}

Utilizing the understanding from the assessment metrics, regression analysis was undertaken to assess the reliability in LCS values. For this analysis, LCS measurement is used as the independent variable and BAM measurement is considered as the dependent variable. We use the LCS measurement to estimate the confidence interval around the mean response of BAM and predict the range of the BAM measurement for a given LCS measurement. The scatter plot between LCS and the reference BAM measurements had a funnel shape (refer Fig.~\ref{fig:Scatterplot Hourly}). The residual error plot also exhibits a funnel shape indicating heteroscedasticity in the data (refer Fig.~\ref{fig:95CI}(a)). To address this issue, various transformations of the independent and dependent variables are considered. Logarithmic transformation of LCS and reference BAM measurements results in the most favorable standardized residual error plot, indicative of homoscedasticity (refer Fig.~\ref{fig:95CI}(b)). Outliers are identified and subsequently removed using a standard residual threshold corresponding to three times the standard deviation. The confidence and prediction intervals are computed and the data is then transformed back to its original form using the antilogarithm function.
Using this analysis, for a given LCS measurement in the field, we can estimate the $95\%$ confidence interval (using Supplementary equation (S1)) and the prediction interval (using~\eqref{eq:CI})~\cite{Montgomery2006} of this measurement to estimate the reliability/uncertainty in the LCS measurement.
\begin{equation}\label{eq:CI}
\begin{split}
    \hat{y}_o - t_{\alpha/2,n-2} & \sqrt{\text{MSE}\left[1+\frac{1}{n}+\frac{(x_o-\Bar{x})^2}{S_{xx}}\right]} \leq y_o \\
    & \leq \hat{y}_o + t_{\alpha/2,n-2} \sqrt{\text{MSE}\left[1+\frac{1}{n}+\frac{(x_o-\Bar{x})^2}{S_{xx}}\right]}
\end{split}
\end{equation}
with
\begin{equation*}\label{eq:MSE_Sxx}
    \text{MSE} = \frac{\sum_{i=1}^{n} (y_i - \hat{y}_{i})^2}{(n-2)} \quad \text{ and } \quad S_{xx} = \sum_{i=1}^{n}(x_i-\Bar{x})^2
\end{equation*}
where $\hat{y}_o$ is the estimated reference $\text{PM}_{2.5}$ value based on the linear fit, $n$ is the number of measurements, $t_ {\alpha/2,n-2}$ is the $t$-critical value with $\alpha$ denoting the confidence level and ($n-2$) the degrees of freedom, $x_o$ is the predictor variable value for which prediction is being made, and $\bar{x}$ is mean of the predictor variable values. Fig.~\ref{fig:95CI}(c)-(e) show the $95\%$ confidence and prediction intervals between the three LCS units and the reference BAM measurements. 
\begin{table*}
\resizebox{\textwidth}{!}{
\centering
\begin{tabular}{|c|c|c|c|c|c|c|c|c|c|c|c|c|}
\hline
Time & \multicolumn{9}{c|}{$R^{2}$ values} & Average & Average & Average\\
\cline{2-10}
Period & Sensirion-1 & Sensirion-2 & Sensirion-3 & Plantower-1 & Plantower-2 & Plantower-3 & Honeywell-1 & Honeywell-2 & Honeywell-3 & BAM & Humidity & Temp.\\
\hline
Nov 2021 & 0.93 & 0.93 & 0.89  & 0.90 & 0.56 & 0.91 & 0.92 & 0.91 & 0.88 & $197~\mu\text{g/m}^{3}$ & $66.63\%$ & $19.85~^{\circ}$C\\ 
\hline
Dec 2021 & 0.81 & 0.76 & 0.44  & 0.82 & 0.78 & 0.79 & 0.86 & 0.86 & 0.84 & $153~\mu\text{g/m}^{3}$ & $72.32\%$ & $15.33~^{\circ}$C\\
\hline
Jan 2022 & 0.77 & 0.78 & 0.86  & 0.78 & 0.81 & 0.80 & 0.82 & 0.83 & 0.83 & $130~\mu\text{g/m}^{3}$ & $82.00\%$ & $12.94~^{\circ}$C\\
\hline
Feb 2022 & 0.86 & 0.85 & 0.80  & 0.82 & 0.80 & 0.79 & 0.82 & 0.80 & 0.77 & $74~\mu\text{g/m}^{3}$ & $69.00\%$ & $17.25~^{\circ}$C\\
\hline
\end{tabular}
}
\caption{Month-wise $R^{2}$ values between the BAM and LCS units, monthly average $\text{PM}_{2.5}$ (BAM), humidity, and temperature values.}
\label{table:monthly_analysis}
\end{table*}

We see that the LCS prediction interval diverges as the LCS measurement value increases. The Sensirion-1 unit has the best prediction interval among the three LCS units compared in Fig.~\ref{fig:95CI}. From our regression analysis, we see from Fig.~\ref{fig:95CI} that the $95\%$ confidence interval around LCS is very narrow. For example, from the zoomed in portion of Fig.~\ref{fig:95CI}(c), we can see that for an LCS measurement of $475~\mu\text{g/m}^3$, BAM values lie within the range of approximately $275$-$282~\mu \text{g/m}^3$ indicating a high degree of confidence in this LCS measurement.

\begin{figure}[h]
    \centering
    \includegraphics[width=.95\linewidth]{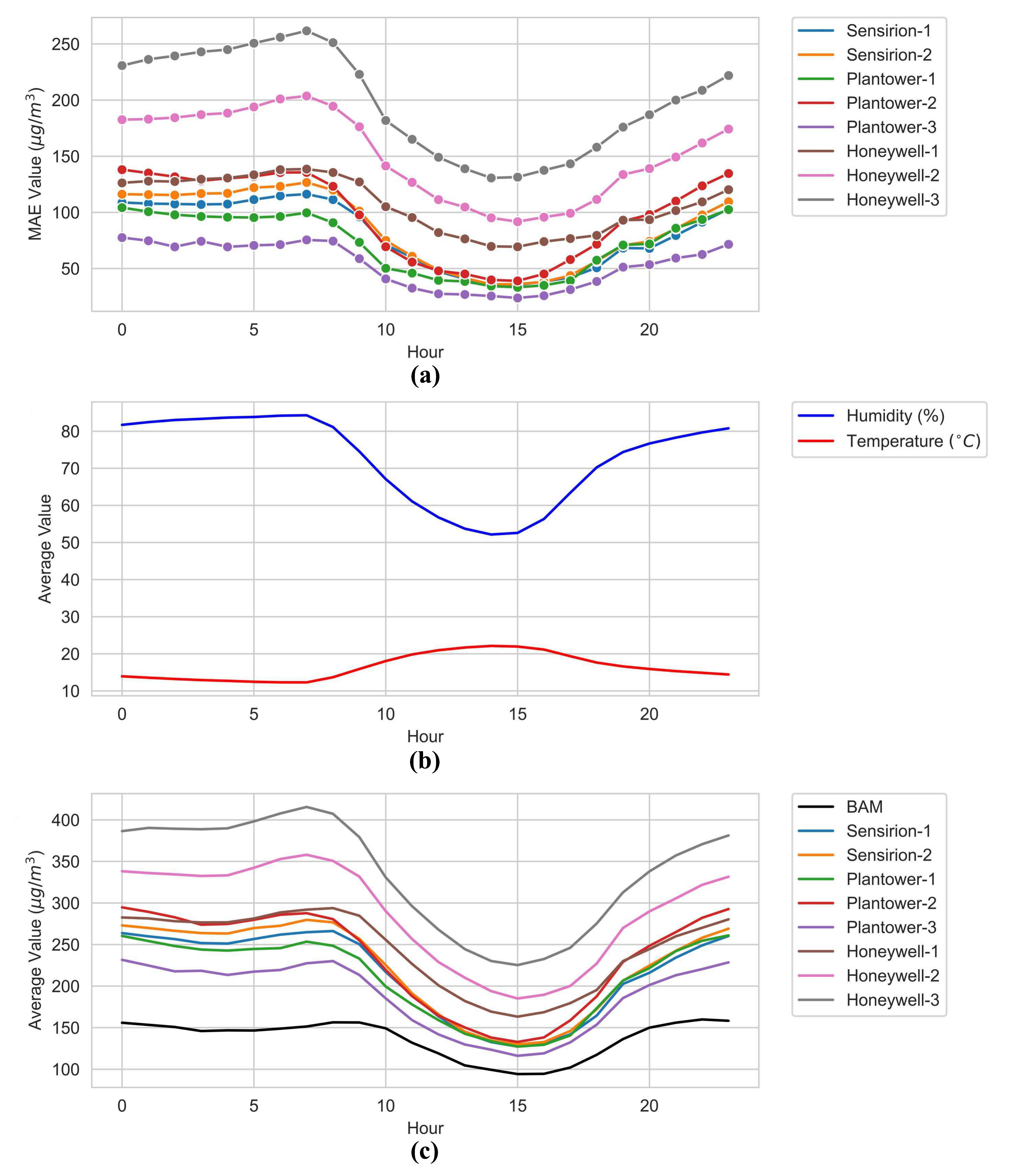}
    \caption{(a) Diurnal variation of MAE between LCS and reference BAM $\text{PM}_{2.5}$, (b) diurnal variation of temperature and RH, and (c) Diurnal variation in $\text{PM}_{2.5}$ levels recorded by LCS and reference BAM.} 
    \label{fig:Diurnal analysis}
\end{figure}
\subsection{Effect of Meteorology}
The reference BAM and LCS are based on different operating principles. BAM utilizes beta ray attenuation and is typically coupled with a heating system to handle higher RH levels, while LCS are based on Mie scattering theory and lack humidity control. Due to the absence of humidity control, LCS are exposed to phenomenon such as hygroscopic growth under high RH conditions. In our study, the BAM filter tape is maintained at a temperature setpoint of $35^{\circ}\text{C}$ and a heater gets activated when the sampled air temperature drops below this threshold~\cite{BAM1022}. This in turn allows some control over the filter tape RH. However, due to difference in operating principles between the BAM and LCS, and the lack of humidity control in LCS can lead to difference in their measurements. To analyze the effect of humidity, we divide the data into bins based on both $\text{PM}_{2.5}$ and RH levels. This will allow us to compute the LCS measurement accuracy and precision for each bin. High $\text{PM}_{2.5}$ measurements were mostly associated with high RH levels for both the reference BAM and LCS (refer Fig.~\ref{fig:DATA count} and Supplementary Fig.~S6). Based on the granular analysis, it is seen that as the $\text{PM}_{2.5}$ and RH levels increase, the accuracy of LCS declines notably. While LCS initially exhibit a trend similar to the reference BAM, the discrepancy between their measurements increases significantly at higher levels of $\text{PM}_{2.5}$ exposure and RH. However, in  Fig.~\ref{fig: Granular Analysis}, a slight improvement in LCS performance is observed for some $\text{PM}_{2.5}$ ranges when the RH is above $90\%$. This is possibly due to the BAM heater unable to completely remove the water content in the PM at such high RH levels, resulting in the BAM also reporting higher mass concentrations~\cite{Shukla2022}.

Furthermore, recognizing the importance of humidity in improving the calibration models for LCS, we compute statistical significance in terms of $p$-value for the hourly averaged data. This statistical analysis is aimed at quantifying the impact of humidity on LCS measurements and its possible role in improving the calibration models. Overall, this analysis will contribute to a nuanced understanding of how humidity influences the reliability of LCS in real-world outdoor conditions. The computed $p$-values show that humidity has a significant role in the calibration model $y_i = f_{\text{linear}}(x_i,\text{RH})$ at RH above $70\%$ (refer Supplementary Tables S7 and S8).

We also use the hourly averaged data to understand the effect of diurnal variations on humidity. As shown in Fig.~\ref{fig:Diurnal analysis}, there is a correlation between increasing temperature during daylight hours, a concurrent decrease in humidity, and a noticeable reduction in $\text{PM}_{2.5}$ levels during these hours. The peak in humidity levels is noted around 7:00 am. Subsequent to this, as the temperatures rise, humidity levels decrease. However, possibly due to the early morning vehicular traffic, there is an evident rise in $\text{PM}_{2.5}$ levels until 9:00 am.

In Table~\ref{table:monthly_analysis}, we present a month-wise analysis of the data.  It is seen that the average RH, temperature, and $\text{PM}_{2.5}$ levels are in the range of $67$-$82\%$, $13$-$20^\circ\text{C}$, and $74$-$197~\mu\text{g/m}^3$, respectively, across the four months of our study. The observed changes in $\text{PM}_{2.5}$ concentration across the months correlates with variations in temperature and RH. The $R^2$ values are highest during the initial month of deployment, with progressively lower values in the subsequent months. Given the changes in pollution sources and meteorological conditions over the four-month period, controlled laboratory experiments before and after the outdoor deployment could provide further insights. Future investigations will include drift analysis to understand the long-term performance of LCS.

\section{Discussion}
In contrast to previous studies, where LCS were chosen based on general performance metrics, we undertook a comprehensive selection process and evaluation involving most of the commonly used LCS models (Refer Table~\ref{Table:sensors}). First, we assessed technical parameters such as the working principle and the method for maintaining constant air flow rate, both crucial for reliable performance. Second, our evaluation considered the ability to operate under extreme $\text{PM}_{2.5}$ exposure and high humidity levels which are characteristic of Delhi’s winter and can significantly impact sensor accuracy and life. Finally, cost considerations are critical for identifying models that could be scaled affordably for widespread deployment. This method is aimed to facilitate the large-scale deployment of LCS across Delhi and other areas with comparable meteorological conditions as part of future studies, ensuring both performance efficiency and cost-effectiveness.

Most outdoor collocation studies of multiple LCS have been conducted in environments with relatively low pollution levels, where $\text{PM}_{2.5}$ concentrations remain below $150~\mu\text{g/m}^3$~\cite{molina2023size, levy2018, shittu2024, patel2024, zheng2018}. Although these evaluations are valuable for those specific contexts, they do not sufficiently address the challenges posed by Delhi's highly polluted winter conditions. In Delhi, $\text{PM}_{2.5}$ levels can surpass $900~\mu\text{g/m}^3$ for long periods, requiring a more customized approach to LCS evaluation. Furthermore, Delhi's pollution scenario is affected by various local factors, including agricultural stubble burning, reduced boundary layer height, and reduced air circulation during winter months. These distinctive characteristics highlight the necessity for in situ evaluation, as results from studies in different meteorological and pollution conditions may not be reliably adapted to Delhi’s extreme pollution conditions.

In our study, we recorded $\text{PM}_{2.5}$ levels peaking at $611~\mu\text{g/m}^3$, according to the reference BAM, with an average over four months of $137~\mu\text{g/m}^3$ and a maximum monthly average of $197~\mu\text{g/m}^3$ for November. Previous studies in Delhi often focused on a single LCS model without thorough comparative analysis~\cite{campmier2023, sahu2020validation}. Our study addresses this gap by evaluating multiple LCS models, offering insights into their performance during the winter months with high pollution levels and providing guidance for future deployments in similar environments.

In contrast to most previous studies that evaluate LCS performance based on USEPA guidelines using the entire dataset, our study employs a granular analysis by binning data according to humidity and $\text{PM}_{2.5}$ exposure levels. This approach reveals critical insights often missed during whole-dataset evaluations. For example, while Sensirion-1 and Sensirion-2 displayed $R^2$ values of $0.92$ and $0.91$ for the entire dataset, granular analysis showed significantly lower $R^2$ values of $0.56$ and $0.60$ under exposure levels of $0-100~\mu\text{g/m}^3$ and $80-90\%$ humidity. This  analysis underscores the importance of going beyond broad evaluations to better understand sensor performance. Although USEPA recommendations are based on daily-averaged performance metrics, our granular analysis, using hourly-averaged data, provides deeper insights that would not have been evident from daily averages alone, further emphasizing the value of this approach.

The time-series plots (Fig.~\ref{fig:Daily Average Comparison}) show a similar trend between LCS and the reference BAM measurements, however, the scatter plots (Fig.~\ref{fig:Scatterplot Hourly}) indicate a greater variance in LCS measurements as the $\text{PM}_{2.5}$ levels increase. Fig.~\ref{fig:DATA count} illustrates that while this study was predominantly conducted under high humidity conditions, higher $\text{PM}_{2.5}$ concentrations ($>400~\mu\text{g/m}^3$) were observed exclusively when humidity exceeded $70\%$. The impact of humidity becomes evident through granular analysis, with increase in error metrics and reduction in $R^2$ values at higher $\text{PM}_{2.5}$ and RH levels.

In Fig.~\ref{fig:95CI}, the $95\%$ confidence intervals show that the LCS concentration has a very narrow band, indicating good confidence in the LCS measurement. However, the $95\%$ prediction interval is much wider and there is a need to work on methods to reduce this prediction uncertainty. One of the possible methods would be to add a heating element to the LCS and study if the impact of increased RH and hygroscopic growth on LCS measurement can be reduced.

Over the four-month study, potential shifts were observed in the measurements which are likely due to a change in the major particulate sources~\cite{sharma2022long,shukla2023spatio}. Different pollution sources may affect LCS measurement as the shape and density of particles from different sources results in different scattering patterns leading to difference in the signals captured by LCS. Laboratory experiments with known particle sources and controlled conditions could provide further insights into the effect of high humidity and high $\text{PM}_{2.5}$ concentration on LCS measurement. The precision of each LCS group remained well within the USEPA-recommended limits, with the Sensirion group exhibiting the highest precision. Through granular analysis, we found that $\text{PM}_{2.5}$ concentration and humidity levels did not notably impact sensor precision. However, the reduced performance observed in Sensirion-3 unit, caused by airflow path blockage due to cobweb, highlighted the need for an improved airflow path design, particularly for outdoor deployments. This underscores the importance of regular maintenance to ensure continued accuracy and reliability of LCS.

The primary objective of this study was to assess the performance of collocated LCS under high $\text{PM}_{2.5}$ concentrations and high humidity. Our findings can serve as a valuable reference for researchers and users in selecting suitable sensors for high-exposure conditions in similar environments.  
Furthermore, the limitations of this study include the absence of particle size distribution and chemical composition data, which are crucial for further insights into the factors affecting sensor accuracy. Our future research plans include long-term studies with LCS and studies aimed at understanding sensor drift over time. These studies are expected to improve the understanding and performance of LCS in challenging environmental conditions.

This collocated study of selected LCS shows that these sensors are effective at capturing general $\text{PM}_{2.5}$ trends. However, data accuracy becomes a concern at humidity levels above $80\%$ and $\text{PM}_{2.5}$ concentrations exceeding $100~\mu\text{g/m}^3$. The precision of LCS, particularly Sensirion SPS30, remained within acceptable limits set by the USEPA. Overall, this study emphasizes the importance of considering environmental conditions, such as humidity and $\text{PM}_{2.5}$ levels, when calibrating and deploying LCS as a large network. Given the growing global concern over air pollution, there is an urgent need for reliable LCS to effectively monitor pollution exposure.
\appendices
\section*{Acknowledgment}
This research was partly funded by the Industrial Research and Development (IRD) unit of IIT Delhi through research project no. MI02086G. We also acknowledge the support of Prof. Sagnik Dey and Prof. Ravi Kumar Kunchala of the Centre of Atmospheric Sciences, IIT Delhi for providing access to the BAM data and the two sites. We thank Bavath D. and Saran Raj K. for their assistance in building the sensing monitors and setting up the GBAM site. We also thank Pankaj Pathak and Sonal Gangrade for proofreading the manuscript.
\section*{Data Availability}
All data generated or analysed during this study are included in this published article and supplementary information files.

\bibliographystyle{IEEEtran}
\bibliography{IEEEabrv,bibs}

\end{document}


\maketitle
%
%
\renewcommand{\figurename}{Figure}
\renewcommand{\thefigure}{S\arabic{figure}}

\begin{figure}[H]
\centering
\includegraphics[width=.9\linewidth]{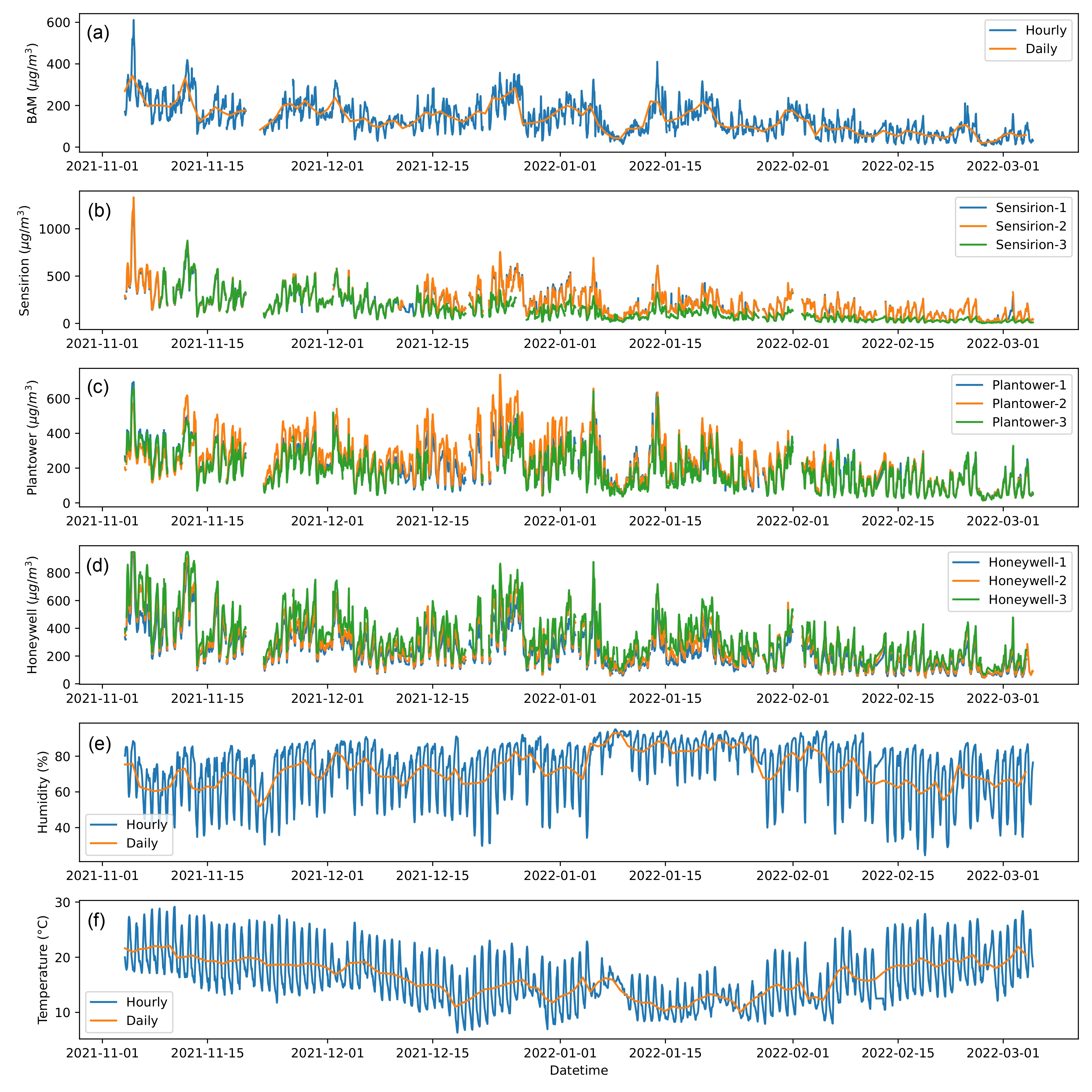}
\caption{Time-series plots depicting various parameters: (a) Reference BAM $\text{PM}_{2.5}$ concentration, (b) Sensirion $\text{PM}_{2.5}$, (c) Plantower $\text{PM}_{2.5}$, (d) Honeywell $\text{PM}_{2.5}$, (e) Humidity (\%), (f) Temperature ($^\circ \text{C}$). Missing data points are attributed to power failure and maintenance.}
\label{fig:time_series_plot}
\end{figure}
\clearpage
\begin{figure}
    \centering
    \includegraphics[width=1\textwidth]{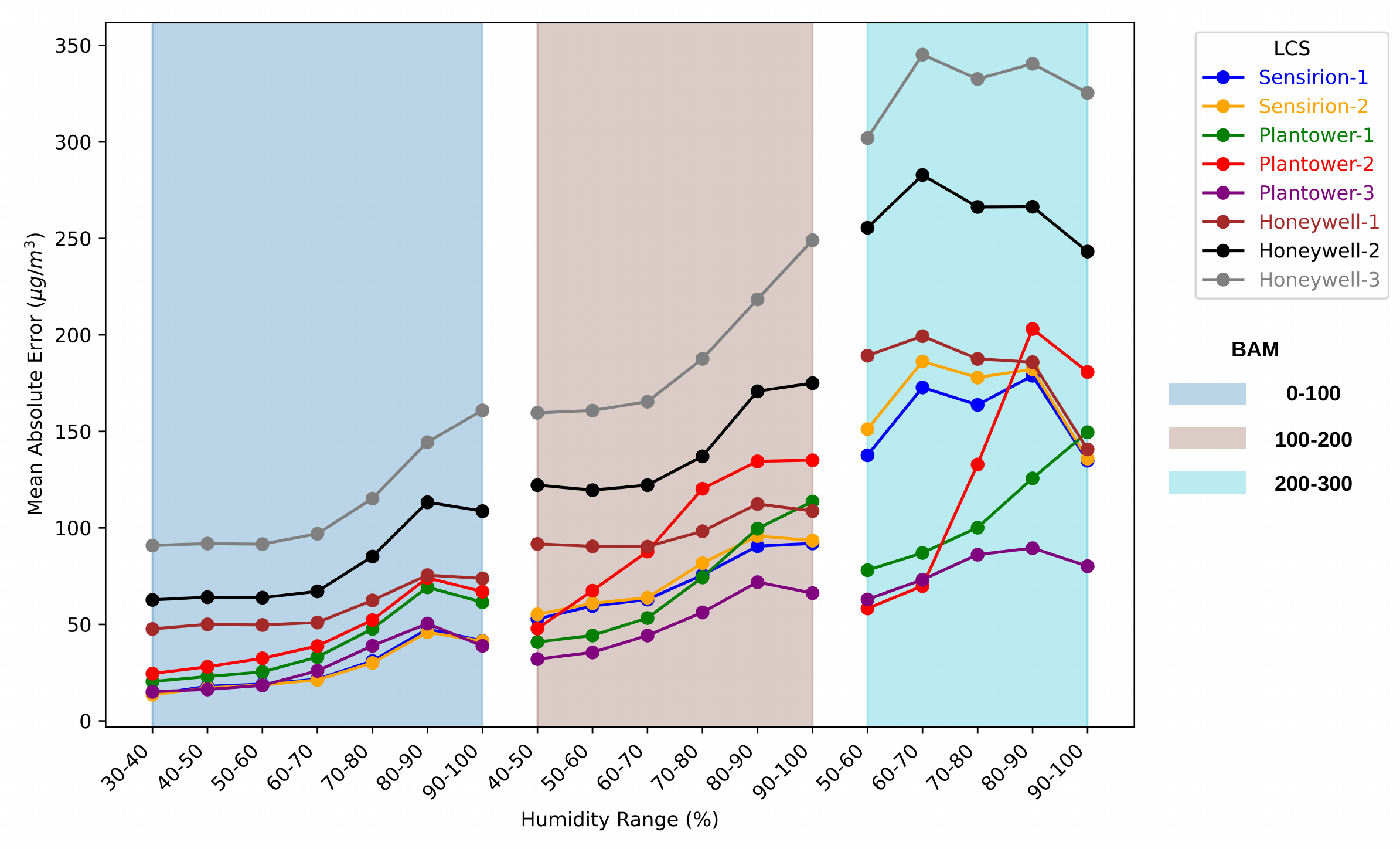}
    \caption{Mean absolute error (MAE) for hourly averaged granular data. With higher humidity levels under the same exposure, as well as higher $\text{PM}_{2.5}$ levels with similar humidity levels, an increase in MAE is observed. }
    \label{fig:Supplementary Figure 2}
\end{figure}
\begin{figure}
    \centering
    \includegraphics[width=.8\textwidth]{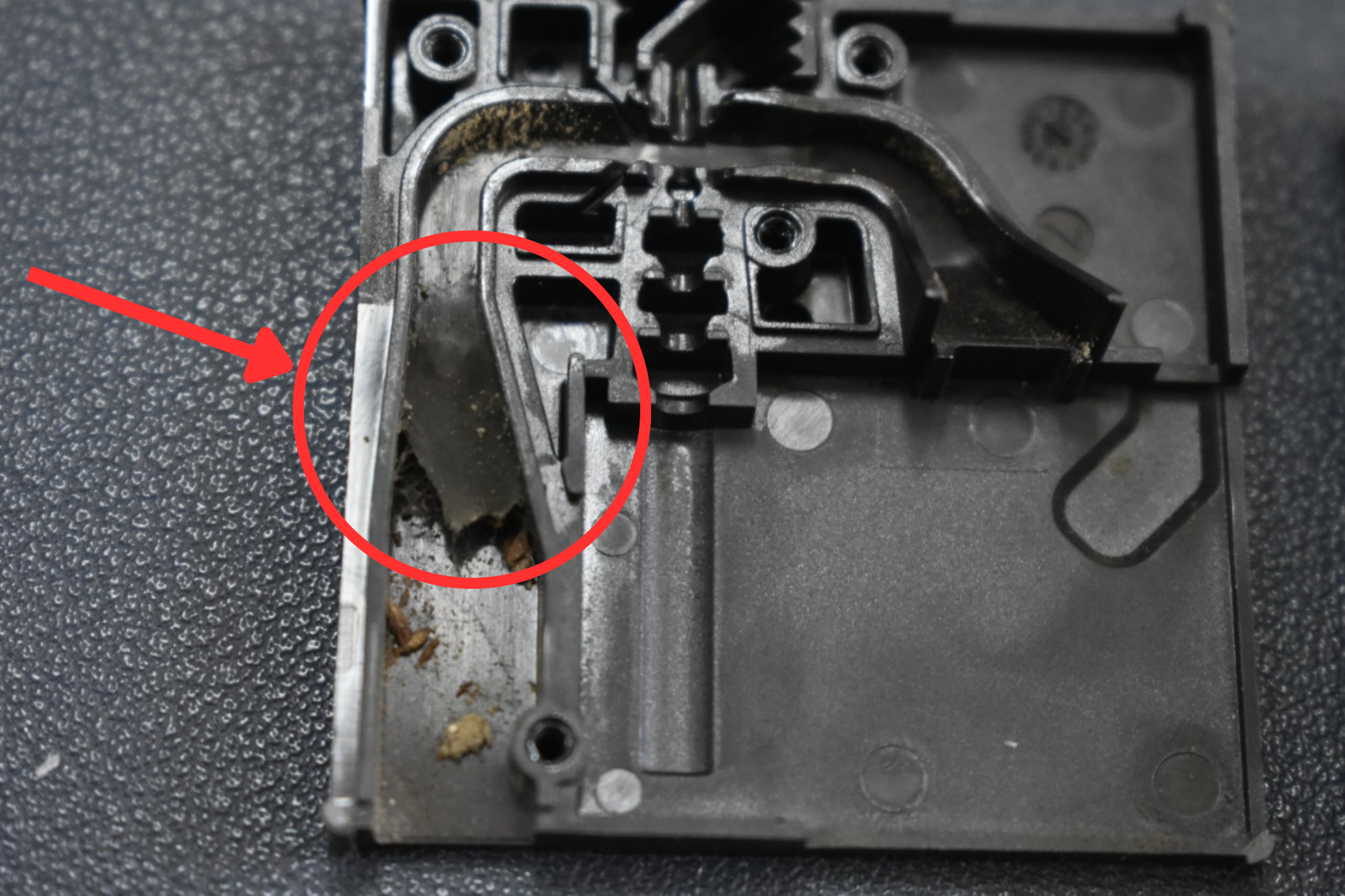}
    \caption{Sensirion-3 unit after being opened. Red circle shows the region where cobweb was found inside the Sensirion-3 unit.}
    \label{fig:Supplementary Figure 3}
\end{figure}
\begin{figure}
    \centering
    \includegraphics[width=1\textwidth]{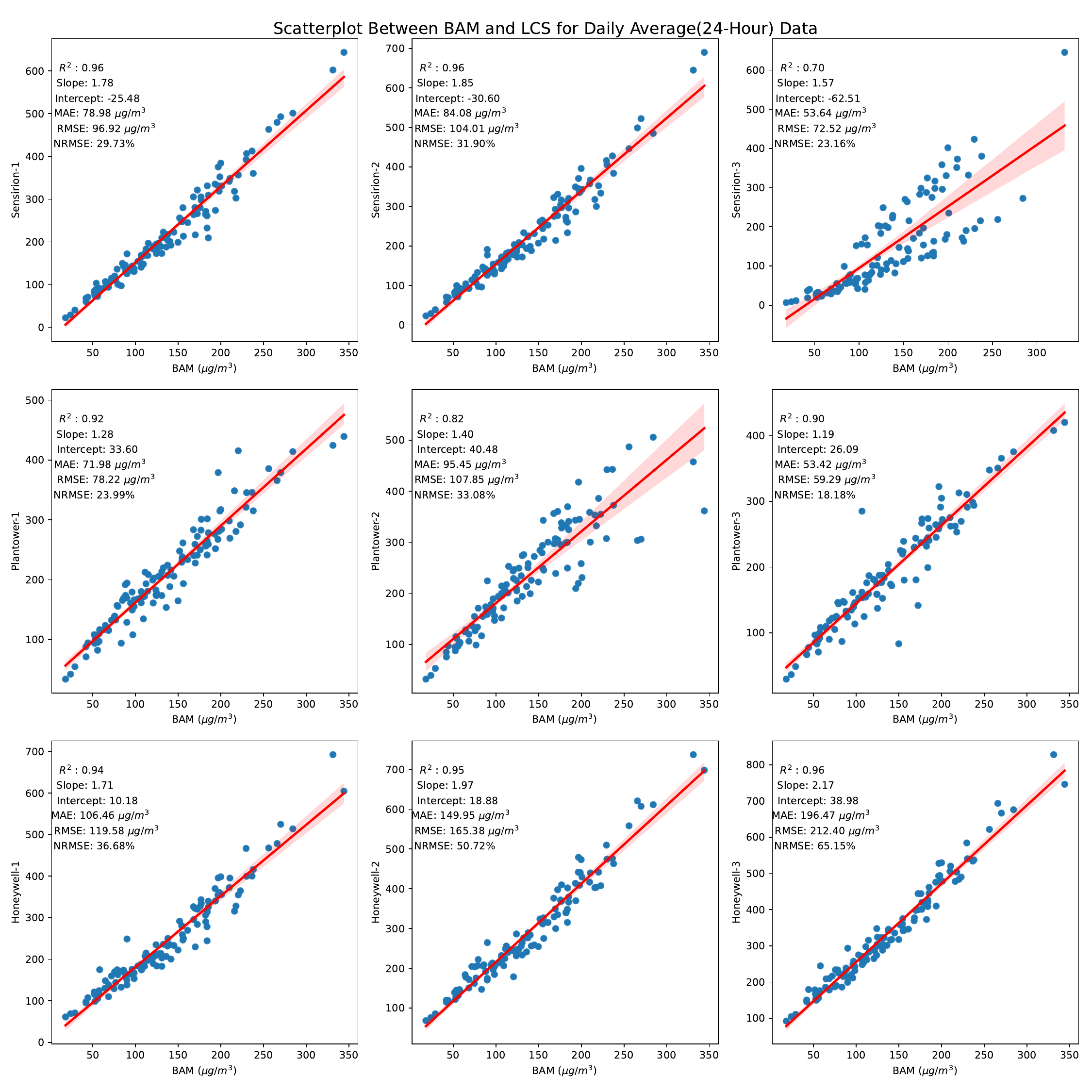}
    \caption{ Scatterplot between LCS and reference BAM for daily averaged data. The solid red line represents the linear regression fit to the daily average data points, with shaded regions indicating the $95\%$ confidence interval. Additionally, parameters such as linearity and error are displayed within each subplot corresponding to the LCS-reference pair.}
    \label{fig:Supplementary Figure 4}
\end{figure}
\begin{figure}
    \centering
    \includegraphics[width=1\textwidth]{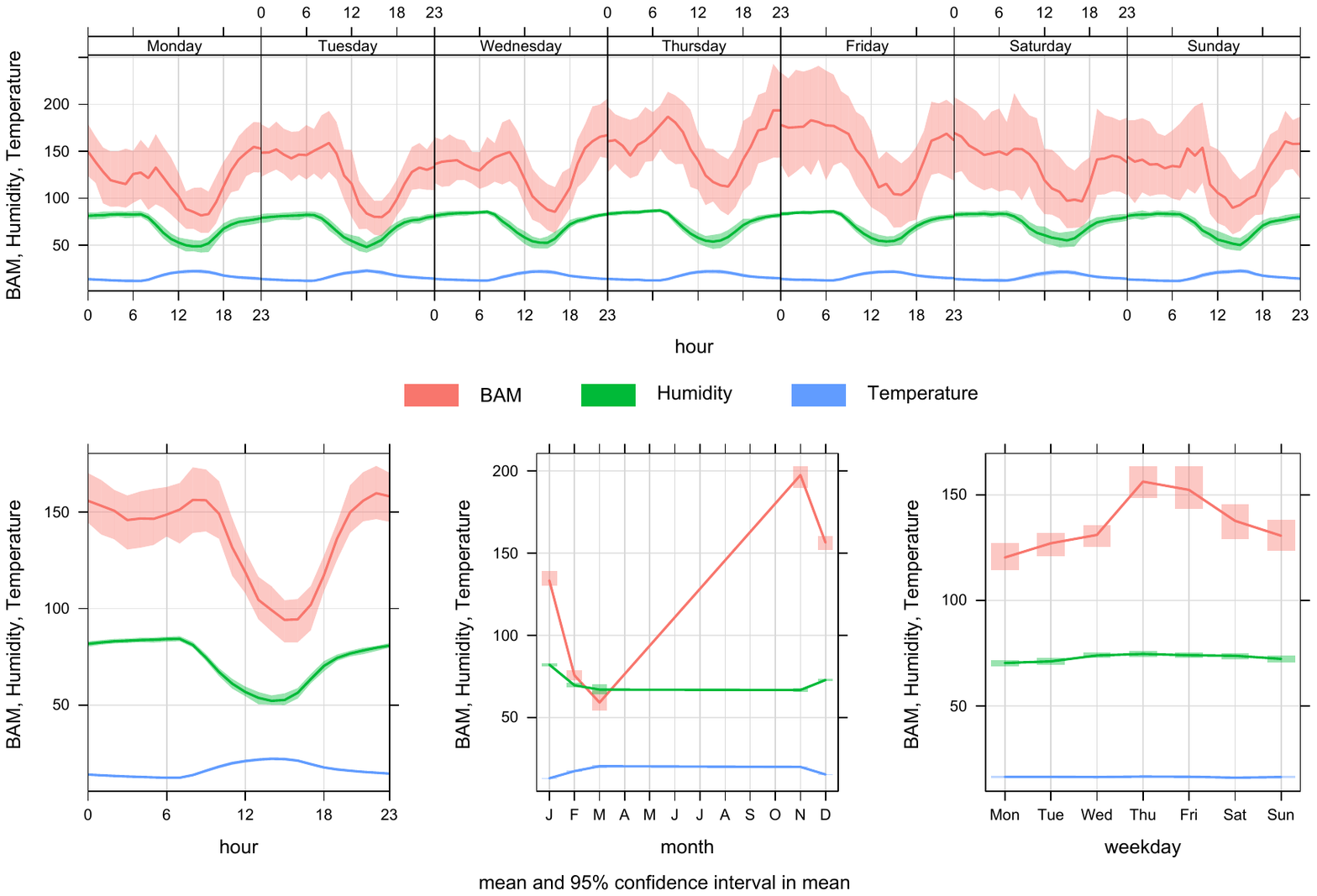}
    \caption{Time-series plot of reference BAM ($\text{PM}_{2.5}$), humidity and temperature based on day of the week. The subplots show Weekly, hourly, and monthly variation of these parameters.}
    \label{fig:Supplementary Figure 5}
\end{figure}
\begin{figure}
    \centering
    \includegraphics[width=.8\textwidth]{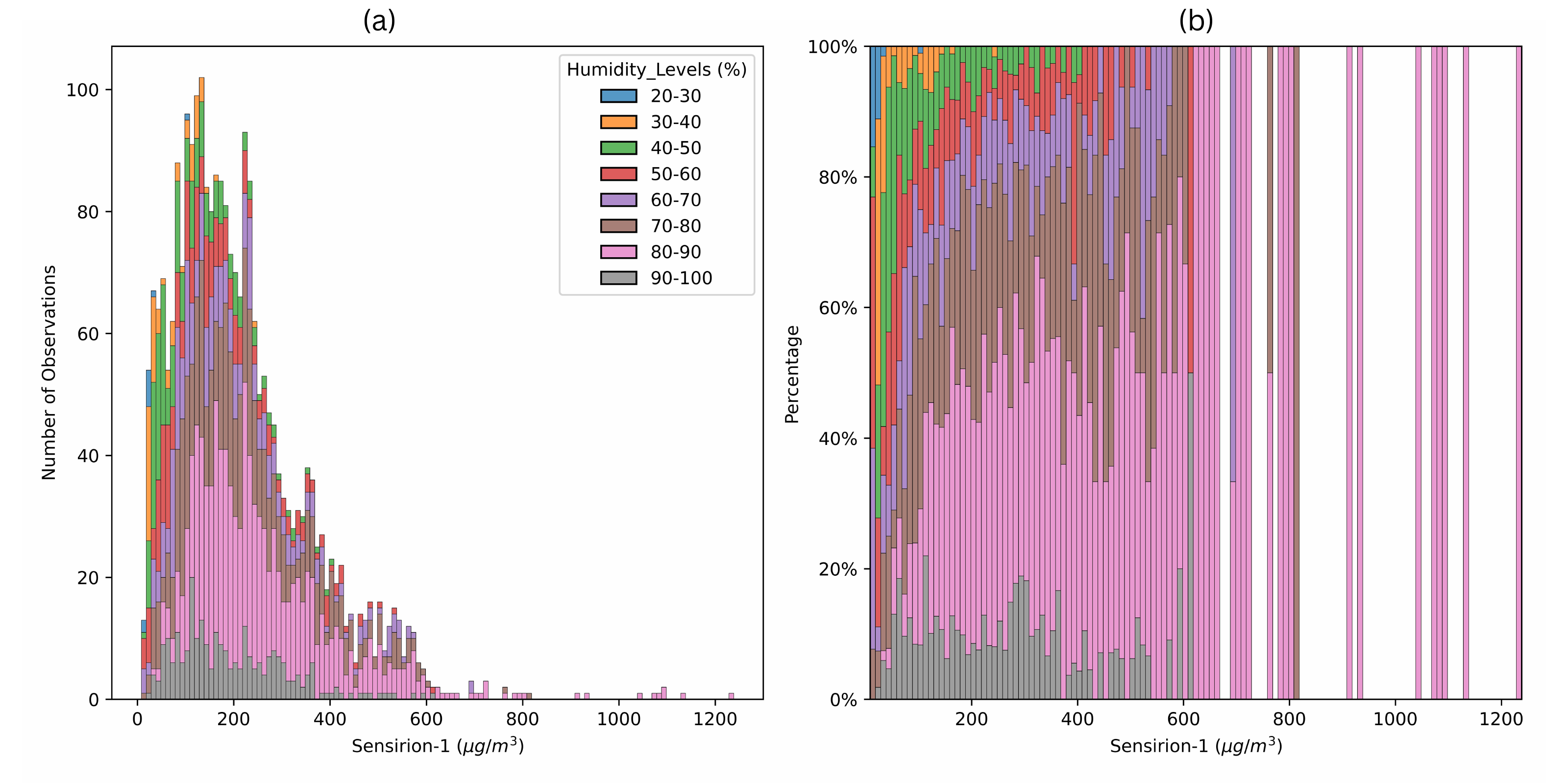}
    \caption{
    Distribution of $\text{PM}_{2.5}$ measurements across different humidity levels for the Sensirion-1 data. (a) Number of observations for different BAM $\text{PM}_{2.5}$ concentrations, stacked by RH levels, (b) Percentage distribution of Sensirion-1 concentrations within each humidity level. In these plots, the $\text{PM}_{2.5}$ concentration is divided into bins of size $10~\mu \text{g/m}^{3}$. It is seen that high $\text{PM}_{2.5}$ concentration episodes were predominantly accompanied by high RH levels.}
    \label{fig:Supplementary Figure 6}
\end{figure}
\begin{figure}
    \centering
    \includegraphics[width=\textwidth]{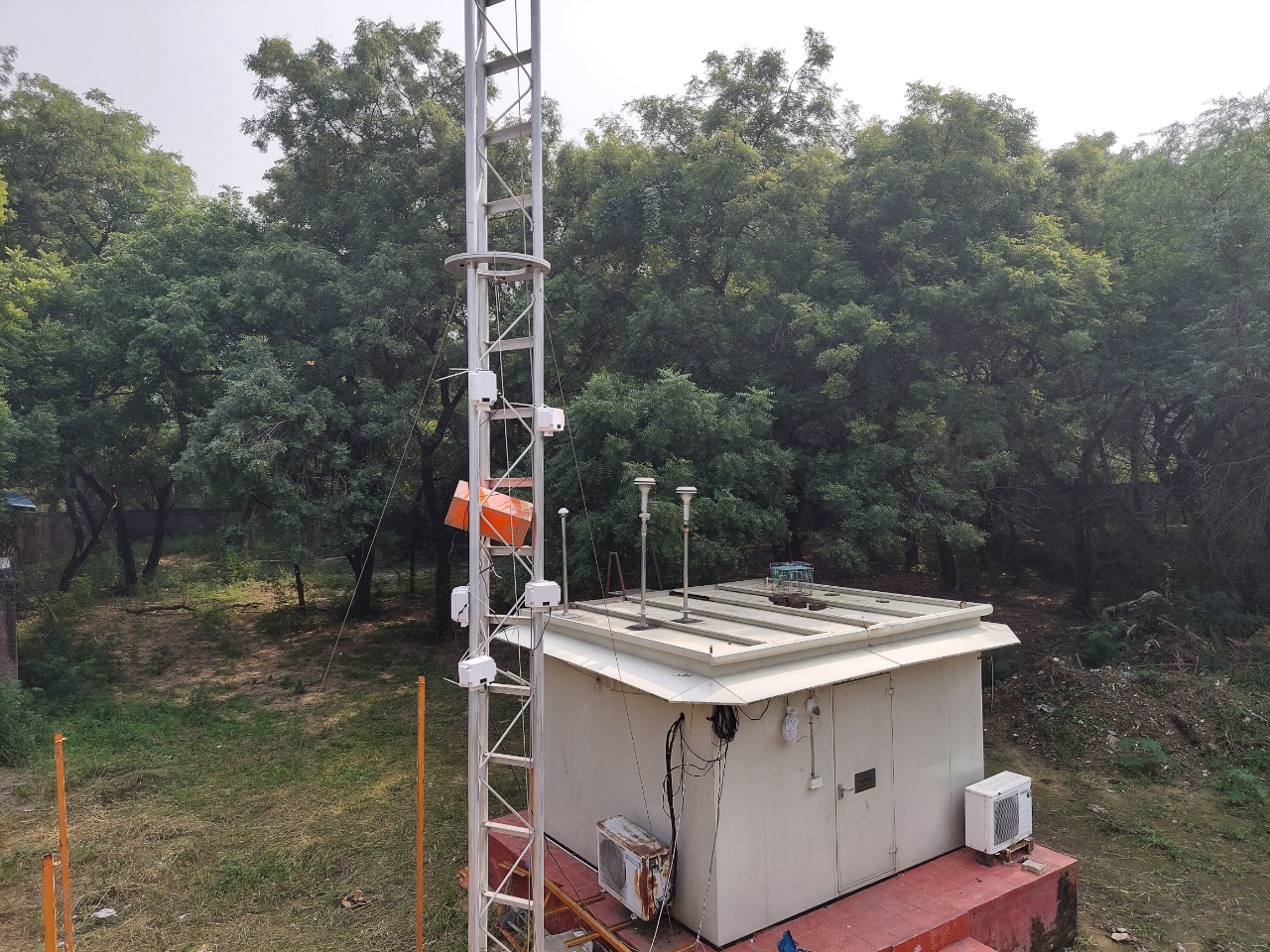}
    \caption{Sensor setup at the monitoring site. The low-cost sensing monitors are mounted on an aluminum pole.}
    \label{fig:Supplementary Figure 7}
\end{figure}
\FloatBarrier
%
%
\renewcommand{\tablename}{Table}
\renewcommand{\thetable}{S\arabic{table}}
\begin{table}[H]
\resizebox{1\textwidth}{!}{%
\begin{tabular}{|cccccccccc|}
\hline
\multicolumn{10}{|c|}{\textbf{$R^2$}}\\ 
\hline
\multicolumn{1}{|c|}{\textbf{\begin{tabular}[c]{@{}c@{}}BAM\\ $(\mu \text{g/m}^3)$\end{tabular}}} & \multicolumn{1}{c|}{\textbf{\begin{tabular}[c]{@{}c@{}}Humidity\\ $(\%)$\end{tabular}}} & \multicolumn{1}{c|}{\textbf{Sensirion-1}}           & \multicolumn{1}{c|}{\textbf{Sensirion-2}}           & \multicolumn{1}{c|}{\textbf{Plantower-1}}           & \multicolumn{1}{c|}{\textbf{Plantower-2}}           & \multicolumn{1}{c|}{\textbf{Plantower-3}}           & \multicolumn{1}{c|}{\textbf{Honeywell-1}}           & \multicolumn{1}{c|}{\textbf{Honeywell-2}}           & \textbf{Honeywell-3}           \\ \hline
\multicolumn{1}{|c|}{\cellcolor[HTML]{FDE9D9}}                                             & \multicolumn{1}{c|}{\cellcolor[HTML]{FDE9D9}20-30}                                      & \multicolumn{1}{c|}{\cellcolor[HTML]{6DC17C}0.9679} & \multicolumn{1}{c|}{\cellcolor[HTML]{6DC17C}0.9673} & \multicolumn{1}{c|}{\cellcolor[HTML]{6BC17C}0.9716} & \multicolumn{1}{c|}{\cellcolor[HTML]{66BF7C}0.9835} & \multicolumn{1}{c|}{\cellcolor[HTML]{63BE7B}0.9901} & \multicolumn{1}{c|}{\cellcolor[HTML]{6DC17C}0.9662} & \multicolumn{1}{c|}{\cellcolor[HTML]{8ECB7E}0.8877} & \cellcolor[HTML]{68C07C}0.9783 \\ \cline{2-10} 
\multicolumn{1}{|c|}{\cellcolor[HTML]{FDE9D9}}                                             & \multicolumn{1}{c|}{\cellcolor[HTML]{FDE9D9}30-40}                                      & \multicolumn{1}{c|}{\cellcolor[HTML]{75C37C}0.9485} & \multicolumn{1}{c|}{\cellcolor[HTML]{73C37C}0.9518} & \multicolumn{1}{c|}{\cellcolor[HTML]{81C77D}0.919}  & \multicolumn{1}{c|}{\cellcolor[HTML]{80C77D}0.9215} & \multicolumn{1}{c|}{\cellcolor[HTML]{9DCF7F}0.8503} & \multicolumn{1}{c|}{\cellcolor[HTML]{7CC57D}0.9317} & \multicolumn{1}{c|}{\cellcolor[HTML]{88C97E}0.9027} & \cellcolor[HTML]{89C97E}0.8999 \\ \cline{2-10} 
\multicolumn{1}{|c|}{\cellcolor[HTML]{FDE9D9}}                                             & \multicolumn{1}{c|}{\cellcolor[HTML]{FDE9D9}40-50}                                      & \multicolumn{1}{c|}{\cellcolor[HTML]{91CC7E}0.8788} & \multicolumn{1}{c|}{\cellcolor[HTML]{8ACA7E}0.896}  & \multicolumn{1}{c|}{\cellcolor[HTML]{A8D27F}0.8245} & \multicolumn{1}{c|}{\cellcolor[HTML]{86C87D}0.9072} & \multicolumn{1}{c|}{\cellcolor[HTML]{AFD480}0.8077} & \multicolumn{1}{c|}{\cellcolor[HTML]{9DCF7F}0.8498} & \multicolumn{1}{c|}{\cellcolor[HTML]{A5D27F}0.83}   & \cellcolor[HTML]{A6D27F}0.8295 \\ \cline{2-10} 
\multicolumn{1}{|c|}{\cellcolor[HTML]{FDE9D9}}                                             & \multicolumn{1}{c|}{\cellcolor[HTML]{FDE9D9}50-60}                                      & \multicolumn{1}{c|}{\cellcolor[HTML]{77C47D}0.9427} & \multicolumn{1}{c|}{\cellcolor[HTML]{74C37C}0.9502} & \multicolumn{1}{c|}{\cellcolor[HTML]{85C87D}0.9083} & \multicolumn{1}{c|}{\cellcolor[HTML]{7CC57D}0.9318} & \multicolumn{1}{c|}{\cellcolor[HTML]{90CB7E}0.8824} & \multicolumn{1}{c|}{\cellcolor[HTML]{8ECB7E}0.8864} & \multicolumn{1}{c|}{\cellcolor[HTML]{86C97E}0.9055} & \cellcolor[HTML]{A0D07F}0.8427 \\ \cline{2-10} 
\multicolumn{1}{|c|}{\cellcolor[HTML]{FDE9D9}}                                             & \multicolumn{1}{c|}{\cellcolor[HTML]{FDE9D9}60-70}                                      & \multicolumn{1}{c|}{\cellcolor[HTML]{95CD7E}0.8694} & \multicolumn{1}{c|}{\cellcolor[HTML]{9BCE7F}0.856}  & \multicolumn{1}{c|}{\cellcolor[HTML]{D3DF82}0.7188} & \multicolumn{1}{c|}{\cellcolor[HTML]{B1D580}0.802}  & \multicolumn{1}{c|}{\cellcolor[HTML]{CADC81}0.7406} & \multicolumn{1}{c|}{\cellcolor[HTML]{9CCF7F}0.8521} & \multicolumn{1}{c|}{\cellcolor[HTML]{ACD380}0.8141} & \cellcolor[HTML]{E9E583}0.6671 \\ \cline{2-10} 
\multicolumn{1}{|c|}{\cellcolor[HTML]{FDE9D9}}                                             & \multicolumn{1}{c|}{\cellcolor[HTML]{FDE9D9}70-80}                                      & \multicolumn{1}{c|}{\cellcolor[HTML]{C4DA81}0.7552} & \multicolumn{1}{c|}{\cellcolor[HTML]{BDD881}0.7718} & \multicolumn{1}{c|}{\cellcolor[HTML]{FEEB84}0.616}  & \multicolumn{1}{c|}{\cellcolor[HTML]{DFE283}0.6914} & \multicolumn{1}{c|}{\cellcolor[HTML]{FEE382}0.5767} & \multicolumn{1}{c|}{\cellcolor[HTML]{ECE683}0.6592} & \multicolumn{1}{c|}{\cellcolor[HTML]{FFEB84}0.6114} & \cellcolor[HTML]{FEE883}0.598  \\ \cline{2-10} 
\multicolumn{1}{|c|}{\cellcolor[HTML]{FDE9D9}}                                             & \multicolumn{1}{c|}{\cellcolor[HTML]{FDE9D9}80-90}                                      & \multicolumn{1}{c|}{\cellcolor[HTML]{FEE081}0.5624} & \multicolumn{1}{c|}{\cellcolor[HTML]{FEE983}0.604}  & \multicolumn{1}{c|}{\cellcolor[HTML]{FDD27F}0.4984} & \multicolumn{1}{c|}{\cellcolor[HTML]{FDD47F}0.5063} & \multicolumn{1}{c|}{\cellcolor[HTML]{FCC17C}0.416}  & \multicolumn{1}{c|}{\cellcolor[HTML]{FDD57F}0.5083} & \multicolumn{1}{c|}{\cellcolor[HTML]{FDC57C}0.4371} & \cellcolor[HTML]{FDD780}0.5185 \\ \cline{2-10} 
\multicolumn{1}{|c|}{\multirow{-8}{*}{\cellcolor[HTML]{FDE9D9}0-100}}                      & \multicolumn{1}{c|}{\cellcolor[HTML]{FDE9D9}90-100}                                     & \multicolumn{1}{c|}{\cellcolor[HTML]{B0D580}0.8033} & \multicolumn{1}{c|}{\cellcolor[HTML]{9CCF7F}0.8534} & \multicolumn{1}{c|}{\cellcolor[HTML]{DBE182}0.6996} & \multicolumn{1}{c|}{\cellcolor[HTML]{B8D780}0.7861} & \multicolumn{1}{c|}{\cellcolor[HTML]{EAE583}0.6637} & \multicolumn{1}{c|}{\cellcolor[HTML]{F3E884}0.6426} & \multicolumn{1}{c|}{\cellcolor[HTML]{D4DF82}0.718}  & \cellcolor[HTML]{FEEA83}0.6084 \\ \hline
\multicolumn{1}{|c|}{\cellcolor[HTML]{FCD5B4}}                                             & \multicolumn{1}{c|}{\cellcolor[HTML]{FCD5B4}40-50}                                      & \multicolumn{1}{c|}{\cellcolor[HTML]{D1DE82}0.724}  & \multicolumn{1}{c|}{\cellcolor[HTML]{B8D780}0.7849} & \multicolumn{1}{c|}{\cellcolor[HTML]{BDD881}0.7736} & \multicolumn{1}{c|}{\cellcolor[HTML]{FDCC7E}0.4695} & \multicolumn{1}{c|}{\cellcolor[HTML]{BED981}0.7709} & \multicolumn{1}{c|}{\cellcolor[HTML]{C4DA81}0.756}  & \multicolumn{1}{c|}{\cellcolor[HTML]{CEDD82}0.7326} & \cellcolor[HTML]{C4DA81}0.7561 \\ \cline{2-10} 
\multicolumn{1}{|c|}{\cellcolor[HTML]{FCD5B4}}                                             & \multicolumn{1}{c|}{\cellcolor[HTML]{FCD5B4}50-60}                                      & \multicolumn{1}{c|}{\cellcolor[HTML]{98CE7F}0.8638} & \multicolumn{1}{c|}{\cellcolor[HTML]{9ACE7F}0.8581} & \multicolumn{1}{c|}{\cellcolor[HTML]{A7D27F}0.8253} & \multicolumn{1}{c|}{\cellcolor[HTML]{FEDF81}0.5582} & \multicolumn{1}{c|}{\cellcolor[HTML]{C1D981}0.7631} & \multicolumn{1}{c|}{\cellcolor[HTML]{9BCE7F}0.856}  & \multicolumn{1}{c|}{\cellcolor[HTML]{9BCF7F}0.8551} & \cellcolor[HTML]{9ECF7F}0.8481 \\ \cline{2-10} 
\multicolumn{1}{|c|}{\cellcolor[HTML]{FCD5B4}}                                             & \multicolumn{1}{c|}{\cellcolor[HTML]{FCD5B4}60-70}                                      & \multicolumn{1}{c|}{\cellcolor[HTML]{F7E984}0.6331} & \multicolumn{1}{c|}{\cellcolor[HTML]{E7E583}0.6699} & \multicolumn{1}{c|}{\cellcolor[HTML]{FEE482}0.5806} & \multicolumn{1}{c|}{\cellcolor[HTML]{FCB379}0.3513} & \multicolumn{1}{c|}{\cellcolor[HTML]{FDC97D}0.4532} & \multicolumn{1}{c|}{\cellcolor[HTML]{E4E483}0.6781} & \multicolumn{1}{c|}{\cellcolor[HTML]{E3E383}0.6807} & \cellcolor[HTML]{D1DE82}0.7247 \\ \cline{2-10} 
\multicolumn{1}{|c|}{\cellcolor[HTML]{FCD5B4}}                                             & \multicolumn{1}{c|}{\cellcolor[HTML]{FCD5B4}70-80}                                      & \multicolumn{1}{c|}{\cellcolor[HTML]{FEE382}0.5743} & \multicolumn{1}{c|}{\cellcolor[HTML]{FED880}0.5258} & \multicolumn{1}{c|}{\cellcolor[HTML]{FDC87D}0.447}  & \multicolumn{1}{c|}{\cellcolor[HTML]{FCB479}0.3551} & \multicolumn{1}{c|}{\cellcolor[HTML]{FCBC7A}0.3913} & \multicolumn{1}{c|}{\cellcolor[HTML]{FEE582}0.5858} & \multicolumn{1}{c|}{\cellcolor[HTML]{FEE582}0.5864} & \cellcolor[HTML]{FEE282}0.5717 \\ \cline{2-10} 
\multicolumn{1}{|c|}{\cellcolor[HTML]{FCD5B4}}                                             & \multicolumn{1}{c|}{\cellcolor[HTML]{FCD5B4}80-90}                                      & \multicolumn{1}{c|}{\cellcolor[HTML]{FEDB81}0.5404} & \multicolumn{1}{c|}{\cellcolor[HTML]{FEDF81}0.5588} & \multicolumn{1}{c|}{\cellcolor[HTML]{FCC27C}0.419}  & \multicolumn{1}{c|}{\cellcolor[HTML]{FCC47C}0.4311} & \multicolumn{1}{c|}{\cellcolor[HTML]{FCB479}0.3545} & \multicolumn{1}{c|}{\cellcolor[HTML]{FDCA7D}0.46}   & \multicolumn{1}{c|}{\cellcolor[HTML]{FDC97D}0.4554} & \cellcolor[HTML]{FED980}0.5272 \\ \cline{2-10} 
\multicolumn{1}{|c|}{\multirow{-6}{*}{\cellcolor[HTML]{FCD5B4}100-200}}                    & \multicolumn{1}{c|}{\cellcolor[HTML]{FCD5B4}90-100}                                     & \multicolumn{1}{c|}{\cellcolor[HTML]{EAE583}0.6633} & \multicolumn{1}{c|}{\cellcolor[HTML]{CEDD82}0.7315} & \multicolumn{1}{c|}{\cellcolor[HTML]{FBA376}0.2767} & \multicolumn{1}{c|}{\cellcolor[HTML]{FDD37F}0.5003} & \multicolumn{1}{c|}{\cellcolor[HTML]{FDCC7E}0.4699} & \multicolumn{1}{c|}{\cellcolor[HTML]{FEDE81}0.5512} & \multicolumn{1}{c|}{\cellcolor[HTML]{FEDD81}0.5461} & \cellcolor[HTML]{FEE182}0.567  \\ \hline
\multicolumn{1}{|c|}{\cellcolor[HTML]{FABF8F}}                                             & \multicolumn{1}{c|}{\cellcolor[HTML]{FABF8F}50-60}                                      & \multicolumn{1}{c|}{\cellcolor[HTML]{FEDB81}0.5391} & \multicolumn{1}{c|}{\cellcolor[HTML]{F1E784}0.6475} & \multicolumn{1}{c|}{\cellcolor[HTML]{FEE983}0.6038} & \multicolumn{1}{c|}{\cellcolor[HTML]{F98570}0.1321} & \multicolumn{1}{c|}{\cellcolor[HTML]{E1E383}0.6843} & \multicolumn{1}{c|}{\cellcolor[HTML]{F6E984}0.6333} & \multicolumn{1}{c|}{\cellcolor[HTML]{FEDD81}0.5491} & \cellcolor[HTML]{FEE282}0.5727 \\ \cline{2-10} 
\multicolumn{1}{|c|}{\cellcolor[HTML]{FABF8F}}                                             & \multicolumn{1}{c|}{\cellcolor[HTML]{FABF8F}60-70}                                      & \multicolumn{1}{c|}{\cellcolor[HTML]{C0D981}0.765}  & \multicolumn{1}{c|}{\cellcolor[HTML]{C5DB81}0.7536} & \multicolumn{1}{c|}{\cellcolor[HTML]{F1E784}0.6468} & \multicolumn{1}{c|}{\cellcolor[HTML]{FA9473}0.2053} & \multicolumn{1}{c|}{\cellcolor[HTML]{E5E483}0.6747} & \multicolumn{1}{c|}{\cellcolor[HTML]{D1DE82}0.7251} & \multicolumn{1}{c|}{\cellcolor[HTML]{D9E082}0.7043} & \cellcolor[HTML]{D5DF82}0.714  \\ \cline{2-10} 
\multicolumn{1}{|c|}{\cellcolor[HTML]{FABF8F}}                                             & \multicolumn{1}{c|}{\cellcolor[HTML]{FABF8F}70-80}                                      & \multicolumn{1}{c|}{\cellcolor[HTML]{FDD37F}0.501}  & \multicolumn{1}{c|}{\cellcolor[HTML]{FDC77D}0.4462} & \multicolumn{1}{c|}{\cellcolor[HTML]{FBA676}0.2911} & \multicolumn{1}{c|}{\cellcolor[HTML]{F8696B}0.0004} & \multicolumn{1}{c|}{\cellcolor[HTML]{FA9F75}0.2579} & \multicolumn{1}{c|}{\cellcolor[HTML]{FCC47C}0.4293} & \multicolumn{1}{c|}{\cellcolor[HTML]{FDD37F}0.5022} & \cellcolor[HTML]{FDD880}0.5232 \\ \cline{2-10} 
\multicolumn{1}{|c|}{\cellcolor[HTML]{FABF8F}}                                             & \multicolumn{1}{c|}{\cellcolor[HTML]{FABF8F}80-90}                                      & \multicolumn{1}{c|}{\cellcolor[HTML]{FBB078}0.3358} & \multicolumn{1}{c|}{\cellcolor[HTML]{FCB579}0.3588} & \multicolumn{1}{c|}{\cellcolor[HTML]{FCB87A}0.3745} & \multicolumn{1}{c|}{\cellcolor[HTML]{FA9172}0.1901} & \multicolumn{1}{c|}{\cellcolor[HTML]{FBAB77}0.3132} & \multicolumn{1}{c|}{\cellcolor[HTML]{FCBE7B}0.4042} & \multicolumn{1}{c|}{\cellcolor[HTML]{FCBE7B}0.4018} & \cellcolor[HTML]{FCC37C}0.4241 \\ \cline{2-10} 
\multicolumn{1}{|c|}{\multirow{-5}{*}{\cellcolor[HTML]{FABF8F}200-300}}                    & \multicolumn{1}{c|}{\cellcolor[HTML]{FABF8F}90-100}                                     & \multicolumn{1}{c|}{\cellcolor[HTML]{FCB679}0.3645} & \multicolumn{1}{c|}{\cellcolor[HTML]{FCB579}0.3583} & \multicolumn{1}{c|}{\cellcolor[HTML]{FA9072}0.1876} & \multicolumn{1}{c|}{\cellcolor[HTML]{FCB77A}0.3681} & \multicolumn{1}{c|}{\cellcolor[HTML]{FA9072}0.187}  & \multicolumn{1}{c|}{\cellcolor[HTML]{FBA376}0.275}  & \multicolumn{1}{c|}{\cellcolor[HTML]{FCB479}0.3552} & \cellcolor[HTML]{FA9B74}0.24   \\ \hline
\multicolumn{1}{|c|}{\cellcolor[HTML]{E26B0A}}                                             & \multicolumn{1}{c|}{\cellcolor[HTML]{E26B0A}70-80}                                      & \multicolumn{1}{c|}{\cellcolor[HTML]{F86B6B}0.0132} & \multicolumn{1}{c|}{\cellcolor[HTML]{F86A6B}0.0069} & \multicolumn{1}{c|}{\cellcolor[HTML]{F97F6F}0.104}  & \multicolumn{1}{c|}{\cellcolor[HTML]{F86B6B}0.0144} & \multicolumn{1}{c|}{\cellcolor[HTML]{F8706C}0.0355} & \multicolumn{1}{c|}{\cellcolor[HTML]{F8696B}0.0029} & \multicolumn{1}{c|}{\cellcolor[HTML]{F8696B}0.0007} & \cellcolor[HTML]{FFFFFF}       \\ \cline{2-10} 
\multicolumn{1}{|c|}{\multirow{-2}{*}{\cellcolor[HTML]{E26B0A}300-400}}                    & \multicolumn{1}{c|}{\cellcolor[HTML]{E26B0A}80-90}                                      & \multicolumn{1}{c|}{\cellcolor[HTML]{FBA276}0.2693} & \multicolumn{1}{c|}{\cellcolor[HTML]{FBA777}0.2945} & \multicolumn{1}{c|}{\cellcolor[HTML]{F86B6B}0.0107} & \multicolumn{1}{c|}{\cellcolor[HTML]{F8696B}0.0012} & \multicolumn{1}{c|}{\cellcolor[HTML]{F8796E}0.0783} & \multicolumn{1}{c|}{\cellcolor[HTML]{FBB279}0.3451} & \multicolumn{1}{c|}{\cellcolor[HTML]{FBA676}0.2907} & \cellcolor[HTML]{FA9974}0.2273 \\ \hline
\end{tabular}%
}
\caption{$R^2$ values for different humidity and reference $\text{PM}_{2.5}$ values/bins (hourly averaged data). A graded color scheme has been applied to the $R^2$ values in the table with red, yellow, and green representing the lowest, middle, and highest $R^2$ values, respectively.}
\label{tab:my-table}
\end{table}

\begin{table}[H]
\resizebox{1\textwidth}{!}{
\centering
\begin{tabular}{|cccccccccc|}
\hline
\multicolumn{10}{|c|}{\textbf{Mean Absolute Error (MAE)}}\\ \hline
\multicolumn{1}{|c|}{\begin{tabular}[c]{@{}c@{}}\textbf{BAM}\\ $(\mu \text{g/m}^3)$\end{tabular}} & \multicolumn{1}{c|}{\begin{tabular}[c]{@{}c@{}}\textbf{Humidity}\\ $(\%)$\end{tabular}} & \multicolumn{1}{c|}{\begin{tabular}[c]{@{}c@{}}\textbf{Sensirion-1}\\ $(\mu \text{g/m}^3)$\end{tabular}} & \multicolumn{1}{c|}{\begin{tabular}[c]{@{}c@{}}\textbf{Sensirion-2}\\ $(\mu \text{g/m}^3)$\end{tabular}} & \multicolumn{1}{c|}{\begin{tabular}[c]{@{}c@{}}\textbf{Plantower-1}\\ $(\mu \text{g/m}^3)$\end{tabular}} & \multicolumn{1}{c|}{\begin{tabular}[c]{@{}c@{}}\textbf{Plantower-2}\\ $(\mu \text{g/m}^3)$\end{tabular}} & \multicolumn{1}{c|}{\begin{tabular}[c]{@{}c@{}}\textbf{Plantower-3}\\ $(\mu \text{g/m}^3)$\end{tabular}} & \multicolumn{1}{c|}{\begin{tabular}[c]{@{}c@{}}\textbf{Honeywell-1}\\ $(\mu \text{g/m}^3)$\end{tabular}} & \multicolumn{1}{c|}{\begin{tabular}[c]{@{}c@{}}\textbf{Honeywell-2}\\ $(\mu \text{g/m}^3)$\end{tabular}} & \begin{tabular}[c]{@{}c@{}}\textbf{Honeywell-3}\\ $(\mu \text{g/m}^3)$\end{tabular} \\ \hline
\multicolumn{1}{|c|}{\cellcolor[HTML]{FDE9D9}}                                            & \multicolumn{1}{c|}{\cellcolor[HTML]{FDE9D9}20-30}                                      & \multicolumn{1}{c|}{\cellcolor[HTML]{00B050}5.37}                                               & \multicolumn{1}{c|}{\cellcolor[HTML]{00B050}5.36}                                                 & \multicolumn{1}{c|}{\cellcolor[HTML]{21B756}16.01}                                                & \multicolumn{1}{c|}{\cellcolor[HTML]{20B756}15.54}                                                & \multicolumn{1}{c|}{\cellcolor[HTML]{17B554}12.89}                                                & \multicolumn{1}{c|}{\cellcolor[HTML]{73CA67}41.67}                                                & \multicolumn{1}{c|}{\cellcolor[HTML]{8CD06C}49.70}                                                & \cellcolor[HTML]{F2E881}81.85                                                \\ \cline{2-10} 
\multicolumn{1}{|c|}{\cellcolor[HTML]{FDE9D9}}                                            & \multicolumn{1}{c|}{\cellcolor[HTML]{FDE9D9}30-40}                                      & \multicolumn{1}{c|}{\cellcolor[HTML]{1AB655}13.72}                                              & \multicolumn{1}{c|}{\cellcolor[HTML]{19B555}13.50}                                                & \multicolumn{1}{c|}{\cellcolor[HTML]{2FBB59}20.38}                                                & \multicolumn{1}{c|}{\cellcolor[HTML]{3CBE5C}24.52}                                                & \multicolumn{1}{c|}{\cellcolor[HTML]{1EB756}15.06}                                                & \multicolumn{1}{c|}{\cellcolor[HTML]{86CF6B}47.64}                                                & \multicolumn{1}{c|}{\cellcolor[HTML]{B6DA75}62.72}                                                & \cellcolor[HTML]{FFE883}90.82                                                \\ \cline{2-10} 
\multicolumn{1}{|c|}{\cellcolor[HTML]{FDE9D9}}                                            & \multicolumn{1}{c|}{\cellcolor[HTML]{FDE9D9}40-50}                                      & \multicolumn{1}{c|}{\cellcolor[HTML]{27B958}17.87}                                              & \multicolumn{1}{c|}{\cellcolor[HTML]{25B857}17.10}                                                & \multicolumn{1}{c|}{\cellcolor[HTML]{37BC5B}22.98}                                                & \multicolumn{1}{c|}{\cellcolor[HTML]{48C05E}28.03}                                                & \multicolumn{1}{c|}{\cellcolor[HTML]{22B857}16.31}                                                & \multicolumn{1}{c|}{\cellcolor[HTML]{8DD06C}50.02}                                                & \multicolumn{1}{c|}{\cellcolor[HTML]{BADB76}64.10}                                                & \cellcolor[HTML]{FFE882}91.85                                                \\ \cline{2-10} 
\multicolumn{1}{|c|}{\cellcolor[HTML]{FDE9D9}}                                            & \multicolumn{1}{c|}{\cellcolor[HTML]{FDE9D9}50-60}                                      & \multicolumn{1}{c|}{\cellcolor[HTML]{2BBA58}19.02}                                              & \multicolumn{1}{c|}{\cellcolor[HTML]{29B958}18.55}                                                & \multicolumn{1}{c|}{\cellcolor[HTML]{3FBE5C}25.34}                                                & \multicolumn{1}{c|}{\cellcolor[HTML]{55C361}32.40}                                                & \multicolumn{1}{c|}{\cellcolor[HTML]{29B958}18.38}                                                & \multicolumn{1}{c|}{\cellcolor[HTML]{8CD06C}49.69}                                                & \multicolumn{1}{c|}{\cellcolor[HTML]{B9DA75}63.84}                                                & \cellcolor[HTML]{FFE882}91.54                                                \\ \cline{2-10} 
\multicolumn{1}{|c|}{\cellcolor[HTML]{FDE9D9}}                                            & \multicolumn{1}{c|}{\cellcolor[HTML]{FDE9D9}60-70}                                      & \multicolumn{1}{c|}{\cellcolor[HTML]{33BB5A}21.62}                                              & \multicolumn{1}{c|}{\cellcolor[HTML]{32BB5A}21.22}                                                & \multicolumn{1}{c|}{\cellcolor[HTML]{57C461}33.06}                                                & \multicolumn{1}{c|}{\cellcolor[HTML]{6AC865}38.80}                                                & \multicolumn{1}{c|}{\cellcolor[HTML]{41BF5D}25.92}                                                & \multicolumn{1}{c|}{\cellcolor[HTML]{90D16D}50.97}                                                & \multicolumn{1}{c|}{\cellcolor[HTML]{C4DD77}67.09}                                                & \cellcolor[HTML]{FFE480}96.94                                                \\ \cline{2-10} 
\multicolumn{1}{|c|}{\cellcolor[HTML]{FDE9D9}}                                            & \multicolumn{1}{c|}{\cellcolor[HTML]{FDE9D9}70-80}                                      & \multicolumn{1}{c|}{\cellcolor[HTML]{51C260}31.02}                                              & \multicolumn{1}{c|}{\cellcolor[HTML]{4EC25F}30.04}                                                & \multicolumn{1}{c|}{\cellcolor[HTML]{86CF6B}47.68}                                                & \multicolumn{1}{c|}{\cellcolor[HTML]{94D26E}52.23}                                                & \multicolumn{1}{c|}{\cellcolor[HTML]{6AC865}38.93}                                                & \multicolumn{1}{c|}{\cellcolor[HTML]{B5D974}62.44}                                                & \multicolumn{1}{c|}{\cellcolor[HTML]{FDEA83}85.15}                                                & \cellcolor[HTML]{FFD97A}115.19                                               \\ \cline{2-10} 
\multicolumn{1}{|c|}{\cellcolor[HTML]{FDE9D9}}                                            & \multicolumn{1}{c|}{\cellcolor[HTML]{FDE9D9}80-90}                                      & \multicolumn{1}{c|}{\cellcolor[HTML]{85CE6B}47.52}                                              & \multicolumn{1}{c|}{\cellcolor[HTML]{80CD6A}45.94}                                                & \multicolumn{1}{c|}{\cellcolor[HTML]{CADE79}69.22}                                                & \multicolumn{1}{c|}{\cellcolor[HTML]{DAE27C}74.10}                                                & \multicolumn{1}{c|}{\cellcolor[HTML]{8FD16D}50.40}                                                & \multicolumn{1}{c|}{\cellcolor[HTML]{DEE37D}75.50}                                                & \multicolumn{1}{c|}{\cellcolor[HTML]{FFDA7B}113.27}                                               & \cellcolor[HTML]{FFC66F}144.42                                               \\ \cline{2-10} 
\multicolumn{1}{|c|}{\multirow{-8}{*}{\cellcolor[HTML]{FDE9D9}0-100}}                     & \multicolumn{1}{c|}{\cellcolor[HTML]{FDE9D9}90-100}                                     & \multicolumn{1}{c|}{\cellcolor[HTML]{72CA67}41.45}                                              & \multicolumn{1}{c|}{\cellcolor[HTML]{72CA67}41.28}                                                & \multicolumn{1}{c|}{\cellcolor[HTML]{B2D974}61.49}                                                & \multicolumn{1}{c|}{\cellcolor[HTML]{C3DD77}66.93}                                                & \multicolumn{1}{c|}{\cellcolor[HTML]{6AC865}38.89}                                                & \multicolumn{1}{c|}{\cellcolor[HTML]{D9E27C}73.81}                                                & \multicolumn{1}{c|}{\cellcolor[HTML]{FFDD7C}108.69}                                               & \cellcolor[HTML]{FFBB69}160.91                                               \\ \hline
\multicolumn{1}{|c|}{\cellcolor[HTML]{FCD5B4}}                                            & \multicolumn{1}{c|}{\cellcolor[HTML]{FCD5B4}40-50}                                      & \multicolumn{1}{c|}{\cellcolor[HTML]{96D26E}52.82}                                              & \multicolumn{1}{c|}{\cellcolor[HTML]{9ED470}55.19}                                                & \multicolumn{1}{c|}{\cellcolor[HTML]{70CA67}40.88}                                                & \multicolumn{1}{c|}{\cellcolor[HTML]{86CF6B}47.84}                                                & \multicolumn{1}{c|}{\cellcolor[HTML]{54C361}32.01}                                                & \multicolumn{1}{c|}{\cellcolor[HTML]{FFE882}91.66}                                                & \multicolumn{1}{c|}{\cellcolor[HTML]{FFD477}122.16}                                               & \cellcolor[HTML]{FFBC6A}159.59                                               \\ \cline{2-10} 
\multicolumn{1}{|c|}{\cellcolor[HTML]{FCD5B4}}                                            & \multicolumn{1}{c|}{\cellcolor[HTML]{FCD5B4}50-60}                                      & \multicolumn{1}{c|}{\cellcolor[HTML]{ACD773}59.52}                                              & \multicolumn{1}{c|}{\cellcolor[HTML]{B0D873}60.90}                                                & \multicolumn{1}{c|}{\cellcolor[HTML]{7BCC69}44.26}                                                & \multicolumn{1}{c|}{\cellcolor[HTML]{C5DD78}67.48}                                                & \multicolumn{1}{c|}{\cellcolor[HTML]{5FC663}35.47}                                                & \multicolumn{1}{c|}{\cellcolor[HTML]{FFE883}90.42}                                                & \multicolumn{1}{c|}{\cellcolor[HTML]{FFD678}119.52}                                               & \cellcolor[HTML]{FFBB69}160.78                                               \\ \cline{2-10} 
\multicolumn{1}{|c|}{\cellcolor[HTML]{FCD5B4}}                                            & \multicolumn{1}{c|}{\cellcolor[HTML]{FCD5B4}60-70}                                      & \multicolumn{1}{c|}{\cellcolor[HTML]{B6DA75}62.88}                                              & \multicolumn{1}{c|}{\cellcolor[HTML]{B9DA75}63.85}                                                & \multicolumn{1}{c|}{\cellcolor[HTML]{98D36F}53.31}                                                & \multicolumn{1}{c|}{\cellcolor[HTML]{FFEA84}87.74}                                                & \multicolumn{1}{c|}{\cellcolor[HTML]{7BCC69}44.24}                                                & \multicolumn{1}{c|}{\cellcolor[HTML]{FFE983}90.30}                                                & \multicolumn{1}{c|}{\cellcolor[HTML]{FFD477}122.15}                                               & \cellcolor[HTML]{FFB868}165.40                                               \\ \cline{2-10} 
\multicolumn{1}{|c|}{\cellcolor[HTML]{FCD5B4}}                                            & \multicolumn{1}{c|}{\cellcolor[HTML]{FCD5B4}70-80}                                      & \multicolumn{1}{c|}{\cellcolor[HTML]{DEE37D}75.56}                                              & \multicolumn{1}{c|}{\cellcolor[HTML]{F2E881}81.72}                                                & \multicolumn{1}{c|}{\cellcolor[HTML]{DBE27C}74.35}                                                & \multicolumn{1}{c|}{\cellcolor[HTML]{FFD578}120.27}                                               & \multicolumn{1}{c|}{\cellcolor[HTML]{A1D570}56.21}                                                & \multicolumn{1}{c|}{\cellcolor[HTML]{FFE380}98.24}                                                & \multicolumn{1}{c|}{\cellcolor[HTML]{FFCB72}137.07}                                               & \cellcolor[HTML]{FFAA60}187.56                                               \\ \cline{2-10} 
\multicolumn{1}{|c|}{\cellcolor[HTML]{FCD5B4}}                                            & \multicolumn{1}{c|}{\cellcolor[HTML]{FCD5B4}80-90}                                      & \multicolumn{1}{c|}{\cellcolor[HTML]{FFE883}90.49}                                              & \multicolumn{1}{c|}{\cellcolor[HTML]{FFE581}95.81}                                                & \multicolumn{1}{c|}{\cellcolor[HTML]{FFE37F}99.59}                                                & \multicolumn{1}{c|}{\cellcolor[HTML]{FFCC73}134.50}                                               & \multicolumn{1}{c|}{\cellcolor[HTML]{D3E07B}71.89}                                                & \multicolumn{1}{c|}{\cellcolor[HTML]{FFDA7B}112.40}                                               & \multicolumn{1}{c|}{\cellcolor[HTML]{FFB566}170.80}                                               & \cellcolor[HTML]{FF9655}218.41                                               \\ \cline{2-10} 
\multicolumn{1}{|c|}{\multirow{-6}{*}{\cellcolor[HTML]{FCD5B4}100-200}}                   & \multicolumn{1}{c|}{\cellcolor[HTML]{FCD5B4}90-100}                                     & \multicolumn{1}{c|}{\cellcolor[HTML]{FFE782}91.95}                                              & \multicolumn{1}{c|}{\cellcolor[HTML]{FFE782}93.42}                                                & \multicolumn{1}{c|}{\cellcolor[HTML]{FFDA7A}113.65}                                               & \multicolumn{1}{c|}{\cellcolor[HTML]{FFCC73}135.04}                                               & \multicolumn{1}{c|}{\cellcolor[HTML]{C1DC77}66.16}                                                & \multicolumn{1}{c|}{\cellcolor[HTML]{FFDD7C}108.73}                                               & \multicolumn{1}{c|}{\cellcolor[HTML]{FFB264}174.97}                                               & \cellcolor[HTML]{FF834A}249.03                                               \\ \hline
\multicolumn{1}{|c|}{\cellcolor[HTML]{FABF8F}}                                            & \multicolumn{1}{c|}{\cellcolor[HTML]{FABF8F}50-60}                                      & \multicolumn{1}{c|}{\cellcolor[HTML]{FFCA72}137.63}                                             & \multicolumn{1}{c|}{\cellcolor[HTML]{FFC26D}151.16}                                               & \multicolumn{1}{c|}{\cellcolor[HTML]{E6E57F}78.07}                                                & \multicolumn{1}{c|}{\cellcolor[HTML]{A8D672}58.28}                                                & \multicolumn{1}{c|}{\cellcolor[HTML]{B6DA75}62.96}                                                & \multicolumn{1}{c|}{\cellcolor[HTML]{FFA95F}189.24}                                               & \multicolumn{1}{c|}{\cellcolor[HTML]{FF7F47}255.47}                                               & \cellcolor[HTML]{FF6137}301.94                                               \\ \cline{2-10} 
\multicolumn{1}{|c|}{\cellcolor[HTML]{FABF8F}}                                            & \multicolumn{1}{c|}{\cellcolor[HTML]{FABF8F}60-70}                                      & \multicolumn{1}{c|}{\cellcolor[HTML]{FFB465}172.75}                                             & \multicolumn{1}{c|}{\cellcolor[HTML]{FFAB60}186.22}                                               & \multicolumn{1}{c|}{\cellcolor[HTML]{FFEB84}87.04}                                                & \multicolumn{1}{c|}{\cellcolor[HTML]{CCDF79}69.85}                                                & \multicolumn{1}{c|}{\cellcolor[HTML]{D7E17B}73.13}                                                & \multicolumn{1}{c|}{\cellcolor[HTML]{FFA35C}199.36}                                               & \multicolumn{1}{c|}{\cellcolor[HTML]{FF6D3E}282.82}                                               & \cellcolor[HTML]{FF4527}345.21                                               \\ \cline{2-10} 
\multicolumn{1}{|c|}{\cellcolor[HTML]{FABF8F}}                                            & \multicolumn{1}{c|}{\cellcolor[HTML]{FABF8F}70-80}                                      & \multicolumn{1}{c|}{\cellcolor[HTML]{FFB968}163.72}                                             & \multicolumn{1}{c|}{\cellcolor[HTML]{FFB063}177.90}                                               & \multicolumn{1}{c|}{\cellcolor[HTML]{FFE27F}100.12}                                               & \multicolumn{1}{c|}{\cellcolor[HTML]{FFCD74}132.84}                                               & \multicolumn{1}{c|}{\cellcolor[HTML]{FFEB84}86.12}                                                & \multicolumn{1}{c|}{\cellcolor[HTML]{FFAA60}187.54}                                               & \multicolumn{1}{c|}{\cellcolor[HTML]{FF7844}266.26}                                               & \cellcolor[HTML]{FF4D2C}332.53                                               \\ \cline{2-10} 
\multicolumn{1}{|c|}{\cellcolor[HTML]{FABF8F}}                                            & \multicolumn{1}{c|}{\cellcolor[HTML]{FABF8F}80-90}                                      & \multicolumn{1}{c|}{\cellcolor[HTML]{FFB063}178.77}                                             & \multicolumn{1}{c|}{\cellcolor[HTML]{FFAE62}182.18}                                               & \multicolumn{1}{c|}{\cellcolor[HTML]{FFD276}125.65}                                               & \multicolumn{1}{c|}{\cellcolor[HTML]{FFA05A}203.08}                                               & \multicolumn{1}{c|}{\cellcolor[HTML]{FFE983}89.51}                                                & \multicolumn{1}{c|}{\cellcolor[HTML]{FFAB60}185.86}                                               & \multicolumn{1}{c|}{\cellcolor[HTML]{FF7843}266.40}                                               & \cellcolor[HTML]{FF4829}340.44                                               \\ \cline{2-10} 
\multicolumn{1}{|c|}{\multirow{-5}{*}{\cellcolor[HTML]{FABF8F}200-300}}                   & \multicolumn{1}{c|}{\cellcolor[HTML]{FABF8F}90-100}                                     & \multicolumn{1}{c|}{\cellcolor[HTML]{FFCC73}134.83}                                             & \multicolumn{1}{c|}{\cellcolor[HTML]{FFCB72}135.88}                                               & \multicolumn{1}{c|}{\cellcolor[HTML]{FFC36D}149.57}                                               & \multicolumn{1}{c|}{\cellcolor[HTML]{FFAF62}180.77}                                               & \multicolumn{1}{c|}{\cellcolor[HTML]{EDE680}80.15}                                                & \multicolumn{1}{c|}{\cellcolor[HTML]{FFC871}140.64}                                               & \multicolumn{1}{c|}{\cellcolor[HTML]{FF874C}243.19}                                               & \cellcolor[HTML]{FF522E}325.36                                               \\ \hline
\multicolumn{1}{|c|}{\cellcolor[HTML]{E26B0A}}                                            & \multicolumn{1}{c|}{\cellcolor[HTML]{E26B0A}70-80}                                      & \multicolumn{1}{c|}{\cellcolor[HTML]{FF9554}219.93}                                             & \multicolumn{1}{c|}{\cellcolor[HTML]{FF8B4E}236.70}                                               & \multicolumn{1}{c|}{\cellcolor[HTML]{E3E47E}77.06}                                                & \multicolumn{1}{c|}{\cellcolor[HTML]{F6E982}82.92}                                                & \multicolumn{1}{c|}{\cellcolor[HTML]{AAD772}59.03}                                                & \multicolumn{1}{c|}{\cellcolor[HTML]{FF8A4E}237.51}                                               & \multicolumn{1}{c|}{\cellcolor[HTML]{FF4125}351.68}                                               & \cellcolor[HTML]{FF0000}452.56                                               \\ \cline{2-10} 
\multicolumn{1}{|c|}{\multirow{-2}{*}{\cellcolor[HTML]{E26B0A}300-400}}                   & \multicolumn{1}{c|}{\cellcolor[HTML]{E26B0A}80-90}                                      & \multicolumn{1}{c|}{\cellcolor[HTML]{FF7F47}255.77}                                             & \multicolumn{1}{c|}{\cellcolor[HTML]{FF7341}273.18}                                               & \multicolumn{1}{c|}{\cellcolor[HTML]{FFC871}141.13}                                               & \multicolumn{1}{c|}{\cellcolor[HTML]{FF9F5A}205.05}                                               & \multicolumn{1}{c|}{\cellcolor[HTML]{FFE681}94.02}                                                & \multicolumn{1}{c|}{\cellcolor[HTML]{FF7442}271.53}                                               & \multicolumn{1}{c|}{\cellcolor[HTML]{FF3A21}362.59}                                               & \cellcolor[HTML]{FF0D08}432.49                                               \\ \hline
\end{tabular}
}
\caption{Mean absolute error (MAE) for different humidity and reference $\text{PM}_{2.5}$ values/bins (hourly averaged data). A graded color scheme has been applied to the MAE values in the table with red, yellow, and green representing the highest, middle, and lowest MAE values, respectively.}
\label{tab:Mean Absolute Error}
\end{table}

\begin{table}[]
\resizebox{1\textwidth}{!}{
\centering
\begin{tabular}{|cccccccccc|}
\hline
\multicolumn{10}{|c|}{\textbf{Root Mean Square Error (RMSE)}}\\ \hline
\multicolumn{1}{|c|}{\begin{tabular}[c]{@{}c@{}}\textbf{BAM}\\ $(\mu \text{g/m}^3)$\end{tabular}} & \multicolumn{1}{c|}{\begin{tabular}[c]{@{}c@{}}\textbf{Humidity}\\ $(\%)$\end{tabular}} & \multicolumn{1}{c|}{\textbf{\begin{tabular}[c]{@{}c@{}}Sensirion-1\\ $(\mu \text{g/m}^3)$\end{tabular}}} & \multicolumn{1}{c|}{\textbf{\begin{tabular}[c]{@{}c@{}}Sensirion-2\\ $(\mu \text{g/m}^3)$\end{tabular}}} & \multicolumn{1}{c|}{\textbf{\begin{tabular}[c]{@{}c@{}}Plantower-1\\ $(\mu \text{g/m}^3)$\end{tabular}}} & \multicolumn{1}{c|}{\textbf{\begin{tabular}[c]{@{}c@{}}Plantower-2\\ $(\mu \text{g/m}^3)$\end{tabular}}} & \multicolumn{1}{c|}{\textbf{\begin{tabular}[c]{@{}c@{}}Plantower-3\\ $(\mu \text{g/m}^3)$\end{tabular}}} & \multicolumn{1}{c|}{\textbf{\begin{tabular}[c]{@{}c@{}}Honeywell-1\\ $(\mu \text{g/m}^3)$\end{tabular}}} & \multicolumn{1}{c|}{\textbf{\begin{tabular}[c]{@{}c@{}}Honeywell-2\\ $(\mu \text{g/m}^3)$\end{tabular}}} & \textbf{\begin{tabular}[c]{@{}c@{}}Honeywell-3\\ $(\mu \text{g/m}^3)$\end{tabular}} \\ \hline
\multicolumn{1}{|c|}{\cellcolor[HTML]{FDE9D9}}                                            & \multicolumn{1}{c|}{\cellcolor[HTML]{FDE9D9}20-30}                                      & \multicolumn{1}{c|}{\cellcolor[HTML]{00B050}7.50}                                               & \multicolumn{1}{c|}{\cellcolor[HTML]{00B050}7.51}                                                 & \multicolumn{1}{c|}{\cellcolor[HTML]{1AB655}16.36}                                                & \multicolumn{1}{c|}{\cellcolor[HTML]{1BB655}16.88}                                                & \multicolumn{1}{c|}{\cellcolor[HTML]{18B555}15.84}                                                & \multicolumn{1}{c|}{\cellcolor[HTML]{67C765}42.54}                                                & \multicolumn{1}{c|}{\cellcolor[HTML]{81CD6A}51.33}                                                & \cellcolor[HTML]{DEE37D}82.97                                                \\ \cline{2-10} 
\multicolumn{1}{|c|}{\cellcolor[HTML]{FDE9D9}}                                            & \multicolumn{1}{c|}{\cellcolor[HTML]{FDE9D9}30-40}                                      & \multicolumn{1}{c|}{\cellcolor[HTML]{1FB756}18.31}                                              & \multicolumn{1}{c|}{\cellcolor[HTML]{20B756}18.47}                                                & \multicolumn{1}{c|}{\cellcolor[HTML]{2BBA58}22.31}                                                & \multicolumn{1}{c|}{\cellcolor[HTML]{3DBE5C}28.46}                                                & \multicolumn{1}{c|}{\cellcolor[HTML]{23B857}19.48}                                                & \multicolumn{1}{c|}{\cellcolor[HTML]{7BCC69}49.29}                                                & \multicolumn{1}{c|}{\cellcolor[HTML]{AAD772}65.25}                                                & \cellcolor[HTML]{FEEA83}93.63                                                \\ \cline{2-10} 
\multicolumn{1}{|c|}{\cellcolor[HTML]{FDE9D9}}                                            & \multicolumn{1}{c|}{\cellcolor[HTML]{FDE9D9}40-50}                                      & \multicolumn{1}{c|}{\cellcolor[HTML]{2FBA59}23.43}                                              & \multicolumn{1}{c|}{\cellcolor[HTML]{2DBA59}22.86}                                                & \multicolumn{1}{c|}{\cellcolor[HTML]{38BD5B}26.66}                                                & \multicolumn{1}{c|}{\cellcolor[HTML]{49C15E}32.41}                                                & \multicolumn{1}{c|}{\cellcolor[HTML]{27B958}20.97}                                                & \multicolumn{1}{c|}{\cellcolor[HTML]{84CE6A}52.31}                                                & \multicolumn{1}{c|}{\cellcolor[HTML]{AFD873}67.00}                                                & \cellcolor[HTML]{FFEB84}94.55                                                \\ \cline{2-10} 
\multicolumn{1}{|c|}{\cellcolor[HTML]{FDE9D9}}                                            & \multicolumn{1}{c|}{\cellcolor[HTML]{FDE9D9}50-60}                                      & \multicolumn{1}{c|}{\cellcolor[HTML]{2FBA59}23.49}                                              & \multicolumn{1}{c|}{\cellcolor[HTML]{2EBA59}23.22}                                                & \multicolumn{1}{c|}{\cellcolor[HTML]{3CBE5C}28.12}                                                & \multicolumn{1}{c|}{\cellcolor[HTML]{58C461}37.33}                                                & \multicolumn{1}{c|}{\cellcolor[HTML]{2AB958}22.03}                                                & \multicolumn{1}{c|}{\cellcolor[HTML]{81CD6A}51.30}                                                & \multicolumn{1}{c|}{\cellcolor[HTML]{ADD873}66.21}                                                & \cellcolor[HTML]{FFEB84}94.77                                                \\ \cline{2-10} 
\multicolumn{1}{|c|}{\cellcolor[HTML]{FDE9D9}}                                            & \multicolumn{1}{c|}{\cellcolor[HTML]{FDE9D9}60-70}                                      & \multicolumn{1}{c|}{\cellcolor[HTML]{33BC5A}25.08}                                              & \multicolumn{1}{c|}{\cellcolor[HTML]{33BC5A}25.10}                                                & \multicolumn{1}{c|}{\cellcolor[HTML]{59C462}37.73}                                                & \multicolumn{1}{c|}{\cellcolor[HTML]{6AC865}43.70}                                                & \multicolumn{1}{c|}{\cellcolor[HTML]{42BF5D}29.91}                                                & \multicolumn{1}{c|}{\cellcolor[HTML]{84CE6A}52.33}                                                & \multicolumn{1}{c|}{\cellcolor[HTML]{B6DA75}69.45}                                                & \cellcolor[HTML]{FFE782}100.83                                               \\ \cline{2-10} 
\multicolumn{1}{|c|}{\cellcolor[HTML]{FDE9D9}}                                            & \multicolumn{1}{c|}{\cellcolor[HTML]{FDE9D9}70-80}                                      & \multicolumn{1}{c|}{\cellcolor[HTML]{55C361}36.31}                                              & \multicolumn{1}{c|}{\cellcolor[HTML]{51C260}35.15}                                                & \multicolumn{1}{c|}{\cellcolor[HTML]{8CD06C}55.03}                                                & \multicolumn{1}{c|}{\cellcolor[HTML]{9AD36F}59.75}                                                & \multicolumn{1}{c|}{\cellcolor[HTML]{74CA67}47.03}                                                & \multicolumn{1}{c|}{\cellcolor[HTML]{ABD772}65.45}                                                & \multicolumn{1}{c|}{\cellcolor[HTML]{F4E881}90.38}                                                & \cellcolor[HTML]{FFDB7B}119.23                                               \\ \cline{2-10} 
\multicolumn{1}{|c|}{\cellcolor[HTML]{FDE9D9}}                                            & \multicolumn{1}{c|}{\cellcolor[HTML]{FDE9D9}80-90}                                      & \multicolumn{1}{c|}{\cellcolor[HTML]{87CF6B}53.39}                                              & \multicolumn{1}{c|}{\cellcolor[HTML]{81CD6A}51.22}                                                & \multicolumn{1}{c|}{\cellcolor[HTML]{C9DE79}75.75}                                                & \multicolumn{1}{c|}{\cellcolor[HTML]{D8E17C}80.71}                                                & \multicolumn{1}{c|}{\cellcolor[HTML]{98D36F}59.31}                                                & \multicolumn{1}{c|}{\cellcolor[HTML]{D7E17B}80.47}                                                & \multicolumn{1}{c|}{\cellcolor[HTML]{FFDB7B}119.76}                                               & \cellcolor[HTML]{FFC871}149.66                                               \\ \cline{2-10} 
\multicolumn{1}{|c|}{\multirow{-8}{*}{\cellcolor[HTML]{FDE9D9}0-100}}                     & \multicolumn{1}{c|}{\cellcolor[HTML]{FDE9D9}90-100}                                     & \multicolumn{1}{c|}{\cellcolor[HTML]{76CB68}47.61}                                              & \multicolumn{1}{c|}{\cellcolor[HTML]{71CA67}45.86}                                                & \multicolumn{1}{c|}{\cellcolor[HTML]{B3D974}68.45}                                                & \multicolumn{1}{c|}{\cellcolor[HTML]{C1DC77}73.05}                                                & \multicolumn{1}{c|}{\cellcolor[HTML]{71CA67}45.91}                                                & \multicolumn{1}{c|}{\cellcolor[HTML]{CEDF7A}77.59}                                                & \multicolumn{1}{c|}{\cellcolor[HTML]{FFDE7D}114.46}                                               & \cellcolor[HTML]{FFBC6A}167.94                                               \\ \hline
\multicolumn{1}{|c|}{\cellcolor[HTML]{FCD5B4}}                                            & \multicolumn{1}{c|}{\cellcolor[HTML]{FCD5B4}40-50}                                      & \multicolumn{1}{c|}{\cellcolor[HTML]{99D36F}59.54}                                              & \multicolumn{1}{c|}{\cellcolor[HTML]{9FD570}61.70}                                                & \multicolumn{1}{c|}{\cellcolor[HTML]{70CA66}45.65}                                                & \multicolumn{1}{c|}{\cellcolor[HTML]{8FD16D}56.14}                                                & \multicolumn{1}{c|}{\cellcolor[HTML]{56C361}36.76}                                                & \multicolumn{1}{c|}{\cellcolor[HTML]{FFEA84}96.54}                                                & \multicolumn{1}{c|}{\cellcolor[HTML]{FFD678}128.36}                                               & \cellcolor[HTML]{FFBF6B}164.05                                               \\ \cline{2-10} 
\multicolumn{1}{|c|}{\cellcolor[HTML]{FCD5B4}}                                            & \multicolumn{1}{c|}{\cellcolor[HTML]{FCD5B4}50-60}                                      & \multicolumn{1}{c|}{\cellcolor[HTML]{A8D772}64.66}                                              & \multicolumn{1}{c|}{\cellcolor[HTML]{ACD773}66.06}                                                & \multicolumn{1}{c|}{\cellcolor[HTML]{78CB68}48.44}                                                & \multicolumn{1}{c|}{\cellcolor[HTML]{BFDC76}72.25}                                                & \multicolumn{1}{c|}{\cellcolor[HTML]{60C663}40.08}                                                & \multicolumn{1}{c|}{\cellcolor[HTML]{FFEB84}94.91}                                                & \multicolumn{1}{c|}{\cellcolor[HTML]{FFD87A}124.64}                                               & \cellcolor[HTML]{FFBE6B}165.67                                               \\ \cline{2-10} 
\multicolumn{1}{|c|}{\cellcolor[HTML]{FCD5B4}}                                            & \multicolumn{1}{c|}{\cellcolor[HTML]{FCD5B4}60-70}                                      & \multicolumn{1}{c|}{\cellcolor[HTML]{BBDB76}71.18}                                              & \multicolumn{1}{c|}{\cellcolor[HTML]{BCDB76}71.32}                                                & \multicolumn{1}{c|}{\cellcolor[HTML]{9BD36F}60.11}                                                & \multicolumn{1}{c|}{\cellcolor[HTML]{FFE983}98.06}                                                & \multicolumn{1}{c|}{\cellcolor[HTML]{86CF6B}53.00}                                                & \multicolumn{1}{c|}{\cellcolor[HTML]{FFE983}97.29}                                                & \multicolumn{1}{c|}{\cellcolor[HTML]{FFD578}129.32}                                               & \cellcolor[HTML]{FFBB69}170.95                                               \\ \cline{2-10} 
\multicolumn{1}{|c|}{\cellcolor[HTML]{FCD5B4}}                                            & \multicolumn{1}{c|}{\cellcolor[HTML]{FCD5B4}70-80}                                      & \multicolumn{1}{c|}{\cellcolor[HTML]{E6E57E}85.51}                                              & \multicolumn{1}{c|}{\cellcolor[HTML]{FBEA83}92.60}                                                & \multicolumn{1}{c|}{\cellcolor[HTML]{DFE37D}83.17}                                                & \multicolumn{1}{c|}{\cellcolor[HTML]{FFD377}131.75}                                               & \multicolumn{1}{c|}{\cellcolor[HTML]{AAD772}65.31}                                                & \multicolumn{1}{c|}{\cellcolor[HTML]{FFE480}105.15}                                               & \multicolumn{1}{c|}{\cellcolor[HTML]{FFCB72}145.16}                                               & \cellcolor[HTML]{FFAB61}194.72                                               \\ \cline{2-10} 
\multicolumn{1}{|c|}{\cellcolor[HTML]{FCD5B4}}                                            & \multicolumn{1}{c|}{\cellcolor[HTML]{FCD5B4}80-90}                                      & \multicolumn{1}{c|}{\cellcolor[HTML]{FFE682}102.03}                                             & \multicolumn{1}{c|}{\cellcolor[HTML]{FFE380}107.57}                                               & \multicolumn{1}{c|}{\cellcolor[HTML]{FFE380}107.90}                                               & \multicolumn{1}{c|}{\cellcolor[HTML]{FFC971}147.51}                                               & \multicolumn{1}{c|}{\cellcolor[HTML]{D9E27C}81.22}                                                & \multicolumn{1}{c|}{\cellcolor[HTML]{FFDA7B}121.58}                                               & \multicolumn{1}{c|}{\cellcolor[HTML]{FFB566}179.48}                                               & \cellcolor[HTML]{FF9755}226.49                                               \\ \cline{2-10} 
\multicolumn{1}{|c|}{\multirow{-6}{*}{\cellcolor[HTML]{FCD5B4}100-200}}                   & \multicolumn{1}{c|}{\cellcolor[HTML]{FCD5B4}90-100}                                     & \multicolumn{1}{c|}{\cellcolor[HTML]{FFE983}98.44}                                              & \multicolumn{1}{c|}{\cellcolor[HTML]{FFE983}98.58}                                                & \multicolumn{1}{c|}{\cellcolor[HTML]{FFD879}124.79}                                               & \multicolumn{1}{c|}{\cellcolor[HTML]{FFCD74}141.36}                                               & \multicolumn{1}{c|}{\cellcolor[HTML]{C5DD78}74.38}                                                & \multicolumn{1}{c|}{\cellcolor[HTML]{FFDE7D}114.89}                                               & \multicolumn{1}{c|}{\cellcolor[HTML]{FFB566}179.73}                                               & \cellcolor[HTML]{FF854B}254.58                                               \\ \hline
\multicolumn{1}{|c|}{\cellcolor[HTML]{FABF8F}}                                            & \multicolumn{1}{c|}{\cellcolor[HTML]{FABF8F}50-60}                                      & \multicolumn{1}{c|}{\cellcolor[HTML]{FFCC73}143.65}                                             & \multicolumn{1}{c|}{\cellcolor[HTML]{FFC36E}157.26}                                               & \multicolumn{1}{c|}{\cellcolor[HTML]{D8E27C}80.91}                                                & \multicolumn{1}{c|}{\cellcolor[HTML]{C3DD77}73.64}                                                & \multicolumn{1}{c|}{\cellcolor[HTML]{ACD873}66.09}                                                & \multicolumn{1}{c|}{\cellcolor[HTML]{FFAC61}194.45}                                               & \multicolumn{1}{c|}{\cellcolor[HTML]{FF8149}261.01}                                               & \cellcolor[HTML]{FF6438}307.27                                               \\ \cline{2-10} 
\multicolumn{1}{|c|}{\cellcolor[HTML]{FABF8F}}                                            & \multicolumn{1}{c|}{\cellcolor[HTML]{FABF8F}60-70}                                      & \multicolumn{1}{c|}{\cellcolor[HTML]{FFB365}183.00}                                             & \multicolumn{1}{c|}{\cellcolor[HTML]{FFAA60}197.16}                                               & \multicolumn{1}{c|}{\cellcolor[HTML]{F8E982}91.69}                                                & \multicolumn{1}{c|}{\cellcolor[HTML]{FFEB84}94.17}                                                & \multicolumn{1}{c|}{\cellcolor[HTML]{CFE07A}77.88}                                                & \multicolumn{1}{c|}{\cellcolor[HTML]{FFA45C}206.27}                                               & \multicolumn{1}{c|}{\cellcolor[HTML]{FF6D3D}293.58}                                               & \cellcolor[HTML]{FF4427}357.20                                               \\ \cline{2-10} 
\multicolumn{1}{|c|}{\cellcolor[HTML]{FABF8F}}                                            & \multicolumn{1}{c|}{\cellcolor[HTML]{FABF8F}70-80}                                      & \multicolumn{1}{c|}{\cellcolor[HTML]{FFB867}175.32}                                             & \multicolumn{1}{c|}{\cellcolor[HTML]{FFAD61}192.70}                                               & \multicolumn{1}{c|}{\cellcolor[HTML]{FFE380}107.56}                                               & \multicolumn{1}{c|}{\cellcolor[HTML]{FFC36E}156.90}                                               & \multicolumn{1}{c|}{\cellcolor[HTML]{FFEB84}95.14}                                                & \multicolumn{1}{c|}{\cellcolor[HTML]{FFA95F}198.96}                                               & \multicolumn{1}{c|}{\cellcolor[HTML]{FF7743}276.97}                                               & \cellcolor[HTML]{FF4B2B}345.94                                               \\ \cline{2-10} 
\multicolumn{1}{|c|}{\cellcolor[HTML]{FABF8F}}                                            & \multicolumn{1}{c|}{\cellcolor[HTML]{FABF8F}80-90}                                      & \multicolumn{1}{c|}{\cellcolor[HTML]{FFAC61}193.91}                                             & \multicolumn{1}{c|}{\cellcolor[HTML]{FFAA60}196.84}                                               & \multicolumn{1}{c|}{\cellcolor[HTML]{FFD276}133.66}                                               & \multicolumn{1}{c|}{\cellcolor[HTML]{FF9D58}218.15}                                               & \multicolumn{1}{c|}{\cellcolor[HTML]{FFE882}99.59}                                                & \multicolumn{1}{c|}{\cellcolor[HTML]{FFA95F}198.17}                                               & \multicolumn{1}{c|}{\cellcolor[HTML]{FF7743}277.80}                                               & \cellcolor[HTML]{FF4829}350.64                                               \\ \cline{2-10} 
\multicolumn{1}{|c|}{\multirow{-5}{*}{\cellcolor[HTML]{FABF8F}200-300}}                   & \multicolumn{1}{c|}{\cellcolor[HTML]{FABF8F}90-100}                                     & \multicolumn{1}{c|}{\cellcolor[HTML]{FFC971}147.87}                                             & \multicolumn{1}{c|}{\cellcolor[HTML]{FFCA72}146.84}                                               & \multicolumn{1}{c|}{\cellcolor[HTML]{FFBD6A}166.88}                                               & \multicolumn{1}{c|}{\cellcolor[HTML]{FFAF63}189.12}                                               & \multicolumn{1}{c|}{\cellcolor[HTML]{F5E882}90.59}                                                & \multicolumn{1}{c|}{\cellcolor[HTML]{FFC56F}153.90}                                               & \multicolumn{1}{c|}{\cellcolor[HTML]{FF864C}253.59}                                               & \cellcolor[HTML]{FF532F}333.30                                               \\ \hline
\multicolumn{1}{|c|}{\cellcolor[HTML]{E26B0A}}                                            & \multicolumn{1}{c|}{\cellcolor[HTML]{E26B0A}70-80}                                      & \multicolumn{1}{c|}{\cellcolor[HTML]{FF9855}225.96}                                             & \multicolumn{1}{c|}{\cellcolor[HTML]{FF8A4E}247.15}                                               & \multicolumn{1}{c|}{\cellcolor[HTML]{D6E17B}80.10}                                                & \multicolumn{1}{c|}{\cellcolor[HTML]{FFE480}106.25}                                               & \multicolumn{1}{c|}{\cellcolor[HTML]{A5D671}63.56}                                                & \multicolumn{1}{c|}{\cellcolor[HTML]{FF8B4E}246.22}                                               & \multicolumn{1}{c|}{\cellcolor[HTML]{FF4326}359.79}                                               & \cellcolor[HTML]{FF0000}463.91                                               \\ \cline{2-10} 
\multicolumn{1}{|c|}{\multirow{-2}{*}{\cellcolor[HTML]{E26B0A}300-400}}                   & \multicolumn{1}{c|}{\cellcolor[HTML]{E26B0A}80-90}                                      & \multicolumn{1}{c|}{\cellcolor[HTML]{FF7B45}270.79}                                             & \multicolumn{1}{c|}{\cellcolor[HTML]{FF6D3D}293.48}                                               & \multicolumn{1}{c|}{\cellcolor[HTML]{FFC56F}154.70}                                               & \multicolumn{1}{c|}{\cellcolor[HTML]{FF9454}231.18}                                               & \multicolumn{1}{c|}{\cellcolor[HTML]{FFE17E}110.83}                                               & \multicolumn{1}{c|}{\cellcolor[HTML]{FF703F}288.34}                                               & \multicolumn{1}{c|}{\cellcolor[HTML]{FF371F}377.72}                                               & \cellcolor[HTML]{FF0805}451.40                                               \\ \hline
\end{tabular}
}
\caption{RMSE for different humidity and reference $\text{PM}_{2.5}$ values/bins (Hourly averaged data). A graded color scheme has been applied to the RAMSE values in the table with red, yellow, and green representing the highest, middle, and lowest RMSE values, respectively.}
\label{tab:RMSE}
\end{table}

\begin{table}[h]
\centering
\resizebox{1\textwidth}{!}{

\begin{tabular}{|cccccccccc|}
\hline
\multicolumn{10}{|c|}{\textbf{NRMSE (\%)}}\\ \hline
\multicolumn{1}{|c|}{\begin{tabular}[c]{@{}c@{}}\textbf{BAM}\\ $(\mu \text{g/m}^3)$\end{tabular}} & \multicolumn{1}{c|}{\begin{tabular}[c]{@{}c@{}}\textbf{Humidity}\\ $(\%)$\end{tabular}} & \multicolumn{1}{c|}{\begin{tabular}[c]{@{}c@{}}\textbf{Sensirion-1}\\ $(\%)$\end{tabular}} & \multicolumn{1}{c|}{\begin{tabular}[c]{@{}c@{}}\textbf{Sensirion-2}\\ $(\%)$\end{tabular}} & \multicolumn{1}{c|}{\begin{tabular}[c]{@{}c@{}}\textbf{Plantower-1}\\ $(\%)$\end{tabular}} & \multicolumn{1}{c|}{\begin{tabular}[c]{@{}c@{}}\textbf{Plantower-2}\\ $(\%)$\end{tabular}} & \multicolumn{1}{c|}{\begin{tabular}[c]{@{}c@{}}\textbf{Plantower-3}\\ $(\%)$\end{tabular}} & \multicolumn{1}{c|}{\begin{tabular}[c]{@{}c@{}}\textbf{Honeywell-1}\\ $(\%)$\end{tabular}} & \multicolumn{1}{c|}{\begin{tabular}[c]{@{}c@{}}\textbf{Honeywell-2}\\ $(\%)$\end{tabular}} & \begin{tabular}[c]{@{}c@{}}\textbf{Honeywell-3}\\ $(\%)$\end{tabular} \\ \hline
\multicolumn{1}{|c|}{\cellcolor[HTML]{FDE9D9}}                                             & \multicolumn{1}{c|}{\cellcolor[HTML]{FDE9D9}20-30}                                      & \multicolumn{1}{c|}{\cellcolor[HTML]{22B757}27}                                            & \multicolumn{1}{c|}{\cellcolor[HTML]{22B857}27}                                            & \multicolumn{1}{c|}{\cellcolor[HTML]{BDDB76}59}                                            & \multicolumn{1}{c|}{\cellcolor[HTML]{C6DD78}61}                                            & \multicolumn{1}{c|}{\cellcolor[HTML]{B4D974}57}                                            & \multicolumn{1}{c|}{\cellcolor[HTML]{FF9855}153}                                           & \multicolumn{1}{c|}{\cellcolor[HTML]{FF7743}185}                                           & \cellcolor[HTML]{FF0000}299                                           \\ \cline{2-10} 
\multicolumn{1}{|c|}{\cellcolor[HTML]{FDE9D9}}                                             & \multicolumn{1}{c|}{\cellcolor[HTML]{FDE9D9}30-40}                                      & \multicolumn{1}{c|}{\cellcolor[HTML]{5FC663}40}                                            & \multicolumn{1}{c|}{\cellcolor[HTML]{61C663}40}                                            & \multicolumn{1}{c|}{\cellcolor[HTML]{8FD16D}49}                                            & \multicolumn{1}{c|}{\cellcolor[HTML]{C3DD77}60}                                            & \multicolumn{1}{c|}{\cellcolor[HTML]{85CE6B}47}                                            & \multicolumn{1}{c|}{\cellcolor[HTML]{FFC670}108}                                           & \multicolumn{1}{c|}{\cellcolor[HTML]{FFA65D}140}                                           & \cellcolor[HTML]{FF5F36}208                                           \\ \cline{2-10} 
\multicolumn{1}{|c|}{\cellcolor[HTML]{FDE9D9}}                                             & \multicolumn{1}{c|}{\cellcolor[HTML]{FDE9D9}40-50}                                      & \multicolumn{1}{c|}{\cellcolor[HTML]{83CE6A}47}                                            & \multicolumn{1}{c|}{\cellcolor[HTML]{7ACC68}45}                                            & \multicolumn{1}{c|}{\cellcolor[HTML]{A1D570}53}                                            & \multicolumn{1}{c|}{\cellcolor[HTML]{DBE27C}65}                                            & \multicolumn{1}{c|}{\cellcolor[HTML]{74CA67}44}                                            & \multicolumn{1}{c|}{\cellcolor[HTML]{FFCA72}104}                                           & \multicolumn{1}{c|}{\cellcolor[HTML]{FFAC61}134}                                           & \cellcolor[HTML]{FF7341}189                                           \\ \cline{2-10} 
\multicolumn{1}{|c|}{\cellcolor[HTML]{FDE9D9}}                                             & \multicolumn{1}{c|}{\cellcolor[HTML]{FDE9D9}50-60}                                      & \multicolumn{1}{c|}{\cellcolor[HTML]{6DC966}42}                                            & \multicolumn{1}{c|}{\cellcolor[HTML]{6BC865}42}                                            & \multicolumn{1}{c|}{\cellcolor[HTML]{98D36F}51}                                            & \multicolumn{1}{c|}{\cellcolor[HTML]{E8E57F}68}                                            & \multicolumn{1}{c|}{\cellcolor[HTML]{69C865}42}                                            & \multicolumn{1}{c|}{\cellcolor[HTML]{FFD97A}91}                                            & \multicolumn{1}{c|}{\cellcolor[HTML]{FFBA69}120}                                           & \cellcolor[HTML]{FF894D}167                                           \\ \cline{2-10} 
\multicolumn{1}{|c|}{\cellcolor[HTML]{FDE9D9}}                                             & \multicolumn{1}{c|}{\cellcolor[HTML]{FDE9D9}60-70}                                      & \multicolumn{1}{c|}{\cellcolor[HTML]{68C865}42}                                            & \multicolumn{1}{c|}{\cellcolor[HTML]{69C865}42}                                            & \multicolumn{1}{c|}{\cellcolor[HTML]{D0E07A}63}                                            & \multicolumn{1}{c|}{\cellcolor[HTML]{FFEB84}73}                                            & \multicolumn{1}{c|}{\cellcolor[HTML]{94D26E}51}                                            & \multicolumn{1}{c|}{\cellcolor[HTML]{FFDD7C}86}                                            & \multicolumn{1}{c|}{\cellcolor[HTML]{FFC06C}115}                                           & \cellcolor[HTML]{FF8B4E}166                                           \\ \cline{2-10} 
\multicolumn{1}{|c|}{\cellcolor[HTML]{FDE9D9}}                                             & \multicolumn{1}{c|}{\cellcolor[HTML]{FDE9D9}70-80}                                      & \multicolumn{1}{c|}{\cellcolor[HTML]{ABD772}55}                                            & \multicolumn{1}{c|}{\cellcolor[HTML]{A5D671}54}                                            & \multicolumn{1}{c|}{\cellcolor[HTML]{FFDF7E}84}                                            & \multicolumn{1}{c|}{\cellcolor[HTML]{FFD879}92}                                            & \multicolumn{1}{c|}{\cellcolor[HTML]{FFEA84}74}                                            & \multicolumn{1}{c|}{\cellcolor[HTML]{FFD075}99}                                            & \multicolumn{1}{c|}{\cellcolor[HTML]{FFA95F}136}                                           & \cellcolor[HTML]{FF7E47}178                                           \\ \cline{2-10} 
\multicolumn{1}{|c|}{\cellcolor[HTML]{FDE9D9}}                                             & \multicolumn{1}{c|}{\cellcolor[HTML]{FDE9D9}80-90}                                      & \multicolumn{1}{c|}{\cellcolor[HTML]{F1E781}70}                                            & \multicolumn{1}{c|}{\cellcolor[HTML]{E3E47E}67}                                            & \multicolumn{1}{c|}{\cellcolor[HTML]{FFD175}98}                                            & \multicolumn{1}{c|}{\cellcolor[HTML]{FFC870}107}                                           & \multicolumn{1}{c|}{\cellcolor[HTML]{FFE581}78}                                            & \multicolumn{1}{c|}{\cellcolor[HTML]{FFC971}106}                                           & \multicolumn{1}{c|}{\cellcolor[HTML]{FF9453}157}                                           & \cellcolor[HTML]{FF6B3C}196                                           \\ \cline{2-10} 
\multicolumn{1}{|c|}{\multirow{-8}{*}{\cellcolor[HTML]{FDE9D9}0-100}}                      & \multicolumn{1}{c|}{\cellcolor[HTML]{FDE9D9}90-100}                                     & \multicolumn{1}{c|}{\cellcolor[HTML]{FFEB84}73}                                            & \multicolumn{1}{c|}{\cellcolor[HTML]{F3E881}70}                                            & \multicolumn{1}{c|}{\cellcolor[HTML]{FFC971}106}                                           & \multicolumn{1}{c|}{\cellcolor[HTML]{FFC06C}114}                                           & \multicolumn{1}{c|}{\cellcolor[HTML]{FFEB84}73}                                            & \multicolumn{1}{c|}{\cellcolor[HTML]{FFB867}122}                                           & \multicolumn{1}{c|}{\cellcolor[HTML]{FF8149}175}                                           & \cellcolor[HTML]{FF311C}252                                           \\ \hline
\multicolumn{1}{|c|}{\cellcolor[HTML]{FCD5B4}}                                             & \multicolumn{1}{c|}{\cellcolor[HTML]{FCD5B4}40-50}                                      & \multicolumn{1}{c|}{\cellcolor[HTML]{69C865}42}                                            & \multicolumn{1}{c|}{\cellcolor[HTML]{71CA67}43}                                            & \multicolumn{1}{c|}{\cellcolor[HTML]{3ABD5C}32}                                            & \multicolumn{1}{c|}{\cellcolor[HTML]{5EC563}39}                                            & \multicolumn{1}{c|}{\cellcolor[HTML]{1BB655}26}                                            & \multicolumn{1}{c|}{\cellcolor[HTML]{E8E57F}68}                                            & \multicolumn{1}{c|}{\cellcolor[HTML]{FFD97A}90}                                            & \cellcolor[HTML]{FFBF6C}115                                           \\ \cline{2-10} 
\multicolumn{1}{|c|}{\cellcolor[HTML]{FCD5B4}}                                             & \multicolumn{1}{c|}{\cellcolor[HTML]{FCD5B4}50-60}                                      & \multicolumn{1}{c|}{\cellcolor[HTML]{87CF6B}48}                                            & \multicolumn{1}{c|}{\cellcolor[HTML]{88CF6B}48}                                            & \multicolumn{1}{c|}{\cellcolor[HTML]{4AC15F}35}                                            & \multicolumn{1}{c|}{\cellcolor[HTML]{9FD470}53}                                            & \multicolumn{1}{c|}{\cellcolor[HTML]{2AB958}29}                                            & \multicolumn{1}{c|}{\cellcolor[HTML]{EFE780}69}                                            & \multicolumn{1}{c|}{\cellcolor[HTML]{FFD87A}91}                                            & \cellcolor[HTML]{FFB968}121                                           \\ \cline{2-10} 
\multicolumn{1}{|c|}{\cellcolor[HTML]{FCD5B4}}                                             & \multicolumn{1}{c|}{\cellcolor[HTML]{FCD5B4}60-70}                                      & \multicolumn{1}{c|}{\cellcolor[HTML]{8FD16D}49}                                            & \multicolumn{1}{c|}{\cellcolor[HTML]{8DD06C}49}                                            & \multicolumn{1}{c|}{\cellcolor[HTML]{69C865}42}                                            & \multicolumn{1}{c|}{\cellcolor[HTML]{ECE680}69}                                            & \multicolumn{1}{c|}{\cellcolor[HTML]{50C260}37}                                            & \multicolumn{1}{c|}{\cellcolor[HTML]{E6E57F}67}                                            & \multicolumn{1}{c|}{\cellcolor[HTML]{FFD97A}90}                                            & \cellcolor[HTML]{FFBB6A}119                                           \\ \cline{2-10} 
\multicolumn{1}{|c|}{\cellcolor[HTML]{FCD5B4}}                                             & \multicolumn{1}{c|}{\cellcolor[HTML]{FCD5B4}70-80}                                      & \multicolumn{1}{c|}{\cellcolor[HTML]{BADB75}58}                                            & \multicolumn{1}{c|}{\cellcolor[HTML]{CDDF79}62}                                            & \multicolumn{1}{c|}{\cellcolor[HTML]{B2D974}57}                                            & \multicolumn{1}{c|}{\cellcolor[HTML]{FFDA7A}90}                                            & \multicolumn{1}{c|}{\cellcolor[HTML]{75CB67}44}                                            & \multicolumn{1}{c|}{\cellcolor[HTML]{FBEA83}72}                                            & \multicolumn{1}{c|}{\cellcolor[HTML]{FFD075}99}                                            & \cellcolor[HTML]{FFAD61}133                                           \\ \cline{2-10} 
\multicolumn{1}{|c|}{\cellcolor[HTML]{FCD5B4}}                                             & \multicolumn{1}{c|}{\cellcolor[HTML]{FCD5B4}80-90}                                      & \multicolumn{1}{c|}{\cellcolor[HTML]{EDE780}69}                                            & \multicolumn{1}{c|}{\cellcolor[HTML]{FEEA83}72}                                            & \multicolumn{1}{c|}{\cellcolor[HTML]{FFEB84}73}                                            & \multicolumn{1}{c|}{\cellcolor[HTML]{FFCF74}100}                                           & \multicolumn{1}{c|}{\cellcolor[HTML]{ACD773}55}                                            & \multicolumn{1}{c|}{\cellcolor[HTML]{FFE17F}82}                                            & \multicolumn{1}{c|}{\cellcolor[HTML]{FFB968}122}                                           & \cellcolor[HTML]{FF9755}153                                           \\ \cline{2-10} 
\multicolumn{1}{|c|}{\multirow{-6}{*}{\cellcolor[HTML]{FCD5B4}100-200}}                    & \multicolumn{1}{c|}{\cellcolor[HTML]{FCD5B4}90-100}                                     & \multicolumn{1}{c|}{\cellcolor[HTML]{EBE680}68}                                            & \multicolumn{1}{c|}{\cellcolor[HTML]{EBE680}69}                                            & \multicolumn{1}{c|}{\cellcolor[HTML]{FFDD7C}87}                                            & \multicolumn{1}{c|}{\cellcolor[HTML]{FFD175}98}                                            & \multicolumn{1}{c|}{\cellcolor[HTML]{99D36F}52}                                            & \multicolumn{1}{c|}{\cellcolor[HTML]{FFE480}80}                                            & \multicolumn{1}{c|}{\cellcolor[HTML]{FFB566}125}                                           & \cellcolor[HTML]{FF7F48}177                                           \\ \hline
\multicolumn{1}{|c|}{\cellcolor[HTML]{FABF8F}}                                             & \multicolumn{1}{c|}{\cellcolor[HTML]{FABF8F}50-60}                                      & \multicolumn{1}{c|}{\cellcolor[HTML]{BFDC77}59}                                            & \multicolumn{1}{c|}{\cellcolor[HTML]{DAE27C}65}                                            & \multicolumn{1}{c|}{\cellcolor[HTML]{41BF5D}33}                                            & \multicolumn{1}{c|}{\cellcolor[HTML]{33BB5A}30}                                            & \multicolumn{1}{c|}{\cellcolor[HTML]{23B857}27}                                            & \multicolumn{1}{c|}{\cellcolor[HTML]{FFE380}80}                                            & \multicolumn{1}{c|}{\cellcolor[HTML]{FFC770}108}                                           & \cellcolor[HTML]{FFB365}127                                           \\ \cline{2-10} 
\multicolumn{1}{|c|}{\cellcolor[HTML]{FABF8F}}                                             & \multicolumn{1}{c|}{\cellcolor[HTML]{FABF8F}60-70}                                      & \multicolumn{1}{c|}{\cellcolor[HTML]{FFE983}75}                                            & \multicolumn{1}{c|}{\cellcolor[HTML]{FFE380}81}                                            & \multicolumn{1}{c|}{\cellcolor[HTML]{55C361}38}                                            & \multicolumn{1}{c|}{\cellcolor[HTML]{59C462}38}                                            & \multicolumn{1}{c|}{\cellcolor[HTML]{3ABD5B}32}                                            & \multicolumn{1}{c|}{\cellcolor[HTML]{FFDF7E}84}                                            & \multicolumn{1}{c|}{\cellcolor[HTML]{FFBA69}120}                                           & \cellcolor[HTML]{FF9F5A}146                                           \\ \cline{2-10} 
\multicolumn{1}{|c|}{\cellcolor[HTML]{FABF8F}}                                             & \multicolumn{1}{c|}{\cellcolor[HTML]{FABF8F}70-80}                                      & \multicolumn{1}{c|}{\cellcolor[HTML]{FDEA83}72}                                            & \multicolumn{1}{c|}{\cellcolor[HTML]{FFE581}79}                                            & \multicolumn{1}{c|}{\cellcolor[HTML]{75CB68}44}                                            & \multicolumn{1}{c|}{\cellcolor[HTML]{D6E17B}64}                                            & \multicolumn{1}{c|}{\cellcolor[HTML]{5CC562}39}                                            & \multicolumn{1}{c|}{\cellcolor[HTML]{FFE27F}82}                                            & \multicolumn{1}{c|}{\cellcolor[HTML]{FFC16C}114}                                           & \cellcolor[HTML]{FFA35C}142                                           \\ \cline{2-10} 
\multicolumn{1}{|c|}{\cellcolor[HTML]{FABF8F}}                                             & \multicolumn{1}{c|}{\cellcolor[HTML]{FABF8F}80-90}                                      & \multicolumn{1}{c|}{\cellcolor[HTML]{FFE37F}81}                                            & \multicolumn{1}{c|}{\cellcolor[HTML]{FFE17F}83}                                            & \multicolumn{1}{c|}{\cellcolor[HTML]{B0D873}56}                                            & \multicolumn{1}{c|}{\cellcolor[HTML]{FFD87A}91}                                            & \multicolumn{1}{c|}{\cellcolor[HTML]{6AC865}42}                                            & \multicolumn{1}{c|}{\cellcolor[HTML]{FFE07E}83}                                            & \multicolumn{1}{c|}{\cellcolor[HTML]{FFBD6B}117}                                           & \cellcolor[HTML]{FF9E59}147                                           \\ \cline{2-10} 
\multicolumn{1}{|c|}{\multirow{-5}{*}{\cellcolor[HTML]{FABF8F}200-300}}                    & \multicolumn{1}{c|}{\cellcolor[HTML]{FABF8F}90-100}                                     & \multicolumn{1}{c|}{\cellcolor[HTML]{CFE07A}63}                                            & \multicolumn{1}{c|}{\cellcolor[HTML]{CDDF79}62}                                            & \multicolumn{1}{c|}{\cellcolor[HTML]{F7E982}71}                                            & \multicolumn{1}{c|}{\cellcolor[HTML]{FFE380}80}                                            & \multicolumn{1}{c|}{\cellcolor[HTML]{59C462}38}                                            & \multicolumn{1}{c|}{\cellcolor[HTML]{DCE27C}65}                                            & \multicolumn{1}{c|}{\cellcolor[HTML]{FFC770}107}                                           & \cellcolor[HTML]{FFA45C}142                                           \\ \hline
\multicolumn{1}{|c|}{\cellcolor[HTML]{E26B0A}}                                             & \multicolumn{1}{c|}{\cellcolor[HTML]{E26B0A}70-80}                                      & \multicolumn{1}{c|}{\cellcolor[HTML]{F7E982}71}                                            & \multicolumn{1}{c|}{\cellcolor[HTML]{FFE682}77}                                            & \multicolumn{1}{c|}{\cellcolor[HTML]{19B555}25}                                            & \multicolumn{1}{c|}{\cellcolor[HTML]{40BF5D}33}                                            & \multicolumn{1}{c|}{\cellcolor[HTML]{00B050}20}                                            & \multicolumn{1}{c|}{\cellcolor[HTML]{FFE782}77}                                            & \multicolumn{1}{c|}{\cellcolor[HTML]{FFC26D}113}                                           & \cellcolor[HTML]{FFA05A}145                                           \\ \cline{2-10} 
\multicolumn{1}{|c|}{\multirow{-2}{*}{\cellcolor[HTML]{E26B0A}300-400}}                    & \multicolumn{1}{c|}{\cellcolor[HTML]{E26B0A}80-90}                                      & \multicolumn{1}{c|}{\cellcolor[HTML]{FFE27F}82}                                            & \multicolumn{1}{c|}{\cellcolor[HTML]{FFDA7B}89}                                            & \multicolumn{1}{c|}{\cellcolor[HTML]{82CE6A}47}                                            & \multicolumn{1}{c|}{\cellcolor[HTML]{F2E881}70}                                            & \multicolumn{1}{c|}{\cellcolor[HTML]{41BF5D}34}                                            & \multicolumn{1}{c|}{\cellcolor[HTML]{FFDC7C}87}                                            & \multicolumn{1}{c|}{\cellcolor[HTML]{FFC06C}114}                                           & \cellcolor[HTML]{FFA95F}136                                           \\ \hline
\end{tabular}
}
\caption{NRMSE for different humidity and reference $\text{PM}_{2.5}$ values/bins (Hourly averaged data). A graded color scheme has been applied to the NRMSE values in the table with red, yellow, and green representing the highest, middle, and lowest NRMSE values, respectively.}
\label{tab:Supplementary NRMSE}
\end{table}
\begin{table}[]
\centering
\begin{tabular}{|ccccc|}
\hline
\multicolumn{5}{|c|}{\textbf{Standard deviation (SD) for LCS groups}}\\ \hline
\multicolumn{1}{|c|}{\begin{tabular}[c]{@{}c@{}}\textbf{BAM}\\ $(\mu \text{g/m}^3)$\end{tabular}} & \multicolumn{1}{c|}{\begin{tabular}[c]{@{}c@{}}\textbf{Humidity}\\ $(\%)$\end{tabular}} & \multicolumn{1}{c|}{\begin{tabular}[c]{@{}c@{}}\textbf{Sensirion}\\ $(\mu \text{g/m}^3)$\end{tabular}} & \multicolumn{1}{c|}{\begin{tabular}[c]{@{}c@{}}\textbf{Plantower}\\ $(\mu \text{g/m}^3)$\end{tabular}} & \begin{tabular}[c]{@{}c@{}}\textbf{Honeywell}\\ $(\mu \text{g/m}^3)$\end{tabular} \\ \hline
\multicolumn{1}{|c|}{\cellcolor[HTML]{FDE9D9}}                                             & \multicolumn{1}{c|}{\cellcolor[HTML]{FDE9D9}20-30}                                      & \multicolumn{1}{c|}{\cellcolor[HTML]{00B050}0.67}                                               & \multicolumn{1}{c|}{\cellcolor[HTML]{31BB59}4.12}                                               & \cellcolor[HTML]{FFEB84}18.63                                              \\ \cline{2-5} 
\multicolumn{1}{|c|}{\cellcolor[HTML]{FDE9D9}}                                             & \multicolumn{1}{c|}{\cellcolor[HTML]{FDE9D9}30-40}                                      & \multicolumn{1}{c|}{\cellcolor[HTML]{05B151}1.04}                                               & \multicolumn{1}{c|}{\cellcolor[HTML]{51C260}6.37}                                               & \cellcolor[HTML]{FFE983}19.48                                              \\ \cline{2-5} 
\multicolumn{1}{|c|}{\cellcolor[HTML]{FDE9D9}}                                             & \multicolumn{1}{c|}{\cellcolor[HTML]{FDE9D9}40-50}                                      & \multicolumn{1}{c|}{\cellcolor[HTML]{0DB352}1.63}                                               & \multicolumn{1}{c|}{\cellcolor[HTML]{5FC563}7.36}                                               & \cellcolor[HTML]{FDEA83}18.49                                              \\ \cline{2-5} 
\multicolumn{1}{|c|}{\cellcolor[HTML]{FDE9D9}}                                             & \multicolumn{1}{c|}{\cellcolor[HTML]{FDE9D9}50-60}                                      & \multicolumn{1}{c|}{\cellcolor[HTML]{12B453}1.94}                                               & \multicolumn{1}{c|}{\cellcolor[HTML]{80CD6A}9.71}                                               & \cellcolor[HTML]{FFEA83}19.25                                              \\ \cline{2-5} 
\multicolumn{1}{|c|}{\cellcolor[HTML]{FDE9D9}}                                             & \multicolumn{1}{c|}{\cellcolor[HTML]{FDE9D9}60-70}                                      & \multicolumn{1}{c|}{\cellcolor[HTML]{10B353}1.85}                                               & \multicolumn{1}{c|}{\cellcolor[HTML]{75CB68}8.96}                                               & \cellcolor[HTML]{FFE782}19.91                                              \\ \cline{2-5} 
\multicolumn{1}{|c|}{\cellcolor[HTML]{FDE9D9}}                                             & \multicolumn{1}{c|}{\cellcolor[HTML]{FDE9D9}70-80}                                      & \multicolumn{1}{c|}{\cellcolor[HTML]{0BB252}1.49}                                               & \multicolumn{1}{c|}{\cellcolor[HTML]{80CD6A}9.74}                                               & \cellcolor[HTML]{FFDC7C}23.60                                              \\ \cline{2-5} 
\multicolumn{1}{|c|}{\cellcolor[HTML]{FDE9D9}}                                             & \multicolumn{1}{c|}{\cellcolor[HTML]{FDE9D9}80-90}                                      & \multicolumn{1}{c|}{\cellcolor[HTML]{44BF5E}5.50}                                               & \multicolumn{1}{c|}{\cellcolor[HTML]{F1E781}17.69}                                              & \cellcolor[HTML]{FFC36E}31.42                                              \\ \cline{2-5} 
\multicolumn{1}{|c|}{\multirow{-8}{*}{\cellcolor[HTML]{FDE9D9}0-100}}                      & \multicolumn{1}{c|}{\cellcolor[HTML]{FDE9D9}90-100}                                     & \multicolumn{1}{c|}{\cellcolor[HTML]{39BD5B}4.69}                                               & \multicolumn{1}{c|}{\cellcolor[HTML]{DEE37D}16.33}                                              & \cellcolor[HTML]{FFAA60}39.35                                              \\ \hline
\multicolumn{1}{|c|}{\cellcolor[HTML]{FCD5B4}}                                             & \multicolumn{1}{c|}{\cellcolor[HTML]{FCD5B4}40-50}                                      & \multicolumn{1}{c|}{\cellcolor[HTML]{3BBD5C}4.83}                                               & \multicolumn{1}{c|}{\cellcolor[HTML]{ADD873}12.85}                                              & \cellcolor[HTML]{FFCB72}29.11                                              \\ \cline{2-5} 
\multicolumn{1}{|c|}{\cellcolor[HTML]{FCD5B4}}                                             & \multicolumn{1}{c|}{\cellcolor[HTML]{FCD5B4}50-60}                                      & \multicolumn{1}{c|}{\cellcolor[HTML]{45C05E}5.58}                                               & \multicolumn{1}{c|}{\cellcolor[HTML]{FFE983}19.55}                                              & \cellcolor[HTML]{FFC770}30.33                                              \\ \cline{2-5} 
\multicolumn{1}{|c|}{\cellcolor[HTML]{FCD5B4}}                                             & \multicolumn{1}{c|}{\cellcolor[HTML]{FCD5B4}60-70}                                      & \multicolumn{1}{c|}{\cellcolor[HTML]{34BC5A}4.35}                                               & \multicolumn{1}{c|}{\cellcolor[HTML]{FFD87A}24.83}                                              & \cellcolor[HTML]{FFBE6B}32.96                                              \\ \cline{2-5} 
\multicolumn{1}{|c|}{\cellcolor[HTML]{FCD5B4}}                                             & \multicolumn{1}{c|}{\cellcolor[HTML]{FCD5B4}70-80}                                      & \multicolumn{1}{c|}{\cellcolor[HTML]{57C461}6.84}                                               & \multicolumn{1}{c|}{\cellcolor[HTML]{FFC36E}31.56}                                              & \cellcolor[HTML]{FFAC61}38.82                                              \\ \cline{2-5} 
\multicolumn{1}{|c|}{\cellcolor[HTML]{FCD5B4}}                                             & \multicolumn{1}{c|}{\cellcolor[HTML]{FCD5B4}80-90}                                      & \multicolumn{1}{c|}{\cellcolor[HTML]{62C664}7.63}                                               & \multicolumn{1}{c|}{\cellcolor[HTML]{FFB868}35.00}                                              & \cellcolor[HTML]{FF9151}47.53                                              \\ \cline{2-5} 
\multicolumn{1}{|c|}{\multirow{-6}{*}{\cellcolor[HTML]{FCD5B4}100-200}}                    & \multicolumn{1}{c|}{\cellcolor[HTML]{FCD5B4}90-100}                                     & \multicolumn{1}{c|}{\cellcolor[HTML]{6AC865}8.14}                                               & \multicolumn{1}{c|}{\cellcolor[HTML]{FFB767}35.46}                                              & \cellcolor[HTML]{FF683B}60.49                                              \\ \hline
\multicolumn{1}{|c|}{\cellcolor[HTML]{FABF8F}}                                             & \multicolumn{1}{c|}{\cellcolor[HTML]{FABF8F}50-60}                                      & \multicolumn{1}{c|}{\cellcolor[HTML]{7FCD6A}9.65}                                               & \multicolumn{1}{c|}{\cellcolor[HTML]{FFDD7C}23.23}                                              & \cellcolor[HTML]{FF8F50}48.08                                              \\ \cline{2-5} 
\multicolumn{1}{|c|}{\cellcolor[HTML]{FABF8F}}                                             & \multicolumn{1}{c|}{\cellcolor[HTML]{FABF8F}60-70}                                      & \multicolumn{1}{c|}{\cellcolor[HTML]{7CCC69}9.44}                                               & \multicolumn{1}{c|}{\cellcolor[HTML]{FFCD73}28.30}                                              & \cellcolor[HTML]{FF5A33}64.78                                              \\ \cline{2-5} 
\multicolumn{1}{|c|}{\cellcolor[HTML]{FABF8F}}                                             & \multicolumn{1}{c|}{\cellcolor[HTML]{FABF8F}70-80}                                      & \multicolumn{1}{c|}{\cellcolor[HTML]{94D26E}11.12}                                              & \multicolumn{1}{c|}{\cellcolor[HTML]{FFAA60}39.31}                                              & \cellcolor[HTML]{FF6036}63.05                                              \\ \cline{2-5} 
\multicolumn{1}{|c|}{\cellcolor[HTML]{FABF8F}}                                             & \multicolumn{1}{c|}{\cellcolor[HTML]{FABF8F}80-90}                                      & \multicolumn{1}{c|}{\cellcolor[HTML]{96D26E}11.26}                                              & \multicolumn{1}{c|}{\cellcolor[HTML]{FF7944}55.10}                                              & \cellcolor[HTML]{FF5731}65.88                                              \\ \cline{2-5} 
\multicolumn{1}{|c|}{\multirow{-5}{*}{\cellcolor[HTML]{FABF8F}200-300}}                    & \multicolumn{1}{c|}{\cellcolor[HTML]{FABF8F}90-100}                                     & \multicolumn{1}{c|}{\cellcolor[HTML]{41BF5D}5.26}                                               & \multicolumn{1}{c|}{\cellcolor[HTML]{FF834A}51.96}                                              & \cellcolor[HTML]{FF311C}77.84                                              \\ \hline
\multicolumn{1}{|c|}{\cellcolor[HTML]{E26B0A}}                                             & \multicolumn{1}{c|}{\cellcolor[HTML]{E26B0A}70-80}                                      & \multicolumn{1}{c|}{\cellcolor[HTML]{F0E780}17.58}                                              & \multicolumn{1}{c|}{\cellcolor[HTML]{FFAB60}39.09}                                              & \cellcolor[HTML]{FF0000}93.34                                              \\ \cline{2-5} 
\multicolumn{1}{|c|}{\multirow{-2}{*}{\cellcolor[HTML]{E26B0A}300-400}}                    & \multicolumn{1}{c|}{\cellcolor[HTML]{E26B0A}80-90}                                      & \multicolumn{1}{c|}{\cellcolor[HTML]{F3E881}17.83}                                              & \multicolumn{1}{c|}{\cellcolor[HTML]{FF4C2B}69.23}                                              & \cellcolor[HTML]{FF331D}77.28                                              \\ \hline
\end{tabular}
\caption{Standard deviation (SD) for different LCS groups (hourly averaged data). A graded color scheme has been applied to the SD values in the table with red, yellow, and green representing the highest, middle, and lowest SD values, respectively.}
\label{tab:Supplementary sd}
\end{table}
\begin{table}[]
\centering
\resizebox{0.6\textwidth}{!}{
\begin{tabular}{|ccccc|}
\hline
\multicolumn{5}{|c|}{\textbf{Coefficient of variation (CV)}}\\ \hline
\multicolumn{1}{|c|}{\begin{tabular}[c]{@{}c@{}}BAM\\ $(\mu \text{g/m}^3)$\end{tabular}} & \multicolumn{1}{c|}{\begin{tabular}[c]{@{}c@{}}Humidity\\ $(\%)$\end{tabular}} & \multicolumn{1}{c|}{\begin{tabular}[c]{@{}c@{}}Sensirion\\ $(\%)$\end{tabular}} & \multicolumn{1}{c|}{\begin{tabular}[c]{@{}c@{}}Plantower\\ $(\%)$\end{tabular}} & \begin{tabular}[c]{@{}c@{}}Honeywell\\ $(\%)$\end{tabular} \\ \hline
\multicolumn{1}{|c|}{\cellcolor[HTML]{FDE9D9}}                                    & \multicolumn{1}{c|}{\cellcolor[HTML]{FDE9D9}20-30}                             & \multicolumn{1}{c|}{\cellcolor[HTML]{13B454}2.14}                               & \multicolumn{1}{c|}{\cellcolor[HTML]{E4E47E}9.68}                               & \cellcolor[HTML]{FF0604}21.79                              \\ \cline{2-5} 
\multicolumn{1}{|c|}{\cellcolor[HTML]{FDE9D9}}                                    & \multicolumn{1}{c|}{\cellcolor[HTML]{FDE9D9}30-40}                             & \multicolumn{1}{c|}{\cellcolor[HTML]{09B251}1.76}                               & \multicolumn{1}{c|}{\cellcolor[HTML]{F4E881}10.27}                              & \cellcolor[HTML]{FF6A3C}16.94                              \\ \cline{2-5} 
\multicolumn{1}{|c|}{\cellcolor[HTML]{FDE9D9}}                                    & \multicolumn{1}{c|}{\cellcolor[HTML]{FDE9D9}40-50}                             & \multicolumn{1}{c|}{\cellcolor[HTML]{1BB655}2.42}                               & \multicolumn{1}{c|}{\cellcolor[HTML]{FFEB84}10.65}                              & \cellcolor[HTML]{FF864B}15.59                              \\ \cline{2-5} 
\multicolumn{1}{|c|}{\cellcolor[HTML]{FDE9D9}}                                    & \multicolumn{1}{c|}{\cellcolor[HTML]{FDE9D9}50-60}                             & \multicolumn{1}{c|}{\cellcolor[HTML]{21B756}2.62}                               & \multicolumn{1}{c|}{\cellcolor[HTML]{FFC56F}12.53}                              & \cellcolor[HTML]{FF8A4E}15.36                              \\ \cline{2-5} 
\multicolumn{1}{|c|}{\cellcolor[HTML]{FDE9D9}}                                    & \multicolumn{1}{c|}{\cellcolor[HTML]{FDE9D9}60-70}                             & \multicolumn{1}{c|}{\cellcolor[HTML]{17B554}2.27}                               & \multicolumn{1}{c|}{\cellcolor[HTML]{E8E57F}9.84}                               & \cellcolor[HTML]{FF9152}15.05                              \\ \cline{2-5} 
\multicolumn{1}{|c|}{\cellcolor[HTML]{FDE9D9}}                                    & \multicolumn{1}{c|}{\cellcolor[HTML]{FDE9D9}70-80}                             & \multicolumn{1}{c|}{\cellcolor[HTML]{03B050}1.56}                               & \multicolumn{1}{c|}{\cellcolor[HTML]{D2E07A}9.05}                               & \cellcolor[HTML]{FF8C4F}15.30                              \\ \cline{2-5} 
\multicolumn{1}{|c|}{\cellcolor[HTML]{FDE9D9}}                                    & \multicolumn{1}{c|}{\cellcolor[HTML]{FDE9D9}80-90}                             & \multicolumn{1}{c|}{\cellcolor[HTML]{54C361}4.47}                               & \multicolumn{1}{c|}{\cellcolor[HTML]{FFC16D}12.69}                              & \cellcolor[HTML]{FF6E3E}16.75                              \\ \cline{2-5} 
\multicolumn{1}{|c|}{\multirow{-8}{*}{\cellcolor[HTML]{FDE9D9}0-100}}             & \multicolumn{1}{c|}{\cellcolor[HTML]{FDE9D9}90-100}                            & \multicolumn{1}{c|}{\cellcolor[HTML]{52C260}4.39}                               & \multicolumn{1}{c|}{\cellcolor[HTML]{FFA95F}13.87}                              & \cellcolor[HTML]{FF0000}22.05                              \\ \hline
\multicolumn{1}{|c|}{\cellcolor[HTML]{FCD5B4}}                                    & \multicolumn{1}{c|}{\cellcolor[HTML]{FCD5B4}40-50}                             & \multicolumn{1}{c|}{\cellcolor[HTML]{1CB655}2.45}                               & \multicolumn{1}{c|}{\cellcolor[HTML]{9BD46F}7.06}                               & \cellcolor[HTML]{FFE681}10.91                              \\ \cline{2-5} 
\multicolumn{1}{|c|}{\cellcolor[HTML]{FCD5B4}}                                    & \multicolumn{1}{c|}{\cellcolor[HTML]{FCD5B4}50-60}                             & \multicolumn{1}{c|}{\cellcolor[HTML]{27B958}2.85}                               & \multicolumn{1}{c|}{\cellcolor[HTML]{F7E982}10.37}                              & \cellcolor[HTML]{FFD779}11.63                              \\ \cline{2-5} 
\multicolumn{1}{|c|}{\cellcolor[HTML]{FCD5B4}}                                    & \multicolumn{1}{c|}{\cellcolor[HTML]{FCD5B4}60-70}                             & \multicolumn{1}{c|}{\cellcolor[HTML]{12B453}2.09}                               & \multicolumn{1}{c|}{\cellcolor[HTML]{FFCA72}12.26}                              & \cellcolor[HTML]{FFCA72}12.25                              \\ \cline{2-5} 
\multicolumn{1}{|c|}{\cellcolor[HTML]{FCD5B4}}                                    & \multicolumn{1}{c|}{\cellcolor[HTML]{FCD5B4}70-80}                             & \multicolumn{1}{c|}{\cellcolor[HTML]{2CBA59}3.02}                               & \multicolumn{1}{c|}{\cellcolor[HTML]{FFAA60}13.80}                              & \cellcolor[HTML]{FFAF62}13.60                              \\ \cline{2-5} 
\multicolumn{1}{|c|}{\cellcolor[HTML]{FCD5B4}}                                    & \multicolumn{1}{c|}{\cellcolor[HTML]{FCD5B4}80-90}                             & \multicolumn{1}{c|}{\cellcolor[HTML]{30BB59}3.17}                               & \multicolumn{1}{c|}{\cellcolor[HTML]{FFA15B}14.26}                              & \cellcolor[HTML]{FF8F51}15.11                              \\ \cline{2-5} 
\multicolumn{1}{|c|}{\multirow{-6}{*}{\cellcolor[HTML]{FCD5B4}100-200}}           & \multicolumn{1}{c|}{\cellcolor[HTML]{FCD5B4}90-100}                            & \multicolumn{1}{c|}{\cellcolor[HTML]{37BC5B}3.44}                               & \multicolumn{1}{c|}{\cellcolor[HTML]{FFA55D}14.04}                              & \cellcolor[HTML]{FF4426}18.78                              \\ \hline
\multicolumn{1}{|c|}{\cellcolor[HTML]{FABF8F}}                                    & \multicolumn{1}{c|}{\cellcolor[HTML]{FABF8F}50-60}                             & \multicolumn{1}{c|}{\cellcolor[HTML]{1DB656}2.50}                               & \multicolumn{1}{c|}{\cellcolor[HTML]{A9D772}7.53}                               & \cellcolor[HTML]{E7E57F}9.80                               \\ \cline{2-5} 
\multicolumn{1}{|c|}{\cellcolor[HTML]{FABF8F}}                                    & \multicolumn{1}{c|}{\cellcolor[HTML]{FABF8F}60-70}                             & \multicolumn{1}{c|}{\cellcolor[HTML]{16B554}2.22}                               & \multicolumn{1}{c|}{\cellcolor[HTML]{CEDF7A}8.88}                               & \cellcolor[HTML]{FFC670}12.45                              \\ \cline{2-5} 
\multicolumn{1}{|c|}{\cellcolor[HTML]{FABF8F}}                                    & \multicolumn{1}{c|}{\cellcolor[HTML]{FABF8F}70-80}                             & \multicolumn{1}{c|}{\cellcolor[HTML]{23B857}2.69}                               & \multicolumn{1}{c|}{\cellcolor[HTML]{FFDD7C}11.37}                              & \cellcolor[HTML]{FFC670}12.46                              \\ \cline{2-5} 
\multicolumn{1}{|c|}{\cellcolor[HTML]{FABF8F}}                                    & \multicolumn{1}{c|}{\cellcolor[HTML]{FABF8F}80-90}                             & \multicolumn{1}{c|}{\cellcolor[HTML]{22B757}2.66}                               & \multicolumn{1}{c|}{\cellcolor[HTML]{FF9554}14.83}                              & \cellcolor[HTML]{FFB968}13.08                              \\ \cline{2-5} 
\multicolumn{1}{|c|}{\multirow{-5}{*}{\cellcolor[HTML]{FABF8F}200-300}}           & \multicolumn{1}{c|}{\cellcolor[HTML]{FABF8F}90-100}                            & \multicolumn{1}{c|}{\cellcolor[HTML]{00B050}1.42}                               & \multicolumn{1}{c|}{\cellcolor[HTML]{FFA65E}14.00}                              & \cellcolor[HTML]{FF7442}16.43                              \\ \hline
\multicolumn{1}{|c|}{\cellcolor[HTML]{E26B0A}}                                    & \multicolumn{1}{c|}{\cellcolor[HTML]{E26B0A}70-80}                             & \multicolumn{1}{c|}{\cellcolor[HTML]{31BB5A}3.19}                               & \multicolumn{1}{c|}{\cellcolor[HTML]{EDE680}10.00}                              & \cellcolor[HTML]{FFA65E}14.01                              \\ \cline{2-5} 
\multicolumn{1}{|c|}{\multirow{-2}{*}{\cellcolor[HTML]{E26B0A}300-400}}           & \multicolumn{1}{c|}{\cellcolor[HTML]{E26B0A}80-90}                             & \multicolumn{1}{c|}{\cellcolor[HTML]{2BBA58}3.00}                               & \multicolumn{1}{c|}{\cellcolor[HTML]{FF9C58}14.51}                              & \cellcolor[HTML]{FFE27F}11.12                              \\ \hline
\end{tabular}
}
\caption{Coefficient of variation (CV) for different humidity and reference $\text{PM}_{2.5}$ values/bins (hourly averaged data). A graded color scheme has been applied to the CV values in the table with red, yellow, and green representing the highest, middle, and lowest CV values, respectively.}
\label{tab:CV}
\end{table}

\begin{table}[]
\resizebox{1\textwidth}{!}{%
\centering
\begin{tabular}{|cccccccccc|}
\hline
\multicolumn{10}{|c|}{\textbf{$p$-value for humidity}}\\ \hline
\multicolumn{1}{|c|}{\textbf{\begin{tabular}[c]{@{}c@{}}BAM\\ $(\mu \text{g/m}^3)$\end{tabular}}} & \multicolumn{1}{c|}{\textbf{\begin{tabular}[c]{@{}c@{}}Humidity\\ $(\%)$\end{tabular}}} & \multicolumn{1}{c|}{\textbf{Sensirion-1}}           & \multicolumn{1}{c|}{\textbf{Sensirion-2}}           & \multicolumn{1}{c|}{\textbf{Plantower-1}}           & \multicolumn{1}{c|}{\textbf{Plantower-2}}           & \multicolumn{1}{c|}{\textbf{Plantower-3}}           & \multicolumn{1}{c|}{\textbf{Honeywell-1}}           & \multicolumn{1}{c|}{\textbf{Honeywell-2}}           & \textbf{Honeywell-3}           \\ \hline
\multicolumn{1}{|c|}{}                                                                    & \multicolumn{1}{c|}{30-40}                                                              & \multicolumn{1}{c|}{0.9817}                         & \multicolumn{1}{c|}{0.9921}                         & \multicolumn{1}{c|}{0.7827}                         & \multicolumn{1}{c|}{0.9383}                         & \multicolumn{1}{c|}{0.3055}                         & \multicolumn{1}{c|}{0.5621}                         & \multicolumn{1}{c|}{0.9464}                         & 0.7562                         \\ \cline{2-10} 
\multicolumn{1}{|c|}{}                                                                    & \multicolumn{1}{c|}{40-50}                                                              & \multicolumn{1}{c|}{0.4845}                         & \multicolumn{1}{c|}{0.5781}                         & \multicolumn{1}{c|}{0.7205}                         & \multicolumn{1}{c|}{0.3022}                         & \multicolumn{1}{c|}{0.8201}                         & \multicolumn{1}{c|}{0.1277}                         & \multicolumn{1}{c|}{0.7147}                         & 0.6530                         \\ \cline{2-10} 
\multicolumn{1}{|c|}{}                                                                    & \multicolumn{1}{c|}{50-60}                                                              & \multicolumn{1}{c|}{0.3363}                         & \multicolumn{1}{c|}{0.3068}                         & \multicolumn{1}{c|}{0.1494}                         & \multicolumn{1}{c|}{0.2509}                         & \multicolumn{1}{c|}{0.1672}                         & \multicolumn{1}{c|}{0.1241}                         & \multicolumn{1}{c|}{0.2050}                         & 0.2063                         \\ \cline{2-10} 
\multicolumn{1}{|c|}{}                                                                    & \multicolumn{1}{c|}{60-70}                                                              & \multicolumn{1}{c|}{0.5503}                         & \multicolumn{1}{c|}{0.9856}                         & \multicolumn{1}{c|}{0.9846}                         & \multicolumn{1}{c|}{0.6210}                         & \multicolumn{1}{c|}{0.5557}                         & \multicolumn{1}{c|}{0.6844}                         & \multicolumn{1}{c|}{0.9599}                         & 0.8538                         \\ \cline{2-10} 
\multicolumn{1}{|c|}{}                                                                    & \multicolumn{1}{c|}{70-80}                                                              & \multicolumn{1}{c|}{0.0759}                         & \multicolumn{1}{c|}{\cellcolor[HTML]{00B050}0.0485} & \multicolumn{1}{c|}{\cellcolor[HTML]{00B050}0.0201} & \multicolumn{1}{c|}{0.1336}                         & \multicolumn{1}{c|}{\cellcolor[HTML]{00B050}0.0368} & \multicolumn{1}{c|}{0.1205}                         & \multicolumn{1}{c|}{\cellcolor[HTML]{00B050}0.0281} & \cellcolor[HTML]{00B050}0.0473 \\ \cline{2-10} 
\multicolumn{1}{|c|}{}                                                                    & \multicolumn{1}{c|}{80-90}                                                              & \multicolumn{1}{c|}{0.5953}                         & \multicolumn{1}{c|}{0.8000}                         & \multicolumn{1}{c|}{0.3731}                         & \multicolumn{1}{c|}{0.2894}                         & \multicolumn{1}{c|}{0.8588}                         & \multicolumn{1}{c|}{0.6322}                         & \multicolumn{1}{c|}{0.2358}                         & \cellcolor[HTML]{00B050}0.0256 \\ \cline{2-10} 
\multicolumn{1}{|c|}{\multirow{-7}{*}{0-100}}                                             & \multicolumn{1}{c|}{90-100}                                                             & \multicolumn{1}{c|}{\cellcolor[HTML]{00B050}0.0000} & \multicolumn{1}{c|}{\cellcolor[HTML]{00B050}0.0002} & \multicolumn{1}{c|}{\cellcolor[HTML]{00B050}0.0001} & \multicolumn{1}{c|}{\cellcolor[HTML]{00B050}0.0002} & \multicolumn{1}{c|}{\cellcolor[HTML]{00B050}0.0000} & \multicolumn{1}{c|}{\cellcolor[HTML]{00B050}0.0000} & \multicolumn{1}{c|}{\cellcolor[HTML]{00B050}0.0000} & \cellcolor[HTML]{00B050}0.0000 \\ \hline
\multicolumn{1}{|c|}{}                                                                    & \multicolumn{1}{c|}{40-50}                                                              & \multicolumn{1}{c|}{0.7499}                         & \multicolumn{1}{c|}{0.9628}                         & \multicolumn{1}{c|}{0.7547}                         & \multicolumn{1}{c|}{0.4008}                         & \multicolumn{1}{c|}{0.8637}                         & \multicolumn{1}{c|}{0.7738}                         & \multicolumn{1}{c|}{0.8200}                         & 0.6727                         \\ \cline{2-10} 
\multicolumn{1}{|c|}{}                                                                    & \multicolumn{1}{c|}{50-60}                                                              & \multicolumn{1}{c|}{0.7710}                         & \multicolumn{1}{c|}{0.9918}                         & \multicolumn{1}{c|}{0.5300}                         & \multicolumn{1}{c|}{0.7066}                         & \multicolumn{1}{c|}{0.9709}                         & \multicolumn{1}{c|}{0.2234}                         & \multicolumn{1}{c|}{0.1422}                         & 0.8851                         \\ \cline{2-10} 
\multicolumn{1}{|c|}{}                                                                    & \multicolumn{1}{c|}{60-70}                                                              & \multicolumn{1}{c|}{0.5023}                         & \multicolumn{1}{c|}{0.6723}                         & \multicolumn{1}{c|}{0.1957}                         & \multicolumn{1}{c|}{0.2491}                         & \multicolumn{1}{c|}{0.8019}                         & \multicolumn{1}{c|}{0.2777}                         & \multicolumn{1}{c|}{0.4981}                         & 0.7520                         \\ \cline{2-10} 
\multicolumn{1}{|c|}{}                                                                    & \multicolumn{1}{c|}{70-80}                                                              & \multicolumn{1}{c|}{\cellcolor[HTML]{00B050}0.0052} & \multicolumn{1}{c|}{\cellcolor[HTML]{00B050}0.0242} & \multicolumn{1}{c|}{\cellcolor[HTML]{00B050}0.0054} & \multicolumn{1}{c|}{\cellcolor[HTML]{00B050}0.0128} & \multicolumn{1}{c|}{\cellcolor[HTML]{00B050}0.0497} & \multicolumn{1}{c|}{0.0616}                         & \multicolumn{1}{c|}{\cellcolor[HTML]{00B050}0.0231} & \cellcolor[HTML]{00B050}0.0011 \\ \cline{2-10} 
\multicolumn{1}{|c|}{}                                                                    & \multicolumn{1}{c|}{80-90}                                                              & \multicolumn{1}{c|}{\cellcolor[HTML]{00B050}0.0143} & \multicolumn{1}{c|}{\cellcolor[HTML]{00B050}0.0143} & \multicolumn{1}{c|}{0.6983}                         & \multicolumn{1}{c|}{\cellcolor[HTML]{00B050}0.0179} & \multicolumn{1}{c|}{\cellcolor[HTML]{00B050}0.0006} & \multicolumn{1}{c|}{\cellcolor[HTML]{00B050}0.0275} & \multicolumn{1}{c|}{0.3901}                         & 0.2863                         \\ \cline{2-10} 
\multicolumn{1}{|c|}{\multirow{-6}{*}{100-200}}                                           & \multicolumn{1}{c|}{90-100}                                                             & \multicolumn{1}{c|}{0.0821}                         & \multicolumn{1}{c|}{0.0915}                         & \multicolumn{1}{c|}{0.5793}                         & \multicolumn{1}{c|}{0.6909}                         & \multicolumn{1}{c|}{0.6490}                         & \multicolumn{1}{c|}{0.2100}                         & \multicolumn{1}{c|}{0.2619}                         & 0.0566                         \\ \hline
\multicolumn{1}{|c|}{}                                                                    & \multicolumn{1}{c|}{50-60}                                                              & \multicolumn{1}{c|}{0.5871}                         & \multicolumn{1}{c|}{0.6562}                         & \multicolumn{1}{c|}{0.8594}                         & \multicolumn{1}{c|}{0.7937}                         & \multicolumn{1}{c|}{0.9567}                         & \multicolumn{1}{c|}{0.7195}                         & \multicolumn{1}{c|}{0.3900}                         & 0.6142                         \\ \cline{2-10} 
\multicolumn{1}{|c|}{}                                                                    & \multicolumn{1}{c|}{60-70}                                                              & \multicolumn{1}{c|}{0.6592}                         & \multicolumn{1}{c|}{0.3269}                         & \multicolumn{1}{c|}{0.5098}                         & \multicolumn{1}{c|}{0.0882}                         & \multicolumn{1}{c|}{0.7144}                         & \multicolumn{1}{c|}{0.7990}                         & \multicolumn{1}{c|}{0.7450}                         & 0.3514                         \\ \cline{2-10} 
\multicolumn{1}{|c|}{}                                                                    & \multicolumn{1}{c|}{70-80}                                                              & \multicolumn{1}{c|}{0.1224}                         & \multicolumn{1}{c|}{0.0571}                         & \multicolumn{1}{c|}{\cellcolor[HTML]{00B050}0.0052} & \multicolumn{1}{c|}{0.2957}                         & \multicolumn{1}{c|}{\cellcolor[HTML]{00B050}0.0262} & \multicolumn{1}{c|}{0.1467}                         & \multicolumn{1}{c|}{0.1444}                         & 0.2430                         \\ \cline{2-10} 
\multicolumn{1}{|c|}{}                                                                    & \multicolumn{1}{c|}{80-90}                                                              & \multicolumn{1}{c|}{0.8480}                         & \multicolumn{1}{c|}{0.9488}                         & \multicolumn{1}{c|}{0.1406}                         & \multicolumn{1}{c|}{0.8729}                         & \multicolumn{1}{c|}{0.4502}                         & \multicolumn{1}{c|}{0.7973}                         & \multicolumn{1}{c|}{0.3140}                         & 0.0971                         \\ \cline{2-10} 
\multicolumn{1}{|c|}{\multirow{-5}{*}{200-300}}                                           & \multicolumn{1}{c|}{90-100}                                                             & \multicolumn{1}{c|}{0.3691}                         & \multicolumn{1}{c|}{0.4672}                         & \multicolumn{1}{c|}{0.6449}                         & \multicolumn{1}{c|}{0.8020}                         & \multicolumn{1}{c|}{0.6994}                         & \multicolumn{1}{c|}{0.3544}                         & \multicolumn{1}{c|}{0.4661}                         & 0.0916                         \\ \hline
\multicolumn{1}{|c|}{300-400}                                                             & \multicolumn{1}{c|}{80-90}                                                              & \multicolumn{1}{c|}{\cellcolor[HTML]{00B050}0.0421} & \multicolumn{1}{c|}{0.0688}                         & \multicolumn{1}{c|}{\cellcolor[HTML]{00B050}0.0002} & \multicolumn{1}{c|}{\cellcolor[HTML]{00B050}0.0023} & \multicolumn{1}{c|}{\cellcolor[HTML]{00B050}0.0004} & \multicolumn{1}{c|}{0.2345}                         & \multicolumn{1}{c|}{0.0804}                         & 0.0842                         \\ \hline
\end{tabular}%
}
\caption{$p$-values for humidity (hourly averaged data). Green colored cells indicate $p\leq~0.05$ and represent significant effect of humidity.}
\label{tab:P value for Humidity}
\end{table}

\begin{table}[]
\resizebox{\columnwidth}{!}{%
\begin{tabular}{|
>{\columncolor[HTML]{FFFFFF}}c 
>{\columncolor[HTML]{FFFFFF}}c 
>{\columncolor[HTML]{FFFFFF}}c 
>{\columncolor[HTML]{FFFFFF}}c 
>{\columncolor[HTML]{FFFFFF}}c 
>{\columncolor[HTML]{FFFFFF}}c 
>{\columncolor[HTML]{FFFFFF}}c 
>{\columncolor[HTML]{FFFFFF}}c 
>{\columncolor[HTML]{FFFFFF}}c 
>{\columncolor[HTML]{FFFFFF}}c 
>{\columncolor[HTML]{FFFFFF}}c |}
\hline
\multicolumn{11}{|c|}{\cellcolor[HTML]{FFFFFF}\textbf{$R^2$ values for LCS calibration models with and without humidity}}\\ \hline
\multicolumn{1}{|c|}{\cellcolor[HTML]{FFFFFF}\textbf{\begin{tabular}[c]{@{}c@{}}BAM\\ $(\mu \text{g/m}^3)$\end{tabular}}} & \multicolumn{1}{c|}{\cellcolor[HTML]{FFFFFF}\textbf{\begin{tabular}[c]{@{}c@{}}Humidity\\ $(\%)$\end{tabular}}} & \multicolumn{1}{c|}{\cellcolor[HTML]{FFFFFF}\textbf{Model}}        & \multicolumn{1}{c|}{\cellcolor[HTML]{FFFFFF}\textbf{Sensirion-1}} & \multicolumn{1}{c|}{\cellcolor[HTML]{FFFFFF}\textbf{Sensirion-2}} & \multicolumn{1}{c|}{\cellcolor[HTML]{FFFFFF}\textbf{Plantower-1}} & \multicolumn{1}{c|}{\cellcolor[HTML]{FFFFFF}\textbf{Plantower-2}} & \multicolumn{1}{c|}{\cellcolor[HTML]{FFFFFF}\textbf{Plantower-3}} & \multicolumn{1}{c|}{\cellcolor[HTML]{FFFFFF}\textbf{Honeywell-1}} & \multicolumn{1}{c|}{\cellcolor[HTML]{FFFFFF}\textbf{Honeywell-2}} & \textbf{Honeywell-3} \\ \hline
\multicolumn{1}{|c|}{\cellcolor[HTML]{FFFFFF}}                                                                    & \multicolumn{1}{c|}{\cellcolor[HTML]{FFFFFF}}                                                                   & \multicolumn{1}{c|}{\cellcolor[HTML]{FFFFFF}BAM$\sim$LCS}          & \multicolumn{1}{c|}{\cellcolor[HTML]{FFFFFF}0.9485}               & \multicolumn{1}{c|}{\cellcolor[HTML]{FFFFFF}0.9518}               & \multicolumn{1}{c|}{\cellcolor[HTML]{FFFFFF}0.9190}               & \multicolumn{1}{c|}{\cellcolor[HTML]{FFFFFF}0.9215}               & \multicolumn{1}{c|}{\cellcolor[HTML]{FFFFFF}0.8503}               & \multicolumn{1}{c|}{\cellcolor[HTML]{FFFFFF}0.9317}               & \multicolumn{1}{c|}{\cellcolor[HTML]{FFFFFF}0.9027}               & 0.8999               \\ \cline{3-11} 
\multicolumn{1}{|c|}{\cellcolor[HTML]{FFFFFF}}                                                                    & \multicolumn{1}{c|}{\multirow{-2}{*}{\cellcolor[HTML]{FFFFFF}30-40}}                                            & \multicolumn{1}{c|}{\cellcolor[HTML]{FFFFFF}BAM$\sim$LCS+Humidity} & \multicolumn{1}{c|}{\cellcolor[HTML]{FFFFFF}0.9485}               & \multicolumn{1}{c|}{\cellcolor[HTML]{FFFFFF}0.9518}               & \multicolumn{1}{c|}{\cellcolor[HTML]{FFFFFF}0.9191}               & \multicolumn{1}{c|}{\cellcolor[HTML]{FFFFFF}0.9215}               & \multicolumn{1}{c|}{\cellcolor[HTML]{FFFFFF}0.8529}               & \multicolumn{1}{c|}{\cellcolor[HTML]{FFFFFF}0.9320}               & \multicolumn{1}{c|}{\cellcolor[HTML]{FFFFFF}0.9027}               & 0.9001               \\ \cline{2-11} 
\multicolumn{1}{|c|}{\cellcolor[HTML]{FFFFFF}}                                                                    & \multicolumn{1}{c|}{\cellcolor[HTML]{FFFFFF}}                                                                   & \multicolumn{1}{c|}{\cellcolor[HTML]{FFFFFF}BAM$\sim$LCS}          & \multicolumn{1}{c|}{\cellcolor[HTML]{FFFFFF}0.8788}               & \multicolumn{1}{c|}{\cellcolor[HTML]{FFFFFF}0.8960}               & \multicolumn{1}{c|}{\cellcolor[HTML]{FFFFFF}0.8245}               & \multicolumn{1}{c|}{\cellcolor[HTML]{FFFFFF}0.9072}               & \multicolumn{1}{c|}{\cellcolor[HTML]{FFFFFF}0.8077}               & \multicolumn{1}{c|}{\cellcolor[HTML]{FFFFFF}0.8498}               & \multicolumn{1}{c|}{\cellcolor[HTML]{FFFFFF}0.8300}               & 0.8295               \\ \cline{3-11} 
\multicolumn{1}{|c|}{\cellcolor[HTML]{FFFFFF}}                                                                    & \multicolumn{1}{c|}{\multirow{-2}{*}{\cellcolor[HTML]{FFFFFF}40-50}}                                            & \multicolumn{1}{c|}{\cellcolor[HTML]{FFFFFF}BAM$\sim$LCS+Humidity} & \multicolumn{1}{c|}{\cellcolor[HTML]{FFFFFF}0.8792}               & \multicolumn{1}{c|}{\cellcolor[HTML]{FFFFFF}0.8962}               & \multicolumn{1}{c|}{\cellcolor[HTML]{FFFFFF}0.8246}               & \multicolumn{1}{c|}{\cellcolor[HTML]{FFFFFF}0.9080}               & \multicolumn{1}{c|}{\cellcolor[HTML]{FFFFFF}0.8078}               & \multicolumn{1}{c|}{\cellcolor[HTML]{FFFFFF}0.8521}               & \multicolumn{1}{c|}{\cellcolor[HTML]{FFFFFF}0.8302}               & 0.8297               \\ \cline{2-11} 
\multicolumn{1}{|c|}{\cellcolor[HTML]{FFFFFF}}                                                                    & \multicolumn{1}{c|}{\cellcolor[HTML]{FFFFFF}}                                                                   & \multicolumn{1}{c|}{\cellcolor[HTML]{FFFFFF}BAM$\sim$LCS}          & \multicolumn{1}{c|}{\cellcolor[HTML]{FFFFFF}0.9427}               & \multicolumn{1}{c|}{\cellcolor[HTML]{FFFFFF}0.9502}               & \multicolumn{1}{c|}{\cellcolor[HTML]{FFFFFF}0.9083}               & \multicolumn{1}{c|}{\cellcolor[HTML]{FFFFFF}0.9318}               & \multicolumn{1}{c|}{\cellcolor[HTML]{FFFFFF}0.8824}               & \multicolumn{1}{c|}{\cellcolor[HTML]{FFFFFF}0.8864}               & \multicolumn{1}{c|}{\cellcolor[HTML]{FFFFFF}0.9055}               & 0.8427               \\ \cline{3-11} 
\multicolumn{1}{|c|}{\cellcolor[HTML]{FFFFFF}}                                                                    & \multicolumn{1}{c|}{\multirow{-2}{*}{\cellcolor[HTML]{FFFFFF}50-60}}                                            & \multicolumn{1}{c|}{\cellcolor[HTML]{FFFFFF}BAM$\sim$LCS+Humidity} & \multicolumn{1}{c|}{\cellcolor[HTML]{FFFFFF}0.9432}               & \multicolumn{1}{c|}{\cellcolor[HTML]{FFFFFF}0.9507}               & \multicolumn{1}{c|}{\cellcolor[HTML]{FFFFFF}0.9099}               & \multicolumn{1}{c|}{\cellcolor[HTML]{FFFFFF}0.9326}               & \multicolumn{1}{c|}{\cellcolor[HTML]{FFFFFF}0.8844}               & \multicolumn{1}{c|}{\cellcolor[HTML]{FFFFFF}0.8888}               & \multicolumn{1}{c|}{\cellcolor[HTML]{FFFFFF}0.9068}               & 0.8450               \\ \cline{2-11} 
\multicolumn{1}{|c|}{\cellcolor[HTML]{FFFFFF}}                                                                    & \multicolumn{1}{c|}{\cellcolor[HTML]{FFFFFF}}                                                                   & \multicolumn{1}{c|}{\cellcolor[HTML]{FFFFFF}BAM$\sim$LCS}          & \multicolumn{1}{c|}{\cellcolor[HTML]{FFFFFF}0.8694}               & \multicolumn{1}{c|}{\cellcolor[HTML]{FFFFFF}0.8560}               & \multicolumn{1}{c|}{\cellcolor[HTML]{FFFFFF}0.7188}               & \multicolumn{1}{c|}{\cellcolor[HTML]{FFFFFF}0.8020}               & \multicolumn{1}{c|}{\cellcolor[HTML]{FFFFFF}0.7406}               & \multicolumn{1}{c|}{\cellcolor[HTML]{FFFFFF}0.8521}               & \multicolumn{1}{c|}{\cellcolor[HTML]{FFFFFF}0.8141}               & 0.6671               \\ \cline{3-11} 
\multicolumn{1}{|c|}{\cellcolor[HTML]{FFFFFF}}                                                                    & \multicolumn{1}{c|}{\multirow{-2}{*}{\cellcolor[HTML]{FFFFFF}60-70}}                                            & \multicolumn{1}{c|}{\cellcolor[HTML]{FFFFFF}BAM$\sim$LCS+Humidity} & \multicolumn{1}{c|}{\cellcolor[HTML]{FFFFFF}0.8698}               & \multicolumn{1}{c|}{\cellcolor[HTML]{FFFFFF}0.8560}               & \multicolumn{1}{c|}{\cellcolor[HTML]{FFFFFF}0.7188}               & \multicolumn{1}{c|}{\cellcolor[HTML]{FFFFFF}0.8025}               & \multicolumn{1}{c|}{\cellcolor[HTML]{FFFFFF}0.7414}               & \multicolumn{1}{c|}{\cellcolor[HTML]{FFFFFF}0.8524}               & \multicolumn{1}{c|}{\cellcolor[HTML]{FFFFFF}0.8141}               & 0.6672               \\ \cline{2-11} 
\multicolumn{1}{|c|}{\cellcolor[HTML]{FFFFFF}}                                                                    & \multicolumn{1}{c|}{\cellcolor[HTML]{FFFFFF}}                                                                   & \multicolumn{1}{c|}{\cellcolor[HTML]{FFFFFF}BAM$\sim$LCS}          & \multicolumn{1}{c|}{\cellcolor[HTML]{FFFFFF}0.7552}               & \multicolumn{1}{c|}{\cellcolor[HTML]{FFFFFF}0.7718}               & \multicolumn{1}{c|}{\cellcolor[HTML]{FFFFFF}0.6160}               & \multicolumn{1}{c|}{\cellcolor[HTML]{FFFFFF}0.6914}               & \multicolumn{1}{c|}{\cellcolor[HTML]{FFFFFF}0.5767}               & \multicolumn{1}{c|}{\cellcolor[HTML]{FFFFFF}0.6592}               & \multicolumn{1}{c|}{\cellcolor[HTML]{FFFFFF}0.6114}               & 0.5980               \\ \cline{3-11} 
\multicolumn{1}{|c|}{\cellcolor[HTML]{FFFFFF}}                                                                    & \multicolumn{1}{c|}{\multirow{-2}{*}{\cellcolor[HTML]{FFFFFF}70-80}}                                            & \multicolumn{1}{c|}{\cellcolor[HTML]{FFFFFF}BAM$\sim$LCS+Humidity} & \multicolumn{1}{c|}{\cellcolor[HTML]{FFFFFF}0.7594}               & \multicolumn{1}{c|}{\cellcolor[HTML]{FFFFFF}0.7766}               & \multicolumn{1}{c|}{\cellcolor[HTML]{FFFFFF}0.6272}               & \multicolumn{1}{c|}{\cellcolor[HTML]{FFFFFF}0.6954}               & \multicolumn{1}{c|}{\cellcolor[HTML]{FFFFFF}0.5874}               & \multicolumn{1}{c|}{\cellcolor[HTML]{FFFFFF}0.6639}               & \multicolumn{1}{c|}{\cellcolor[HTML]{FFFFFF}0.6215}               & 0.6068               \\ \cline{2-11} 
\multicolumn{1}{|c|}{\cellcolor[HTML]{FFFFFF}}                                                                    & \multicolumn{1}{c|}{\cellcolor[HTML]{FFFFFF}}                                                                   & \multicolumn{1}{c|}{\cellcolor[HTML]{FFFFFF}BAM$\sim$LCS}          & \multicolumn{1}{c|}{\cellcolor[HTML]{FFFFFF}0.5624}               & \multicolumn{1}{c|}{\cellcolor[HTML]{FFFFFF}0.6040}               & \multicolumn{1}{c|}{\cellcolor[HTML]{FFFFFF}0.4984}               & \multicolumn{1}{c|}{\cellcolor[HTML]{FFFFFF}0.5063}               & \multicolumn{1}{c|}{\cellcolor[HTML]{FFFFFF}0.4160}               & \multicolumn{1}{c|}{\cellcolor[HTML]{FFFFFF}0.5083}               & \multicolumn{1}{c|}{\cellcolor[HTML]{FFFFFF}0.4371}               & 0.5185               \\ \cline{3-11} 
\multicolumn{1}{|c|}{\cellcolor[HTML]{FFFFFF}}                                                                    & \multicolumn{1}{c|}{\multirow{-2}{*}{\cellcolor[HTML]{FFFFFF}80-90}}                                            & \multicolumn{1}{c|}{\cellcolor[HTML]{FFFFFF}BAM$\sim$LCS+Humidity} & \multicolumn{1}{c|}{\cellcolor[HTML]{FFFFFF}0.5630}               & \multicolumn{1}{c|}{\cellcolor[HTML]{FFFFFF}0.6041}               & \multicolumn{1}{c|}{\cellcolor[HTML]{FFFFFF}0.5006}               & \multicolumn{1}{c|}{\cellcolor[HTML]{FFFFFF}0.5095}               & \multicolumn{1}{c|}{\cellcolor[HTML]{FFFFFF}0.4161}               & \multicolumn{1}{c|}{\cellcolor[HTML]{FFFFFF}0.5090}               & \multicolumn{1}{c|}{\cellcolor[HTML]{FFFFFF}0.4412}               & 0.5309               \\ \cline{2-11} 
\multicolumn{1}{|c|}{\cellcolor[HTML]{FFFFFF}}                                                                    & \multicolumn{1}{c|}{\cellcolor[HTML]{FFFFFF}}                                                                   & \multicolumn{1}{c|}{\cellcolor[HTML]{FFFFFF}BAM$\sim$LCS}          & \multicolumn{1}{c|}{\cellcolor[HTML]{FFFFFF}0.8033}               & \multicolumn{1}{c|}{\cellcolor[HTML]{FFFFFF}0.8534}               & \multicolumn{1}{c|}{\cellcolor[HTML]{FFFFFF}0.6996}               & \multicolumn{1}{c|}{\cellcolor[HTML]{FFFFFF}0.7861}               & \multicolumn{1}{c|}{\cellcolor[HTML]{FFFFFF}0.6637}               & \multicolumn{1}{c|}{\cellcolor[HTML]{FFFFFF}0.6426}               & \multicolumn{1}{c|}{\cellcolor[HTML]{FFFFFF}0.7180}               & 0.6084               \\ \cline{3-11} 
\multicolumn{1}{|c|}{\multirow{-14}{*}{\cellcolor[HTML]{FFFFFF}0-100}}                                            & \multicolumn{1}{c|}{\multirow{-2}{*}{\cellcolor[HTML]{FFFFFF}90-100}}                                           & \multicolumn{1}{c|}{\cellcolor[HTML]{FFFFFF}BAM$\sim$LCS+Humidity} & \multicolumn{1}{c|}{\cellcolor[HTML]{FFFFFF}0.8312}               & \multicolumn{1}{c|}{\cellcolor[HTML]{FFFFFF}0.8694}               & \multicolumn{1}{c|}{\cellcolor[HTML]{FFFFFF}0.7402}               & \multicolumn{1}{c|}{\cellcolor[HTML]{FFFFFF}0.8098}               & \multicolumn{1}{c|}{\cellcolor[HTML]{FFFFFF}0.7208}               & \multicolumn{1}{c|}{\cellcolor[HTML]{FFFFFF}0.7379}               & \multicolumn{1}{c|}{\cellcolor[HTML]{FFFFFF}0.7678}               & 0.7034               \\ \hline
\multicolumn{1}{|c|}{\cellcolor[HTML]{FFFFFF}}                                                                    & \multicolumn{1}{c|}{\cellcolor[HTML]{FFFFFF}}                                                                   & \multicolumn{1}{c|}{\cellcolor[HTML]{FFFFFF}BAM$\sim$LCS}          & \multicolumn{1}{c|}{\cellcolor[HTML]{FFFFFF}0.7240}               & \multicolumn{1}{c|}{\cellcolor[HTML]{FFFFFF}0.7849}               & \multicolumn{1}{c|}{\cellcolor[HTML]{FFFFFF}0.7736}               & \multicolumn{1}{c|}{\cellcolor[HTML]{FFFFFF}0.4695}               & \multicolumn{1}{c|}{\cellcolor[HTML]{FFFFFF}0.7709}               & \multicolumn{1}{c|}{\cellcolor[HTML]{FFFFFF}0.7560}               & \multicolumn{1}{c|}{\cellcolor[HTML]{FFFFFF}0.7326}               & 0.7561               \\ \cline{3-11} 
\multicolumn{1}{|c|}{\cellcolor[HTML]{FFFFFF}}                                                                    & \multicolumn{1}{c|}{\multirow{-2}{*}{\cellcolor[HTML]{FFFFFF}40-50}}                                            & \multicolumn{1}{c|}{\cellcolor[HTML]{FFFFFF}BAM$\sim$LCS+Humidity} & \multicolumn{1}{c|}{\cellcolor[HTML]{FFFFFF}0.7245}               & \multicolumn{1}{c|}{\cellcolor[HTML]{FFFFFF}0.7849}               & \multicolumn{1}{c|}{\cellcolor[HTML]{FFFFFF}0.7740}               & \multicolumn{1}{c|}{\cellcolor[HTML]{FFFFFF}0.4756}               & \multicolumn{1}{c|}{\cellcolor[HTML]{FFFFFF}0.7710}               & \multicolumn{1}{c|}{\cellcolor[HTML]{FFFFFF}0.7563}               & \multicolumn{1}{c|}{\cellcolor[HTML]{FFFFFF}0.7328}               & 0.7568               \\ \cline{2-11} 
\multicolumn{1}{|c|}{\cellcolor[HTML]{FFFFFF}}                                                                    & \multicolumn{1}{c|}{\cellcolor[HTML]{FFFFFF}}                                                                   & \multicolumn{1}{c|}{\cellcolor[HTML]{FFFFFF}BAM$\sim$LCS}          & \multicolumn{1}{c|}{\cellcolor[HTML]{FFFFFF}0.8638}               & \multicolumn{1}{c|}{\cellcolor[HTML]{FFFFFF}0.8581}               & \multicolumn{1}{c|}{\cellcolor[HTML]{FFFFFF}0.8253}               & \multicolumn{1}{c|}{\cellcolor[HTML]{FFFFFF}0.5582}               & \multicolumn{1}{c|}{\cellcolor[HTML]{FFFFFF}0.7631}               & \multicolumn{1}{c|}{\cellcolor[HTML]{FFFFFF}0.8560}               & \multicolumn{1}{c|}{\cellcolor[HTML]{FFFFFF}0.8551}               & 0.8481               \\ \cline{3-11} 
\multicolumn{1}{|c|}{\cellcolor[HTML]{FFFFFF}}                                                                    & \multicolumn{1}{c|}{\multirow{-2}{*}{\cellcolor[HTML]{FFFFFF}50-60}}                                            & \multicolumn{1}{c|}{\cellcolor[HTML]{FFFFFF}BAM$\sim$LCS+Humidity} & \multicolumn{1}{c|}{\cellcolor[HTML]{FFFFFF}0.8639}               & \multicolumn{1}{c|}{\cellcolor[HTML]{FFFFFF}0.8581}               & \multicolumn{1}{c|}{\cellcolor[HTML]{FFFFFF}0.8260}               & \multicolumn{1}{c|}{\cellcolor[HTML]{FFFFFF}0.5589}               & \multicolumn{1}{c|}{\cellcolor[HTML]{FFFFFF}0.7631}               & \multicolumn{1}{c|}{\cellcolor[HTML]{FFFFFF}0.8581}               & \multicolumn{1}{c|}{\cellcolor[HTML]{FFFFFF}0.8582}               & 0.8481               \\ \cline{2-11} 
\multicolumn{1}{|c|}{\cellcolor[HTML]{FFFFFF}}                                                                    & \multicolumn{1}{c|}{\cellcolor[HTML]{FFFFFF}}                                                                   & \multicolumn{1}{c|}{\cellcolor[HTML]{FFFFFF}BAM$\sim$LCS}          & \multicolumn{1}{c|}{\cellcolor[HTML]{FFFFFF}0.6331}               & \multicolumn{1}{c|}{\cellcolor[HTML]{FFFFFF}0.6699}               & \multicolumn{1}{c|}{\cellcolor[HTML]{FFFFFF}0.5806}               & \multicolumn{1}{c|}{\cellcolor[HTML]{FFFFFF}0.3513}               & \multicolumn{1}{c|}{\cellcolor[HTML]{FFFFFF}0.4532}               & \multicolumn{1}{c|}{\cellcolor[HTML]{FFFFFF}0.6781}               & \multicolumn{1}{c|}{\cellcolor[HTML]{FFFFFF}0.6807}               & 0.7247               \\ \cline{3-11} 
\multicolumn{1}{|c|}{\cellcolor[HTML]{FFFFFF}}                                                                    & \multicolumn{1}{c|}{\multirow{-2}{*}{\cellcolor[HTML]{FFFFFF}60-70}}                                            & \multicolumn{1}{c|}{\cellcolor[HTML]{FFFFFF}BAM$\sim$LCS+Humidity} & \multicolumn{1}{c|}{\cellcolor[HTML]{FFFFFF}0.6344}               & \multicolumn{1}{c|}{\cellcolor[HTML]{FFFFFF}0.6704}               & \multicolumn{1}{c|}{\cellcolor[HTML]{FFFFFF}0.5857}               & \multicolumn{1}{c|}{\cellcolor[HTML]{FFFFFF}0.3581}               & \multicolumn{1}{c|}{\cellcolor[HTML]{FFFFFF}0.4535}               & \multicolumn{1}{c|}{\cellcolor[HTML]{FFFFFF}0.6809}               & \multicolumn{1}{c|}{\cellcolor[HTML]{FFFFFF}0.6818}               & 0.7249               \\ \cline{2-11} 
\multicolumn{1}{|c|}{\cellcolor[HTML]{FFFFFF}}                                                                    & \multicolumn{1}{c|}{\cellcolor[HTML]{FFFFFF}}                                                                   & \multicolumn{1}{c|}{\cellcolor[HTML]{FFFFFF}BAM$\sim$LCS}          & \multicolumn{1}{c|}{\cellcolor[HTML]{FFFFFF}0.5743}               & \multicolumn{1}{c|}{\cellcolor[HTML]{FFFFFF}0.5258}               & \multicolumn{1}{c|}{\cellcolor[HTML]{FFFFFF}0.4470}               & \multicolumn{1}{c|}{\cellcolor[HTML]{FFFFFF}0.3551}               & \multicolumn{1}{c|}{\cellcolor[HTML]{FFFFFF}0.3913}               & \multicolumn{1}{c|}{\cellcolor[HTML]{FFFFFF}0.5858}               & \multicolumn{1}{c|}{\cellcolor[HTML]{FFFFFF}0.5864}               & 0.5717               \\ \cline{3-11} 
\multicolumn{1}{|c|}{\cellcolor[HTML]{FFFFFF}}                                                                    & \multicolumn{1}{c|}{\multirow{-2}{*}{\cellcolor[HTML]{FFFFFF}70-80}}                                            & \multicolumn{1}{c|}{\cellcolor[HTML]{FFFFFF}BAM$\sim$LCS+Humidity} & \multicolumn{1}{c|}{\cellcolor[HTML]{FFFFFF}0.5861}               & \multicolumn{1}{c|}{\cellcolor[HTML]{FFFFFF}0.5347}               & \multicolumn{1}{c|}{\cellcolor[HTML]{FFFFFF}0.4620}               & \multicolumn{1}{c|}{\cellcolor[HTML]{FFFFFF}0.3695}               & \multicolumn{1}{c|}{\cellcolor[HTML]{FFFFFF}0.4020}               & \multicolumn{1}{c|}{\cellcolor[HTML]{FFFFFF}0.5909}               & \multicolumn{1}{c|}{\cellcolor[HTML]{FFFFFF}0.5942}               & 0.5875               \\ \cline{2-11} 
\multicolumn{1}{|c|}{\cellcolor[HTML]{FFFFFF}}                                                                    & \multicolumn{1}{c|}{\cellcolor[HTML]{FFFFFF}}                                                                   & \multicolumn{1}{c|}{\cellcolor[HTML]{FFFFFF}BAM$\sim$LCS}          & \multicolumn{1}{c|}{\cellcolor[HTML]{FFFFFF}0.5404}               & \multicolumn{1}{c|}{\cellcolor[HTML]{FFFFFF}0.5588}               & \multicolumn{1}{c|}{\cellcolor[HTML]{FFFFFF}0.4190}               & \multicolumn{1}{c|}{\cellcolor[HTML]{FFFFFF}0.4311}               & \multicolumn{1}{c|}{\cellcolor[HTML]{FFFFFF}0.3545}               & \multicolumn{1}{c|}{\cellcolor[HTML]{FFFFFF}0.4600}               & \multicolumn{1}{c|}{\cellcolor[HTML]{FFFFFF}0.4554}               & 0.5272               \\ \cline{3-11} 
\multicolumn{1}{|c|}{\cellcolor[HTML]{FFFFFF}}                                                                    & \multicolumn{1}{c|}{\multirow{-2}{*}{\cellcolor[HTML]{FFFFFF}80-90}}                                            & \multicolumn{1}{c|}{\cellcolor[HTML]{FFFFFF}BAM$\sim$LCS+Humidity} & \multicolumn{1}{c|}{\cellcolor[HTML]{FFFFFF}0.5467}               & \multicolumn{1}{c|}{\cellcolor[HTML]{FFFFFF}0.5648}               & \multicolumn{1}{c|}{\cellcolor[HTML]{FFFFFF}0.4192}               & \multicolumn{1}{c|}{\cellcolor[HTML]{FFFFFF}0.4387}               & \multicolumn{1}{c|}{\cellcolor[HTML]{FFFFFF}0.3728}               & \multicolumn{1}{c|}{\cellcolor[HTML]{FFFFFF}0.4659}               & \multicolumn{1}{c|}{\cellcolor[HTML]{FFFFFF}0.4563}               & 0.5284               \\ \cline{2-11} 
\multicolumn{1}{|c|}{\cellcolor[HTML]{FFFFFF}}                                                                    & \multicolumn{1}{c|}{\cellcolor[HTML]{FFFFFF}}                                                                   & \multicolumn{1}{c|}{\cellcolor[HTML]{FFFFFF}BAM$\sim$LCS}          & \multicolumn{1}{c|}{\cellcolor[HTML]{FFFFFF}0.6633}               & \multicolumn{1}{c|}{\cellcolor[HTML]{FFFFFF}0.7315}               & \multicolumn{1}{c|}{\cellcolor[HTML]{FFFFFF}0.2767}               & \multicolumn{1}{c|}{\cellcolor[HTML]{FFFFFF}0.5003}               & \multicolumn{1}{c|}{\cellcolor[HTML]{FFFFFF}0.4699}               & \multicolumn{1}{c|}{\cellcolor[HTML]{FFFFFF}0.5512}               & \multicolumn{1}{c|}{\cellcolor[HTML]{FFFFFF}0.5461}               & 0.5670               \\ \cline{3-11} 
\multicolumn{1}{|c|}{\multirow{-12}{*}{\cellcolor[HTML]{FFFFFF}100-200}}                                          & \multicolumn{1}{c|}{\multirow{-2}{*}{\cellcolor[HTML]{FFFFFF}90-100}}                                           & \multicolumn{1}{c|}{\cellcolor[HTML]{FFFFFF}BAM$\sim$LCS+Humidity} & \multicolumn{1}{c|}{\cellcolor[HTML]{FFFFFF}0.6743}               & \multicolumn{1}{c|}{\cellcolor[HTML]{FFFFFF}0.7399}               & \multicolumn{1}{c|}{\cellcolor[HTML]{FFFFFF}0.2792}               & \multicolumn{1}{c|}{\cellcolor[HTML]{FFFFFF}0.5012}               & \multicolumn{1}{c|}{\cellcolor[HTML]{FFFFFF}0.4713}               & \multicolumn{1}{c|}{\cellcolor[HTML]{FFFFFF}0.5591}               & \multicolumn{1}{c|}{\cellcolor[HTML]{FFFFFF}0.5523}               & 0.5841               \\ \hline
\multicolumn{1}{|c|}{\cellcolor[HTML]{FFFFFF}}                                                                    & \multicolumn{1}{c|}{\cellcolor[HTML]{FFFFFF}}                                                                   & \multicolumn{1}{c|}{\cellcolor[HTML]{FFFFFF}BAM$\sim$LCS}          & \multicolumn{1}{c|}{\cellcolor[HTML]{FFFFFF}0.5391}               & \multicolumn{1}{c|}{\cellcolor[HTML]{FFFFFF}0.6475}               & \multicolumn{1}{c|}{\cellcolor[HTML]{FFFFFF}0.6038}               & \multicolumn{1}{c|}{\cellcolor[HTML]{FFFFFF}0.1321}               & \multicolumn{1}{c|}{\cellcolor[HTML]{FFFFFF}0.6843}               & \multicolumn{1}{c|}{\cellcolor[HTML]{FFFFFF}0.6333}               & \multicolumn{1}{c|}{\cellcolor[HTML]{FFFFFF}0.5491}               & 0.5727               \\ \cline{3-11} 
\multicolumn{1}{|c|}{\cellcolor[HTML]{FFFFFF}}                                                                    & \multicolumn{1}{c|}{\multirow{-2}{*}{\cellcolor[HTML]{FFFFFF}50-60}}                                            & \multicolumn{1}{c|}{\cellcolor[HTML]{FFFFFF}BAM$\sim$LCS+Humidity} & \multicolumn{1}{c|}{\cellcolor[HTML]{FFFFFF}0.5430}               & \multicolumn{1}{c|}{\cellcolor[HTML]{FFFFFF}0.6495}               & \multicolumn{1}{c|}{\cellcolor[HTML]{FFFFFF}0.6041}               & \multicolumn{1}{c|}{\cellcolor[HTML]{FFFFFF}0.1338}               & \multicolumn{1}{c|}{\cellcolor[HTML]{FFFFFF}0.6843}               & \multicolumn{1}{c|}{\cellcolor[HTML]{FFFFFF}0.6347}               & \multicolumn{1}{c|}{\cellcolor[HTML]{FFFFFF}0.5586}               & 0.5758               \\ \cline{2-11} 
\multicolumn{1}{|c|}{\cellcolor[HTML]{FFFFFF}}                                                                    & \multicolumn{1}{c|}{\cellcolor[HTML]{FFFFFF}}                                                                   & \multicolumn{1}{c|}{\cellcolor[HTML]{FFFFFF}BAM$\sim$LCS}          & \multicolumn{1}{c|}{\cellcolor[HTML]{FFFFFF}0.7650}               & \multicolumn{1}{c|}{\cellcolor[HTML]{FFFFFF}0.7536}               & \multicolumn{1}{c|}{\cellcolor[HTML]{FFFFFF}0.6468}               & \multicolumn{1}{c|}{\cellcolor[HTML]{FFFFFF}0.2053}               & \multicolumn{1}{c|}{\cellcolor[HTML]{FFFFFF}0.6747}               & \multicolumn{1}{c|}{\cellcolor[HTML]{FFFFFF}0.7251}               & \multicolumn{1}{c|}{\cellcolor[HTML]{FFFFFF}0.7043}               & 0.7140               \\ \cline{3-11} 
\multicolumn{1}{|c|}{\cellcolor[HTML]{FFFFFF}}                                                                    & \multicolumn{1}{c|}{\multirow{-2}{*}{\cellcolor[HTML]{FFFFFF}60-70}}                                            & \multicolumn{1}{c|}{\cellcolor[HTML]{FFFFFF}BAM$\sim$LCS+Humidity} & \multicolumn{1}{c|}{\cellcolor[HTML]{FFFFFF}0.7659}               & \multicolumn{1}{c|}{\cellcolor[HTML]{FFFFFF}0.7579}               & \multicolumn{1}{c|}{\cellcolor[HTML]{FFFFFF}0.6496}               & \multicolumn{1}{c|}{\cellcolor[HTML]{FFFFFF}0.2489}               & \multicolumn{1}{c|}{\cellcolor[HTML]{FFFFFF}0.6756}               & \multicolumn{1}{c|}{\cellcolor[HTML]{FFFFFF}0.7254}               & \multicolumn{1}{c|}{\cellcolor[HTML]{FFFFFF}0.7049}               & 0.7185               \\ \cline{2-11} 
\multicolumn{1}{|c|}{\cellcolor[HTML]{FFFFFF}}                                                                    & \multicolumn{1}{c|}{\cellcolor[HTML]{FFFFFF}}                                                                   & \multicolumn{1}{c|}{\cellcolor[HTML]{FFFFFF}BAM$\sim$LCS}          & \multicolumn{1}{c|}{\cellcolor[HTML]{FFFFFF}0.5010}               & \multicolumn{1}{c|}{\cellcolor[HTML]{FFFFFF}0.4462}               & \multicolumn{1}{c|}{\cellcolor[HTML]{FFFFFF}0.2911}               & \multicolumn{1}{c|}{\cellcolor[HTML]{FFFFFF}0.0004}               & \multicolumn{1}{c|}{\cellcolor[HTML]{FFFFFF}0.2579}               & \multicolumn{1}{c|}{\cellcolor[HTML]{FFFFFF}0.4293}               & \multicolumn{1}{c|}{\cellcolor[HTML]{FFFFFF}0.5022}               & 0.5232               \\ \cline{3-11} 
\multicolumn{1}{|c|}{\cellcolor[HTML]{FFFFFF}}                                                                    & \multicolumn{1}{c|}{\multirow{-2}{*}{\cellcolor[HTML]{FFFFFF}70-80}}                                            & \multicolumn{1}{c|}{\cellcolor[HTML]{FFFFFF}BAM$\sim$LCS+Humidity} & \multicolumn{1}{c|}{\cellcolor[HTML]{FFFFFF}0.5116}               & \multicolumn{1}{c|}{\cellcolor[HTML]{FFFFFF}0.4636}               & \multicolumn{1}{c|}{\cellcolor[HTML]{FFFFFF}0.3378}               & \multicolumn{1}{c|}{\cellcolor[HTML]{FFFFFF}0.0104}               & \multicolumn{1}{c|}{\cellcolor[HTML]{FFFFFF}0.2939}               & \multicolumn{1}{c|}{\cellcolor[HTML]{FFFFFF}0.4399}               & \multicolumn{1}{c|}{\cellcolor[HTML]{FFFFFF}0.5115}               & 0.5289               \\ \cline{2-11} 
\multicolumn{1}{|c|}{\cellcolor[HTML]{FFFFFF}}                                                                    & \multicolumn{1}{c|}{\cellcolor[HTML]{FFFFFF}}                                                                   & \multicolumn{1}{c|}{\cellcolor[HTML]{FFFFFF}BAM$\sim$LCS}          & \multicolumn{1}{c|}{\cellcolor[HTML]{FFFFFF}0.3358}               & \multicolumn{1}{c|}{\cellcolor[HTML]{FFFFFF}0.3588}               & \multicolumn{1}{c|}{\cellcolor[HTML]{FFFFFF}0.3745}               & \multicolumn{1}{c|}{\cellcolor[HTML]{FFFFFF}0.1901}               & \multicolumn{1}{c|}{\cellcolor[HTML]{FFFFFF}0.3132}               & \multicolumn{1}{c|}{\cellcolor[HTML]{FFFFFF}0.4042}               & \multicolumn{1}{c|}{\cellcolor[HTML]{FFFFFF}0.4018}               & 0.4241               \\ \cline{3-11} 
\multicolumn{1}{|c|}{\cellcolor[HTML]{FFFFFF}}                                                                    & \multicolumn{1}{c|}{\multirow{-2}{*}{\cellcolor[HTML]{FFFFFF}80-90}}                                            & \multicolumn{1}{c|}{\cellcolor[HTML]{FFFFFF}BAM$\sim$LCS+Humidity} & \multicolumn{1}{c|}{\cellcolor[HTML]{FFFFFF}0.3359}               & \multicolumn{1}{c|}{\cellcolor[HTML]{FFFFFF}0.3588}               & \multicolumn{1}{c|}{\cellcolor[HTML]{FFFFFF}0.3820}               & \multicolumn{1}{c|}{\cellcolor[HTML]{FFFFFF}0.1902}               & \multicolumn{1}{c|}{\cellcolor[HTML]{FFFFFF}0.3157}               & \multicolumn{1}{c|}{\cellcolor[HTML]{FFFFFF}0.4044}               & \multicolumn{1}{c|}{\cellcolor[HTML]{FFFFFF}0.4053}               & 0.4329               \\ \cline{2-11} 
\multicolumn{1}{|c|}{\cellcolor[HTML]{FFFFFF}}                                                                    & \multicolumn{1}{c|}{\cellcolor[HTML]{FFFFFF}}                                                                   & \multicolumn{1}{c|}{\cellcolor[HTML]{FFFFFF}BAM$\sim$LCS}          & \multicolumn{1}{c|}{\cellcolor[HTML]{FFFFFF}0.3645}               & \multicolumn{1}{c|}{\cellcolor[HTML]{FFFFFF}0.3583}               & \multicolumn{1}{c|}{\cellcolor[HTML]{FFFFFF}0.1876}               & \multicolumn{1}{c|}{\cellcolor[HTML]{FFFFFF}0.3681}               & \multicolumn{1}{c|}{\cellcolor[HTML]{FFFFFF}0.1870}               & \multicolumn{1}{c|}{\cellcolor[HTML]{FFFFFF}0.2750}               & \multicolumn{1}{c|}{\cellcolor[HTML]{FFFFFF}0.3552}               & 0.2400               \\ \cline{3-11} 
\multicolumn{1}{|c|}{\multirow{-10}{*}{\cellcolor[HTML]{FFFFFF}200-300}}                                          & \multicolumn{1}{c|}{\multirow{-2}{*}{\cellcolor[HTML]{FFFFFF}90-100}}                                           & \multicolumn{1}{c|}{\cellcolor[HTML]{FFFFFF}BAM$\sim$LCS+Humidity} & \multicolumn{1}{c|}{\cellcolor[HTML]{FFFFFF}0.3792}               & \multicolumn{1}{c|}{\cellcolor[HTML]{FFFFFF}0.3680}               & \multicolumn{1}{c|}{\cellcolor[HTML]{FFFFFF}0.1926}               & \multicolumn{1}{c|}{\cellcolor[HTML]{FFFFFF}0.3692}               & \multicolumn{1}{c|}{\cellcolor[HTML]{FFFFFF}0.1905}               & \multicolumn{1}{c|}{\cellcolor[HTML]{FFFFFF}0.2928}               & \multicolumn{1}{c|}{\cellcolor[HTML]{FFFFFF}0.3663}               & 0.3001               \\ \hline
\end{tabular}%
}
\caption{$R^2$ values for different LCS calibration models (hourly averaged data).}
\label{tab:R2 for different models}
\end{table}
\clearpage
%
%
\noindent\textbf{Confidence Interval for LCS Measurement}
\vspace{3mm}\\
\noindent For a given LCS measurement in the field, the $95\%$ confidence interval of this measurement is given by:
\renewcommand{\theequation}{S\arabic{equation}}
\begin{equation}\label{eq:CI}
    \hat{y}_o - t_{\alpha/2,n-2} \sqrt{\text{MSE}\left[\frac{1}{n}+\frac{(x_o-\Bar{x})^2}{S_{xx}}\right]} \leq y_o \leq \hat{y}_o + t_{\alpha/2,n-2} \sqrt{\text{MSE}\left[\frac{1}{n}+\frac{(x_o-\Bar{x})^2}{S_{xx}}\right]}
\end{equation}
\vspace{2mm}\\
\noindent\textbf{Example Calculation of Prediction Interval and Confidence Interval}
\vspace{3mm}\\
\noindent The example calculation below is carried out for reference BAM $\text{PM}_{2.5}$ level $y_m=93.2~\mu \text{g/m}^3$ and Sensirion-1 measurement $x_m=147~\mu \text{g/m}^3$.\\


\noindent The linear regression fit based on the logarithmically transformed variables $x_o=\log(x)$ and $y_o=\log(y)$ for the data (assuming $y\leq 100~\mu \text{g/m}^3$ as this range represents the majority of ambient $\text{PM}_{2.5}$ concentrations observed in most parts of the world.) is 
\begin{equation*}
y_o= 0.8259 x_o + 0.4058        
\end{equation*}

\noindent Applying logarithm on the BAM and LCS measurement: $\ln(x_m)= x_{o,m}= 4.99043259$, $\ln(y_m)= y_{o,m}= 4.534748$.\\

\noindent The estimated $\hat{y}_{o,m}$ for $x_{o,m}$ based on the linear regression fit is
\begin{equation*}
    \hat{y}_{o,m} = 0.8259\times 4.99043259 + 0.4058 = 4.527398273
\end{equation*}

\noindent With $n= 978$, $\alpha= 0.95$, $\Bar{x}= 4.366082523$, $S_{xx} = 326.1566$, $\text{MSE} = 0.02793705$, $t_{\alpha/2,n-2} = 1.962398$, and using (9) (from main manuscript) \textbf{prediction interval} for the LCS measurement $x_m$ is calculated as
\begin{equation*}
    4.19903208 \leq y_{o,m} \leq 4.85576446
\end{equation*}
Now taking antilogarithm, we obtain the prediction interval as
\begin{align*}
    \exp(4.19903208) & \leq y_{m} \leq \exp(4.85576446)\\
    66.62~\mu \text{g/m}^3 & \leq y_{m} \leq 128.39~\mu \text{g/m}^3
\end{align*}
Using~\eqref{eq:CI}, \textbf{confidence interval} for the LCS measurement $x_m$ as is obtained as
\begin{equation*}
    4.51195127 \leq y_{o,m} \leq 4.54284527    
\end{equation*}
Now taking antilogarithm, we obtain the confidence interval as
\begin{align*}
    \exp(4.51195127) & \leq y_{m} \leq \exp(4.54284527)\\
    91.09~\mu \text{g/m}^3 & \leq y_{m} \leq 93.95~\mu \text{g/m}^3
\end{align*}
\subsection*{Machine Learning Model for Prediction}
We consider three different machine learning (ML) models for the purpose of correcting the LCS measurements: random forest (RF), K-nearest neighbor (KNN), and extreme gradient boosting (XGBoost). The BAM measurement serves as the reference value. A random sub-sampling method was used, employing an $80\%:20\%$ split for the training and test data sets. The results are listed in Table~\ref{table:ml_comparison}.
\renewcommand{\tablename}{Table}
\renewcommand{\thetable}{S\arabic{table}}
\begin{table*}
\centering
\begin{tabular}{|c|cll|cll|cll|}
\hline
& \multicolumn{3}{c|}{Sensirion - 1} & \multicolumn{3}{c|}{Sensirion - 2} & \multicolumn{3}{c|}{Sensirion - 3}\\ 
\hline
\textbf{Model} & \multicolumn{1}{c|}{$R^2$} & \multicolumn{1}{c|}{MAE} & \multicolumn{1}{c|}{RMSE} & \multicolumn{1}{c|}{$R^2$} & \multicolumn{1}{c|}{MAE} & \multicolumn{1}{c|}{RMSE} & \multicolumn{1}{c|}{$R^2$} & \multicolumn{1}{c|}{MAE} & \multicolumn{1}{c|}{RMSE}\\ 
\hline
LR & \multicolumn{1}{c|}{0.919}       & \multicolumn{1}{c|}{17.205}       & \multicolumn{1}{c|}{22.72}         & \multicolumn{1}{c|}{0.913}       & \multicolumn{1}{c|}{17.27}        & \multicolumn{1}{c|}{23.48}         & \multicolumn{1}{c|}{0.697}       & \multicolumn{1}{c|}{31.57}        & \multicolumn{1}{c|}{39.01}         \\ \hline
RF & \multicolumn{1}{c|}{0.926}       & \multicolumn{1}{c|}{15.449}       & \multicolumn{1}{c|}{21.78}         & \multicolumn{1}{c|}{0.925}       & \multicolumn{1}{c|}{15.26}        & \multicolumn{1}{c|}{21.72}         & \multicolumn{1}{c|}{0.785}       & \multicolumn{1}{c|}{24.46}        & \multicolumn{1}{c|}{32.81}         \\ \hline
XGBoost & \multicolumn{1}{c|}{0.923}       & \multicolumn{1}{c|}{15.603}       & \multicolumn{1}{c|}{22.24}         & \multicolumn{1}{c|}{0.920}        & \multicolumn{1}{c|}{15.85}        & \multicolumn{1}{c|}{22.51}         & \multicolumn{1}{c|}{0.763}       & \multicolumn{1}{c|}{25.18}        & \multicolumn{1}{c|}{34.50}          \\ \hline
KNN & \multicolumn{1}{c|}{0.924}       & \multicolumn{1}{c|}{15.610}        & \multicolumn{1}{c|}{22.01}         & \multicolumn{1}{c|}{0.923}       & \multicolumn{1}{c|}{5.61}         & \multicolumn{1}{c|}{22.14}         & \multicolumn{1}{c|}{0.777}       & \multicolumn{1}{c|}{24.76}        & \multicolumn{1}{c|}{33.45}         \\ \hline
\multicolumn{1}{|l|}{} & \multicolumn{3}{c|}{Honeywell- 1} & \multicolumn{3}{c|}{Honeywell- 2} & \multicolumn{3}{c|}{Honeywell- 3}\\ 
\hline
\textbf{Model} & \multicolumn{1}{c|}{$R^2$} & \multicolumn{1}{c|}{MAE} & \multicolumn{1}{c|}{RMSE} & \multicolumn{1}{c|}{$R^2$} & \multicolumn{1}{c|}{MAE} & \multicolumn{1}{c|}{RMSE} & \multicolumn{1}{c|}{$R^2$} & \multicolumn{1}{c|}{MAE} & \multicolumn{1}{c|}{RMSE}\\ 
\hline
LR & \multicolumn{1}{l|}{0.907}  & \multicolumn{1}{l|}{18.1} & 24.58 & \multicolumn{1}{l|}{0.914} & \multicolumn{1}{l|}{18.1} & 23.31 & \multicolumn{1}{l|}{0.9} & \multicolumn{1}{l|}{18.97} & 24.94\\ 
\hline
RF & \multicolumn{1}{l|}{0.915} & \multicolumn{1}{l|}{16.587} & 23.39 & \multicolumn{1}{l|}{0.921} & \multicolumn{1}{l|}{16.96} & 22.44 & \multicolumn{1}{l|}{0.903} & \multicolumn{1}{l|}{18.07} & 24.55\\ 
\hline
XGBoost & \multicolumn{1}{l|}{0.907}       & \multicolumn{1}{l|}{17.47}        & 24.5                               & \multicolumn{1}{l|}{0.91}        & \multicolumn{1}{l|}{17.83}        & 23.91                              & \multicolumn{1}{l|}{0.902}       & \multicolumn{1}{l|}{18.52}        & 24.76  \\ 
\hline
KNN & \multicolumn{1}{l|}{0.906}       & \multicolumn{1}{l|}{17.655}       & 24.61                              & \multicolumn{1}{l|}{0.917}       & \multicolumn{1}{l|}{17.51}        & 22.91                              & \multicolumn{1}{l|}{0.905}       & \multicolumn{1}{l|}{18.09} & 24.37\\ 
\hline
\multicolumn{1}{|l|}{} & \multicolumn{3}{c|}{Plantower - 1} & \multicolumn{3}{c|}{Plantower - 2} & \multicolumn{3}{c|}{Plantower - 3}\\ 
\hline
\textbf{Model}         & \multicolumn{1}{c|}{$R^2$} & \multicolumn{1}{c|}{MAE} & \multicolumn{1}{c|}{RMSE} & \multicolumn{1}{c|}{$R^2$} & \multicolumn{1}{c|}{MAE} & \multicolumn{1}{c|}{RMSE} & \multicolumn{1}{c|}{$R^2$} & \multicolumn{1}{c|}{MAE} & \multicolumn{1}{c|}{RMSE}\\ 
\hline
LR & \multicolumn{1}{l|}{0.857}       & \multicolumn{1}{l|}{21.649}       & 29.65 & \multicolumn{1}{l|}{0.788}       & \multicolumn{1}{l|}{24.57}        & 36.67 & \multicolumn{1}{l|}{0.873}       & \multicolumn{1}{l|}{21.13}        & 29.08\\ 
\hline
RF & \multicolumn{1}{l|}{0.848}       & \multicolumn{1}{l|}{22.151}       & 30.65                              & \multicolumn{1}{l|}{0.789}       & \multicolumn{1}{l|}{24.4}         & 36.63   & \multicolumn{1}{l|}{0.872}       & \multicolumn{1}{l|}{21.27} & 29.12\\ 
\hline
XGBoost & \multicolumn{1}{l|}{0.843}       & \multicolumn{1}{l|}{22.601}       & 31.1  & \multicolumn{1}{l|}{0.779}       & \multicolumn{1}{l|}{24.67}        & 37.47                              & \multicolumn{1}{l|}{0.868}       & \multicolumn{1}{l|}{21.66}        & 29.6                               \\ \hline
KNN & \multicolumn{1}{l|}{0.836}       & \multicolumn{1}{l|}{22.818}       & 31.78                              & \multicolumn{1}{l|}{0.777}       & \multicolumn{1}{l|}{25.13}        & 37.58                              & \multicolumn{1}{l|}{0.868}       & \multicolumn{1}{l|}{21.68} & 29.62\\ 
\hline
\end{tabular}
\caption{Comparison of $R^{2}$, MAE, and RMSE for test dataset with ML models used to correct LCS measurements using reference measurements.}
\label{table:ml_comparison}
\end{table*}

XGBoost uses distributed gradient-boosted decision trees for regression. It was implemented using the open-source xgboost Python library. KNN is a non-parametric method used for classification and regression. In this study, KNeighborsRegressor from sklearn.neighbors was used with hyperparameter tuning for '\text{n\_neighbors}':[3,5,7,9,11], 'weights':['uniform','distance'], and 'p':[1,2]. RF is a supervised ensemble model that uses a combination of decision trees. It is useful as each decision tree theoretically isolates errors. In this study, the RandomForestRegressor from sklearn.ensemble was used and optimized using grid search (maximum features: 1 and maximum depth: 1,2,3,4,5). It is seen from Table~\ref{table:ml_comparison} that XGBoost achieves competitive performance, often close to KNN, with $R^{2}$ values in the range $0.76$-$0.92$. KNN achieves second best performance, with $R^{2}$ values in the range $0.77$-$0.92$. RF outperforms the other ML models, with $R^{2}$ values ranging from $0.78$-$0.92$, showing its robustness across different sensor types.

Implementing ML models on a microcontroller involves varying levels of complexity due to resource constraints. On the other hand, linear regression (LR) has low complexity and requires minimal computational resources, making it suitable for microcontrollers with limited processing power and memory. RF involves multiple decision trees and is computationally intensive, requiring high memory for storing nodes and making it challenging for microcontroller-based implementation. KNN requires storing the entire training dataset and computing distances for each prediction, which is resource-intensive in terms of memory and computation. XGBoost involves gradient-boosted decision trees, which are very computationally and memory intensive. General considerations for deploying these algorithms include memory constraints, limited computational power, energy consumption, and the need for real-time performance. Optimization techniques such as model compression, approximation algorithms, and efficient coding practices are essential for making these algorithms practical for microcontroller deployment.
\section*{Data Availability}
\noindent The raw data set recorded as part of this study has been made available at this link (in excel format): \url{https://t.ly/oXmuN}. 

\bibliography{bibs}